\documentclass[10pt,preprintnumbers,nofootinbib]{revtex4-2}%
\usepackage[latin9]{inputenc}
\usepackage{amsmath}
\usepackage{amssymb}
\usepackage{graphicx}
\usepackage[colorlinks=true, pdfstartview=FitV, linkcolor=blue, citecolor=red, urlcolor=magenta]{hyperref}
\usepackage{amsfonts}
\usepackage{subfigure}
\usepackage{cleveref}
\usepackage{listings}
\usepackage{array}
\usepackage{url}
\usepackage{comment}
\usepackage{todonotes}

\usepackage{tensor}

\usepackage{xcolor}
\usepackage{soul}
\allowdisplaybreaks

\graphicspath{{Figures/}}

\begin{document}

\title{Constructing noncommutative black holes}

\author{Tajron Juri\'c}
\email{tjuric@irb.hr}
\author{A. Naveena Kumara}
\email{nathith@irb.hr}
\author{Filip Po\v{z}ar}
\email{filip.pozar@irb.hr}
\affiliation{Rudjer Bo\v{s}kovi\'c Institute, Bijeni\v cka  c.54, HR-10000 Zagreb, Croatia}

\begin{abstract}

We present a self-contained and consistent formulation of noncommutative (NC) gauge theory of gravity, focusing on spherically symmetric black hole geometries. Our construction starts from the gauge-theoretic viewpoint of Poincar\'{e} (or de Sitter) gravity and introduces noncommutativity through the Moyal star product and the Seiberg-Witten map, retaining NC gauge invariance at each order in the deformation parameter $\Theta$. Working systematically to second order in $\Theta$, we obtain explicit NC corrections to the spin connection, the vierbein, and various geometric objects such as the metric and curvature scalars. Using these results, we compute NC modifications of four-dimensional Schwarzschild and Reissner-Nordstr\"{o}m solutions, including scenarios with a cosmological constant, as well as three-dimensional BTZ-type black holes (both uncharged and charged). For each black hole solution, we explore various possible Moyal twists, each of which generally breaks some symmetries and modifies the horizon structure, surface gravity, and curvature invariants. In particular, we show that while the radial location of horizons in Schwarzschild-like solutions remains unchanged for some twists, other twists introduce important but finite deformations in curvature scalars and can decouple the Killing horizon from the causal horizon. Similar patterns arise in the charged and lower-dimensional cases. Beyond constructing explicit examples, our approach provides a blueprint for systematically incorporating short-distance quantum corrections through noncommutativity in gravitational settings. The methods and expansions we present can be extended to more general geometries including rotating black holes and additional matter fields, offering a broad framework for future studies of NC effects in classical solutions of general relativity.

\end{abstract}
\maketitle

\section{Introduction}

The quest for a unified theoretical framework that encompasses all fundamental interactions remains a central theme in modern physics. In the Standard Model, three of nature's four known forces, electromagnetism, the weak interaction, and the strong interaction, are described by a common gauge-theoretic framework. In contrast, the gravitational interaction, elegantly captured by Einstein's General Relativity (GR), appears in a distinctly geometric form that does not straightforwardly merge with quantum field theory. Although GR's geometrical framework has been remarkably successful in explaining macroscopic gravitational phenomena, including the recent detection of gravitational waves \cite{Abbott:2016blz, TheLIGOScientific:2016src, Abbott:2016nmj} and the imaging of a black hole shadow by the Event Horizon Telescope (EHT) \cite{Akiyama:2019bqs, Akiyama:2019cqa, Akiyama:2019fyp, EventHorizonTelescope:2022wkp, EventHorizonTelescope:2022xqj, Vagnozzi:2022moj}, its disjunction with the principles of quantum mechanics leaves a significant conceptual gap at very short distances or very high energies.

One line of investigation reformulates gravity itself as a gauge theory, aiming to embed Einstein's theory within the broader gauge principle that underpins the Standard Model \cite{Utiyama:1956sy, Kibble:1961ba, Stelle:1979aj, MacDowell:1977jt, Ivanov:1981wn, Ivanov:1981wm, Kaku:1977pa, Fradkin:1985am, Chamseddine:1976bf} (see  \cite{Chamseddine:2005td, Manolakos:2019fle} for a brief overview). In the case of gravity, one faces the challenge of describing the two polarizations of the massless graviton using a symmetric four-dimensional tensor that, at first glance, possesses ten independent components. An elegant way to handle this redundancy was pioneered by Weyl~\cite{weyl1952space} and Cartan~\cite{cartan1981theory}, and later given a concrete gauge-theoretic interpretation by Utiyama~\cite{Utiyama:1956sy} and Kibble~\cite{Kibble:1961ba}. Rather than working solely with the metric, these works introduced the vierbein $e_{\mu}^a$ and the spin connection $\omega_{\mu}^{\,ab}$, imposing both local Lorentz invariance and diffeomorphism invariance. The spin connection acts as the gauge field for local Lorentz transformations, while the vierbein serves as a soldering form between internal indices and spacetime. Imposing that the generalized field strength for translations vanish identifies the usual Christoffel symbols with derivatives of the vierbein, thus recovering the standard metric formulation. This gauge-based viewpoint not only simplifies coupling fermions to gravity but also underpins more advanced frameworks such as loop quantum gravity~\cite{Ashtekar:1987gu,Smolin:2004sx} and supergravity \cite{Freedman:2012zz}. Despite important distinctions from Yang-Mills theories-particularly in how translations are gauged, this approach places gravitational interactions on a footing similar to other gauge forces, offering powerful insights into the geometric and dynamical structure of spacetime.

A separate but related direction explores the possibility that spacetime coordinates fail to commute at very short distances, suggesting a noncommutative (NC) geometry. Traces of this idea date back to the original Snyder model \cite{Snyder:1946qz}, motivated by the hope that quantum uncertainties in spacetime could cure ultraviolet divergences. Modern interest has been reignited by string theory, where open strings moving in a background antisymmetric $B$-field can produce an effective NC structure~\cite{Connes:1997cr, Douglas:1997fm, Cheung:1998nr, Chu:1998qz, Schomerus:1999ug, Ardalan:1998ce, Seiberg:1999vs}. Concretely, the canonical commutator
\begin{equation*}
[x_\mu\stackrel{\star}{,}x_\nu] = i\,\Theta_{\mu\nu}
\end{equation*}
introduces a constant antisymmetric matrix $\Theta_{\mu\nu}$ that controls noncommutativity. In turn, ordinary products of fields $f(x)g(x)$ are replaced by the Moyal (or $\star$) product,
\begin{equation}
    (f \star g)(x) = \lim_{y\to x} 
    \exp\Bigl(-\tfrac{i}{2}\,\Theta^{\mu\nu}
    \tfrac{\partial}{\partial x^\mu}\tfrac{\partial}{\partial y^\nu}\Bigr)
    f(x)\,g(y)\;,
\label{moyalstar}
\end{equation}
To preserve gauge invariance, the \emph{Seiberg--Witten map} systematically expands the NC fields in terms of commutative ones, ensuring that symmetries remain well-defined at each order in $\Theta^{\mu\nu}$~\cite{Seiberg:1999vs}. A similar approach to NC gauge theory of gravity with the Seiberg-Witten map was performed in \cite{Banerjee:2007th} but with a different star product compared to \eqref{moyalstar}.

Constructing a gauge theory of gravity on a NC manifold entails combining the gauge perspective of the de Sitter or Poincar\'{e} group with the $\star$-product formalism. However, formulating gravity on a NC spacetime has proven more challenging than extending flat-space gauge theories to the NC domain. The interplay among local Lorentz invariance, the notion of a metric-compatible connection, and general coordinate transformations becomes more complicated with the $\star$-product. Even so, in \cite{Chamseddine:2000si} it was proposed that one could gauge the de Sitter group $\text{SO}(4,1)$ and then contract to the Poincar\'{e} group, introducing a NC extension via the Seiberg--Witten map. The field strengths, torsion, and curvature tensors acquire corrections in powers of $\Theta^{\mu\nu}$, and the resulting action can be interpreted as a NC extension of Einstein-Hilbert gravity-provided one accounts for additional source terms (e.g., gauge-covariant potentials, possible cosmological constants, and matter couplings) needed to recover standard gravitational phenomena.

Although the underlying mechanism appears elegant, explicit computations of the NC fields at second or higher orders have revealed missing terms or incorrect coefficients. Such omissions affect physical predictions, including the form of the deformed metric, curvature invariants, and horizon structures. In this paper, we revisit this construction to provide a fully self-consistent and corrected formulation of NC gauge theory of gravity in four dimensions, clarifying and resolving various conceptual and computational errors found in the literature. We systematically apply the Seiberg--Witten map to the gauge fields $\omega_\mu^{AB}$ (the spin connection $\omega_\mu^{ab}$ and the tetrad $e_\mu^a$) up to second order in $\Theta^{\mu\nu}$. Concretely, we write \cite{Chamseddine:2000si}
\begin{equation*}
\hat{\omega}_\mu^{AB}(\Theta)
= \omega_\mu^{AB} 
+ \omega_\mu^{AB\,(1)}(\Theta) 
+ \omega_\mu^{AB\,(2)}(\Theta^2)
+ \dots\,,
\end{equation*}
ensuring that local gauge invariance is preserved at each step:
\begin{equation*}
\delta_{\hat{\lambda}} \hat{\omega}_\mu^{AB}
= \partial_\mu \hat{\lambda}^{AB}
+ [\hat{\omega}_\mu \stackrel{\star}{,} \,\hat{\lambda}]^{AB}.
\end{equation*}
Along the way, we highlight a missing term in the second-order NC tetrad and correct a range of faulty factors often cited in the present-day literature \cite{Chamseddine:2000si,Chaichian:2007dr, Chaichian:2007we, Mukherjee:2007fa, Linares:2019gqf, Touati:2023ubi, Touati:2023cxy, Touati:2022kuf, Touati:2022zbm, Touati:2021eem, Zhao:2023uam, Heidari:2023egu, Heidari:2023bww, Touati:2024kmv, AraujoFilho:2025viz}.

To demonstrate the physical relevance of these corrections, we apply the formalism to black hole spacetimes. Following \cite{Zet:2003bv, zet2006computer}, we revisit the Schwarzschild solution (commutative) and treat the vierbein $e_\mu^a$ in spherical coordinates. In these works, the authors adopt an approach in which the de Sitter group acts as an \emph{internal} transformation, leaving the underlying coordinates untouched (see also \cite{Wiesendanger:1995hm}). By choosing a four-dimensional Minkowski spacetime (in spherical coordinates) as the base manifold, they interpret the gravitational field in terms of the gauge potentials associated with the de Sitter group. This viewpoint highlights how gauge transformations are realized without actively transforming spacetime coordinates, thereby maintaining a clear separation between the manifold's geometry and the internal gauge structure. Their results show that Schwarzschild-type solutions can emerge naturally from this internal formulation. We first briefly introduce the $\mathrm{SO}(4,1)$ theory in the spirit of \cite{Zet:2003bv}, reproduce some results of \cite{Zet:2003bv}, and discuss the consequences of contracting $\mathrm{SO}(4,1)$ to $\mathrm{ISO}(3,1)$ via the Wigner--\.{I}n\"on\"u contraction. We discuss that the de Sitter theory action that \cite{zet2006computer} introduces is not invariant to the group of de Sitter gauge transformations. After contraction, the action becomes invariant to $\mathrm{ISO}(3,1)$ gauge transformations and becomes the usual Poincare gauge theory of gravity with Poincare gauge fields. We argue that for constructing the NC gauge theory of gravity using the Seiberg-Witten map, starting with NC de Sitter gauge fields and contracting them to NC Poincar\'{e} is not strictly necessary,\footnote{One could also directly apply the Seiberg--Witten map to the Poincar\'{e} gauge fields.} but it is computationally clearer because in the de Sitter group all the generators are rotations labelled by a single set of indices, whereas in the Poincar\'{e} group there are two distinct sets of generators with their own indices. One conceptual issue in much of the current literature is the incorrect introduction of the cosmological constant $\Lambda$ into gauge theory. References \cite{Zet:2003bv, zet2006computer} (and also \cite{Chaichian:2007dr, Chaichian:2007we}) correctly observe that solutions of de Sitter field theory produce tetrads $e^a_\mu$ which yield metrics
\begin{equation}
    g_{\mu\nu} = \eta_{ab} e^a_\mu e^b_\nu
\end{equation}
that may or may not correspond to spacetimes with a cosmological constant, and also correctly relate
\begin{equation}
    \Lambda = -3k^2
\end{equation}
to the contraction parameter $k$. Hence, once one contracts the de Sitter group to Poincar\'{e} by setting $k\to 0$, one can no longer retain a cosmological constant in the solutions of the pure gauge theory. In this paper, we emphasize that the proper way to introduce a cosmological constant into the (contracted) Poincar\'{e} gauge theory is by adding suitable source terms (i.e., $\Lambda g_{\mu\nu}$), just as is done with $\kappa T_{\mu\nu}$ in electromagnetism \cite{Zet:2003bv, zet2006computer, Chaichian:2007dr} (see \cite{Mukherjee:2007fa} as well).\\

After setting up the commutative Poincare gauge theory of gravity, we move on to its NC extension. A natural approach is to build NC de Sitter gauge fields via the Moyal product and the Seiberg--Witten map, then contract the NC de Sitter gauge fields to obtain deformed gauge fields of an NC Poincar\'{e} gauge theory. The deformed NC Poincare gauge fields are then calculated from the commutative solutions that we presented. Indeed, \cite{Chaichian:2007dr, Chaichian:2007we} followed precisely this approach for NC Schwarzschild (and Schwarzschild--de~Sitter) black holes, by introducing a deformed metric
\begin{equation}\label{introeqn}
    \hat g_{\mu \nu} (x, \Theta)
    = \frac{1}{2}\,\eta_{ab}\,
    \bigl(\hat e_\mu^a \star \hat e_\nu^{b\dagger} 
    + \hat e_\nu^b \star \hat e_\mu^{a\dagger} \bigr)\;.
\end{equation}
However, the missing NC terms in \cite{Chamseddine:2000si}, along with conceptual gaps in \cite{Zet:2003bv, zet2006computer}, were inadvertently propagated into \cite{Chaichian:2007dr, Chaichian:2007we}. Additional errors, such as incorrect overall factors and an inconsistent expansion of the Seiberg--Witten map in \eqref{introeqn}, as well as misinterpretations of horizon structure, also appear in \cite{Chaichian:2007dr, Chaichian:2007we}, eventually filtering into subsequent analyses \cite{Mukherjee:2007fa, Linares:2019gqf, Touati:2023ubi, Touati:2023cxy, Touati:2022kuf, Touati:2022zbm, Touati:2021eem, Zhao:2023uam, Heidari:2023egu, Heidari:2023bww, Touati:2024kmv, AraujoFilho:2025viz}.

This article's scope is summarized as follows (in the order presented in subsequent sections):
\begin{itemize}
    \item We revisit the gauge theory of gravity introduced in \cite{Zet:2003bv, zet2006computer} and discuss how to properly recover the cosmological constant after the Wigner--\.{I}n\"on\"u contraction (Section \ref{sec1}).
    \item We examine the NC extension of Einstein gravity in \cite{Chamseddine:2000si} using the Seiberg--Witten map, fixing previously missing NC corrections in the NC vierbein, field tensors, inverse vierbein, etc.\ (Section \ref{sec2}).
    \item We revisit the NC black hole solutions derived in \cite{Chaichian:2007dr, Chaichian:2007we}, rectifying conceptual and computational errors (NC metrics, horizon structure, etc.; Section \ref{sec3}).
\end{itemize}

In the sections that follow, we provide the algebraic setup for de Sitter theory and Poincare gauge theory in four dimensions, outline the key aspects of the Seiberg--Witten map for $\omega_\mu^{ab}$ and $e_\mu^a$, and show how one contracts from $\mathrm{SO}(4,1)$ to $\mathrm{ISO}(3,1)$. We then compute the NC corrections to the tetrad, spin connection, and curvature components up to second order in $\Theta$. Once our corrected formalism is established, we apply it to black hole solutions, initially focusing on the Schwarzschild metric. We carry out explicit NC corrections to the metric components, the deformed curvature (NC gauge field strength), and the deformed curvature scalar under various Moyal twists (e.g., Abelian twists generated by commuting vector fields like $\partial_t \otimes \partial_\varphi$ or $\partial_r \otimes \partial_\theta$). In each instance, we explore how noncommutativity modifies the horizon structure, surface gravity, curvature singularities, and spherical symmetry (often broken). Our findings show that while the event horizon radius sometimes remains unchanged, other geometric features such as Killing horizons and curvature scalars may exhibit significant NC deformations. Moreover, we observe that only twists not involving the radial coordinate vector field $\partial_r$ yield metrics that remain nonsingular at the event horizon. In Appendices \ref{app1}, \ref{app2}, and \ref{app3}, we extend the formalism to more general solutions, such as Reissner--Nordstr\"{o}m, BTZ, and spacetimes with a cosmological constant.

Overall, this work aims to present a robust, internally consistent framework for NC gauge gravity and to showcase its implications for classical black hole spacetimes. By rectifying longstanding computational and conceptual oversights, we hope to establish a resource for future investigations of quantum-gravity-inspired corrections to black hole solutions, potentially shedding light on new high-energy phenomena near singularities or horizons. Our corrected expansions, along with explicit examples, pave the way for further studies into whether noncommutativity can soften singularities, alter black hole thermodynamics, or otherwise reveal new short-distance features of gravitational physics. The methods presented here can be readily adapted to more general geometries, including rotating black holes and additional matter sources, thus providing a broad platform for exploring noncommutative gravitational physics.

\section{(Commutative) De Sitter theory to Poincar\'{e} Gauge Theory} \label{sec1}

In this section, we recap the well-known formulation of gravity as a gauge theory, aiming to correct typographical errors that appeared in papers extending this theory to a NC gauge theory~\cite{Zet:2003bv, zet2006computer,Chaichian:2007dr, Chaichian:2007we}. As explained in the Introduction, the de Sitter generators are easier to work with in the Seiberg-Witten map than the Poincare group generators so in the NC section we will calculate the Seiberg-Witten map for any de Sitter invariant theory and contract it to the Seiberg-Witten map for Poincare invariant theories. In this section, we will define an attempt at generalizing gauge theory of gravity to this Lie group\footnote{This is an attempt in the context of usual, commutative gauge theory.}, which we will later contract to the Poincar\'{e} gauge theory. We will show that the de Sitter action \eqref{action} is $\mathrm{SO}(4,1)$ gauge variant (but is invariant under its $SO(3,1)$ subgroup) and as such is not a theory with $\mathrm{SO}(4,1)$ gauge symmetry, contrary to the etablished incorrect understanding. Never the less, we will solve its equation of motion and contract it to the solution of the Poincare gauge theory.\\

The de Sitter group $SO(4, 1)$, a 10-dimensional Lie group, has infinitesimal generators denoted by $M_{AB} = -M_{BA}$, where $A, B = 0, 1, 2, 3, 5$. After contraction, the generators $M_{AB}$ correspond to translations $P_a = M_{a5}$ and Lorentz rotations $M_{ab} = -M_{ba}$, where $a, b, \ldots = 0, 1, 2, 3$. The corresponding commutative gauge potentials, $\omega _\mu ^{AB} (x) = - \omega _\mu ^{BA} (x)$, are associated with the spin connection $\omega _\mu ^{ab}(x)=-\omega _\mu ^{ba}(x)$ and the tetrad fields $\omega _\mu ^{a5}(x)=k\, e^a_\mu (x)$, where $k$ is the contraction parameter. As $k \rightarrow 0$, the gauge group reduces to $\mathrm{ISO}(3,1)$, yielding the commutative Poincar\'{e} gauge theory of gravitation.

\medskip

Like any field theory, we define the de Sitter theory over Minkowski spacetime with coordinates $x^\mu = (t,r,\theta,\varphi)$,
\begin{equation}
    ds^2 = -dt^2 + dr^2 + r^2\bigl(d\theta^2 + \sin^2(\theta)\,d\varphi^2\bigr),
\end{equation}
as a theory with fields given by an $\mathfrak{so}(4,1)$-valued 1-form $\omega_\mu$ with components
\begin{equation}
    \omega^{AB}_\mu(x) = -\,\omega^{BA}_\mu(x)
\end{equation}
that transform in the adjoint representation of the de Sitter group. The antisymmetric indices $AB$ label all 10 generators, with each $A, B$ ranging\footnote{We will split the range $0,1,2,3,5$ into $a,5$, with $a\in\{0,1,2,3\}$.} over $\{0,1,2,3\}\cup\{5\}$. The curvature is an $\mathfrak{so}(4,1)$-valued 2-form with components of the usual form
\begin{equation}
    R^{AB}_{\mu\nu} = \partial_\mu \omega^{AB}_\nu \;-\; \partial_\nu \omega^{AB}_\mu 
    \;+\; \eta_{CD}\bigl(\omega^{AC}_\mu \,\omega^{DB}_\nu - \omega^{AC}_\nu \,\omega^{DB}_\mu\bigr)\;,
\label{curv}
\end{equation}
where $\eta_{CD} = \mathrm{diag}(-1,1,1,1,1) = \eta_{cd}\oplus\mathrm{diag}(1)$. Preparing for the contraction, we define
\begin{equation}
    \omega^{a5}_\mu = k\,e^a_\mu
\end{equation}
for $a\in\{0,1,2,3\}$. This rewriting separates the gauge curvature into objects closely related to geometric curvature and torsion:
\begin{equation}
    \begin{split}
        R^{ab}_{\mu\nu} &= \partial_\mu \omega^{ab}_\nu - \partial_\nu \omega^{ab}_\mu 
        + \eta_{cd}\bigl(\omega^{ac}_\mu\,\omega^{db}_\nu \;-\; \omega^{ac}_\nu\,\omega^{db}_\mu\bigr)
        + k^2 \bigl(e^a_\mu\, e^b_\nu - e^a_\nu\, e^b_\mu \bigr),\\
        R^{a5}_{\mu\nu} &= k\,T^{a}_{\mu\nu} 
        \;=\; k\bigl(\partial_\mu e^a_\nu - \partial_\nu e^a_\mu 
        + \bigl(\omega^{ac}_\mu\, e^d_\nu - \omega^{ac}_\nu\, e^d_\mu\bigr)\eta_{cd}\bigr).
    \end{split}
    \label{DS curvature}
\end{equation}
From \eqref{DS curvature}, one sees that taking $k\to 0$ is necessary because, without this contraction, the torsionless curvature $R^{ab}_{\mu\nu}$ would not correspond to the usual physical Riemann tensor.\\ 

The dynamics of the de Sitter theory are governed by the action
\begin{equation}
    S = \frac{1}{16\pi G}\int e\,R = \frac{1}{16\pi G}\int \det\bigl(e^a_\mu\bigr)\; R^{ab}_{\mu\nu}\; e^\mu_a\, e^\nu_b\;,
    \label{action}
\end{equation}
where $e^\mu_a$ is the inverse of $e^a_\mu = \frac{1}{k}\,\omega^{a5}_\mu$, satisfying
\begin{equation}
    e^b_\mu \,e^{\mu}_a = \delta^b_a\;, \qquad
    e^a_\nu \,e^{\mu}_a = \delta^\mu_\nu\;.
\end{equation}
As we see, the action \eqref{action} is built from a fully contracted $\mathfrak{iso}(3,1)$ index structure, and is, therefore, $\mathrm{ISO}(3,1)$ gauge-invariant after the contraction $k \rightarrow 0$. We can also see that the action \eqref{action} is constructed from specifically selected $A,B$ components of the de Sitter gauge fields $\omega^{AB}_\mu$, so a general de Sitter gauge transformation, which mixes the components and can change the notion of the $A=5$ internal index,  will not leave the action invariant in general. As such, \eqref{action} is not invariant under the group of de Sitter gauge transformations\footnote{Before the contraction, the action is not invariant to Poincare gauge transformations because the gauge curvature tensor is not the one for Poincare gauge group.}. Contracted solutions of the action \eqref{action} will correspond to Poincare gauge theory solutions. Finally, the considerations of a NC de Sitter field theory (even though \eqref{action} is not gauge invariant and as such is not a gauge theory) are important because the Seiberg-Witten expansion does not depend on the action, so its contraction will be applicable to the NC Poincare gauge theory whose investigation is our primary goal. It is really the case that we only use NC de Sitter theory for bookkeeping purposes. 
\medskip

Next, we outline how to solve the equations of motion of the de Sitter theory and incorporate source terms that allow solutions (in the contracted Poincar\'{e} gauge theory) carrying electric charge or cosmological constant. Varying \eqref{action} leads to the equations of motion (EOM) 
\begin{equation}
		\begin{split}
        &G^a_\mu \;=\; R^{ab}_{\mu\nu}\,e^{\nu}_b \;-\; \tfrac{1}{2}\,R^{cd}_{\mu\nu}\,e^{\mu}_c\, e^{\nu}_d\, e^a_\mu \;=\; 0 
        \quad\quad\quad\;\;\;\;\;\;\quad\;\;\bigl(\text{EOM for } e^a_\mu\bigr), \\
        &T^a_{\mu\nu} \;=\; \partial_\mu e^a_\nu - \partial_\nu e^a_\mu 
        + \eta_{bc}\bigl(\omega^{ab}_\mu\, e^c_\nu - \omega^{ab}_\nu\, e^c_\mu\bigr) \;=\; 0
        \quad\;(\text{EOM for } \omega^{ab}_\mu).
    \end{split}
    \label{EOM}
\end{equation}
The roles of the de Sitter tetrads and spin connections appear asymmetric in \eqref{EOM} because they enter the action \eqref{action} in different ways. Moreover, \eqref{action} imposes the Einstein equation as the tetrad EOM\footnote{With the Riemann tensor modified by the $k^2$ term in \eqref{DS curvature}.} when $k\to 0$. It also forces torsion to vanish as the spin-connection EOM, indicating that in the limit $k\to 0$ the theory reduces to Einstein's gravity, while for nonzero $k$ it is the de Sitter theory.

Following \cite{Chaichian:2007dr}, one obtains spherically symmetric solutions of the de Sitter theory by choosing the ansatz\footnote{The most general ansatz considered in literature \cite{Chaichian:2007dr} uses $e^2_\theta = rC(r)$ and $e^3_\phi = rC(r)\sin(\theta)$, but the Einstein equation quickly sets $C(r) = 1$ so for simplicity of the calculations, we are using the slightly less general ansatz.}
\begin{align}
    e_{\mu}^{0} &= \bigl(A, 0, 0, 0\bigr),  
    & e_{\mu}^{1} &= \bigl(0, \tfrac{1}{A}, 0, 0\bigr),
    & e_{\mu}^{2} &= \bigl(0, 0, r, 0\bigr),  
    & e_{\mu}^{3} &= \bigl(0, 0, 0, r \sin \theta\bigr),
    \label{ansatz1}
\end{align}
and
\begin{align}
    \omega _{\mu }^{01} &= \bigl(U, 0, 0, 0\bigr), 
    & \omega _{\mu}^{12} &= \bigl(0, 0, W, 0\bigr),  
    & \omega _{\mu}^{13} &= \bigl(0, 0, 0, Z \,\sin \theta \bigr), \nonumber \\
    \omega _{\mu }^{23} &= \bigl(V, 0, 0, -\cos \theta \bigr),  
    & \omega _{\mu}^{02} &= \omega _{\mu }^{03} = \bigl(0, 0, 0, 0\bigr),
    \label{ansatz2}
\end{align}
where $A,\,U,\,V,\,W$ and $Z$  are functions of the radial coordinate $r$ only. Substituting \eqref{ansatz1} and \eqref{ansatz2} into \eqref{DS curvature} yields the components of $T_{\mu \nu}^{a}$ and $R_{\mu \nu }^{ab}$. The nonvanishing components of the torsion are
\begin{align}
    {T}_{10}^{0} &=  A'-\tfrac{U}{A}, 
    & {T}_{21}^{2} &= -\tfrac{W}{A} - 1,\nonumber \\ 
    {T}_{21}^{2} &= -r\,\sin \theta \,V, 
    & {T}_{20}^{3} &= r\,V,\nonumber\\
    {T}_{13}^{3} &= -\tfrac{\sin \theta \,Z}{A} - \sin \theta,
    \label{2.7}
\end{align}
and those of the curvature are
\begin{align}
    R^{10}_{10} &= k^2 - U', 
    & R^{20}_{20} &= k^2\,r\,A + U\,W, \nonumber \\
    R^{21}_{21} &= \tfrac{k^2\, r}{A} + W', 
    & R^{21}_{30} &= V\,Z\,\sin \theta, \nonumber \\
    R^{30}_{30} &= \bigl(k^2\,r\,A + U\,Z\bigr)\sin \theta, 
    & R^{31}_{20} &= -V\,W, \nonumber \\
    R^{31}_{31} &= \Bigl(\tfrac{k^2\,r}{A} + Z'\Bigr)\sin \theta, 
    & R^{31}_{32} &= (Z-W)\,\cos \theta, \nonumber \\
    R^{32}_{10} &= -V', 
    & R^{32}_{32} &= \bigl(k^2\,r^2 - W\,Z + 1\bigr)\,\sin \theta.
\end{align}
Here, the prime denotes the derivative with respect to $r$. 
Solving the null-torsion equations in \eqref{EOM} yields the spin connection $\omega^{ab}_\mu$ in terms of the tetrad fields $e^a_\mu$:
\begin{equation} \label{spin_connection}
    \omega_{\mu}^{\ ab} 
    = \tfrac{1}{2}\, e^{\nu a}\,\bigl(\partial_\mu e_\nu^{\ b}- \partial_\nu e_\mu^{\ b}\bigr) 
    \;-\; \tfrac{1}{2}\, e^{\nu b}\,\bigl(\partial_\mu e_\nu^{\ a}- \partial_\nu e_\mu^{\ a}\bigr) 
    \;-\; \tfrac{1}{2}\,e^{\rho a}\,e^{\sigma b}\,\bigl(\partial_\rho e_{\sigma c}- \partial_\sigma e_{\rho c}\bigr)\,e_\mu^{\ c},
\end{equation}
which constrains the radial functions in \eqref{ansatz2}:
\begin{align}
U &= A\,A', \qquad 
V = 0, \qquad  
W = Z = -\,A.
\label{2.9}
\end{align}
Finally, the remaining unknown function $A(r)$ follows from the Einstein part of the EOM. The nonzero components of the Einstein tensor are:
\begin{equation}
\begin{split}
	G^0_0 &= - 3\,k^{2}\,A 
	\;+\; \frac{2\,A^2\,A'}{r} 
	\;+\; \frac{A^3}{r^{2}} 
	\;-\; \frac{A}{r^{2}},\\
	G^1_1 &= - \frac{3\,k^{2}}{A} 
	\;+\; \frac{2\,A'}{r} 
	\;+\; \frac{A}{r^{2}} 
	\;-\; \frac{1}{r^{2}\,A},\\
	G^2_2 &= - 3\,k^{2}\,r 
	\;+\; r\,A\,A''
	\;+\; r\bigl(A'\bigr)^{2}
	\;+\; 2\,A\,A',\\
	G^3_3 &= - 3\,k^{2}\,r\,\sin \theta
	\;+\; r\,A\,\sin \theta\,A''
	\;+\; r\,\sin \theta\,\bigl(A'\bigr)^{2}
	\;+\; 2\,A\,\sin \theta\,A',
\end{split}
\label{efe}
\end{equation}
and one finds the simultaneous solution of all these components:
\begin{equation}
A(r) \;=\; \sqrt{\,1 \;-\; \frac{\alpha}{r} \;+\; k^2\,r^2}\;,
\label{A(r)}
\end{equation}
where $\alpha\in \mathbb{R}$ is an arbitrary constant. Summarizing, we have derived a spherically symmetric tetrad solution for the de Sitter theory's EOMs. The constant $\alpha$ relates to the black hole mass via $\alpha = 2M$. Likewise, the $r^2$-term in \eqref{A(r)} can be associated with a cosmological constant when one contracts the tetrads into the metric tensor,
\begin{equation}
    g_{\mu\nu} \;=\; \eta_{ab}\,e^a_\mu\, e^b_\nu.
\end{equation}
However, as argued above, we must let $k \to 0$ to reconnect the theory with standard Einstein theory and obtain gauge invariance. Thus, after performing the Seiberg-Witten map of the solution \eqref{ansatz1} and taking $k\to 0$, any cosmological constant dependence disappears, contrary to the claim in \cite{Chaichian:2007dr}. To restore cosmological constant or electric charge, one can add the appropriate extra terms to the action \eqref{action}. Specifically, including
\begin{equation}
    S_M \;=\; \int e \;\bigl(\,-2\,\Lambda \;+\; F_{\mu\nu}\,F_{\rho\sigma}\,e^\mu_a\,e^\rho_b\,e^\nu_c\,e^\sigma_d\;\eta^{ab}\,\eta^{cd}\bigr)
\label{source terms}
\end{equation}
incorporates the cosmological constant $\Lambda$ and the electromagnetic field. In this way, the usual $\Lambda\,e^a_\mu$ and $\mathcal{T}_{\mu\nu}\,e^{a\nu}$ terms\footnote{Where $\mathcal{T}_{\mu\nu}$ is the usual Maxwell energy-momentum tensor stemming from electromagnetism.} appear in the Einstein equation from \eqref{EOM}, without altering the torsion-free condition.

\section{Noncommutative Poincare gauge theory of gravity}\label{sec2}

In order to study the effect of spacetime noncommutativity, we will introduce the de Sitter group gauge fields obeying the NC gauge transformations, which we will contract to Poincare gauge fields obeying Poincare gauge transformations. This means that instead of the regular gauge transformation rule
\begin{equation}
    \omega^{AB}_\mu M_{AB} \longrightarrow U\omega^{AB}_\mu M_{AB} U^\dagger - i \left(\partial_\mu U\right) U^\dagger\;,
\end{equation}
we impose the transformation rule
\begin{equation}
    \hat{\omega}^{AB}_\mu M_{AB}\longrightarrow U_\star\hat{\omega}^{AB}_\mu M_{AB} U_\star - i \left(\partial_\mu U_\star\right) U^\dagger_\star\;,
\label{NCgaugetransf}
\end{equation}
with
\begin{equation}
    U =\exp\left(\alpha^{AB}(x)M_{AB}\right)= \sum_{n=0}^\infty\frac{1}{n!} \left(\alpha(x)^{AB}M_{AB}\right)^{n}
\end{equation}
and
\begin{equation}
    U_\star =\exp_\star\left(\hat{\alpha}^{AB}(x)M_{AB}\right) = \sum_{n=0}^\infty\frac{1}{n!} \left(\hat{\alpha}(x)^{AB}M_{AB}\right)^{\star n}\;.
\end{equation}
In other words, we impose the symmetry (after contraction) to the modified transformations that depend on the NC  $\star$-product  given as the Moyal product \eqref{moyalstar}.

As discovered by Seiberg and Witten in their paper \cite{Seiberg:1999vs}, there exists an equivalence between NC gauge theories and commutative gauge theories with infinitely many interaction vertices whose coupling constants appear as ever-increasing powers of the noncommutativity parameter $\Theta$. The commutative theory in the equivalence is an effective theory, and it is worthwhile to study it truncated to some order. 

In this section, we will find the expansion of the NC de Sitter gauge field up to (and including) the second order of noncommutativity, and identify it as the correction to the corresponding commutative gauge field stemming from quantum effects of spacetime. This expansion will be valid for any theory which is invariant to NC de Sitter gauge transformations. In the following subsections, we will contract the de Sitter group to the Poincar\'{e} group and thus obtain the NC corrections to the gravitational (Poincar\'{e} group) gauge fields appearing in the contracted action \eqref{action}.\\

To formulate a de Sitter gauge invariance within a NC framework, we adopt the explained NC variant \eqref{NCgaugetransf} of the non-abelian gauge theory, as initially proposed by Chamseddine \cite{Chamseddine:2000si}, building upon the Seiberg-Witten map. Utilizing the Seiberg-Witten map, we derive the second-order correction in the NC parameter relative to the corresponding commutative field. Let $\hat{\omega}_\mu ^{AB} \equiv \hat{\omega}_\mu ^{AB} (x , \Theta)$ denote the connection field for the non-abelian gauge group in the NC setting, with $\hat{R}_{\mu\nu}^{AB}$ representing its NC field strength tensor. Throughout our analysis, we distinguish between quantities with the hat symbol ($\hat{X}$), indicative of the NC regime, and those without, signalling the commutative regime. The field strength tensor is defined as
\begin{equation}
	\hat{R}^{AB}_{\mu\nu} = \partial_\mu \hat{\omega}_\nu^{AB} - \partial_\nu \hat{\omega}_\mu^{AB} + \left[\hat{\omega}_\mu \stackrel{\star}{,} \hat{\omega}_\nu\right]^{AB} = \partial_\mu \hat{\omega}_\nu^{AB} - \partial_\nu \hat{\omega}_\mu^{AB} + \eta_{CD}\left(\hat{\omega}_\mu^{AC} \star\hat{\omega}_\nu^{DB} - \hat{\omega}_\nu^{AC} \star \hat{\omega}_\mu^{DB}\right)\;.
	\label{1}
\end{equation}
Note the index convention in the expansion of anticommutator and commutator brackets which we will be using throughout the article,
\begin{equation}
\begin{split}
    \left\{X_\mu, Y_\nu\right\}^{AB}&= \eta_{CD}\left(X_\mu^{AC} Y_\nu^{DB} + Y_\nu^{AC} X_\mu^{DB}\right),\\
   \left\{X_\mu \stackrel{\star}{,} Y_\nu\right\}^{AB}&= \eta_{CD}\left(X_\mu^{AC}\star Y_\nu^{DB} + Y_\nu^{AC}\star X_\mu^{DB}\right),\\
    \left[X_\mu, Y_\nu\right]^{AB}&= \eta_{CD}\left(X_\mu^{AC} Y_\nu^{DB} - Y_\nu^{AC} X_\mu^{DB}\right),\\
   \left[X_\mu \stackrel{\star}{,} Y_\nu\right]^{AB}&= \eta_{CD}\left(X_\mu^{AC}\star Y_\nu^{DB} - Y_\nu^{AC}\star X_\mu^{DB}\right).
\end{split}
\end{equation}
The essence of the Seiberg-Witten map lies in expressing NC gauge fields, governed by the NC gauge algebra, in terms of their commutative counterparts, which adhere to the corresponding commutative gauge algebra, in a manner that preserves compatibility between noncommutative and commutative gauge transformations, i.e.,
\begin{equation}
    \hat \omega _\mu ^{AB} (\omega) + \delta _{\hat \lambda} \hat \omega _\mu ^{AB}(\omega) = \hat \omega _\mu ^{AB} (\omega + \delta _{\lambda} \omega)\;, 
\end{equation}
where $\delta_{\hat \lambda} $ and $\delta _{\lambda}$ denote infinitesimal variations under noncommutative and commutative gauge transformations, respectively. The NC non-abelian infinitesimal gauge transformation is given by \cite{Chamseddine:2000si}
\begin{equation}
	\delta_{\hat{\lambda}}\hat{\omega}_\mu^{AB} = \partial_\mu \hat{\lambda}^{AB} + \left[\hat{\omega}_\mu\stackrel{\star}{,}\hat{\lambda}\right]^{AB} = \partial_\mu \hat{\lambda}^{AB} + \eta_{CD} \left(\hat{\omega}_\mu^{AC}\star\hat{\lambda}^{DB} - \hat{\lambda}^{AC}\star\hat{\omega}_\mu^{DB}\right)\;,
	\label{2}
\end{equation}
which is simply the analogue of the commutative infinitesimal transformation\footnote{The proofs of validity of this analogy (defining natural transformation rules and invariant Lagrangians) are collected in the review paper \cite{Hersent:2022gry}.} and can be obtained as the infinitesimal limit of \eqref{NCgaugetransf}. As demonstrated in \cite{Chamseddine:2000si}, it is feasible to establish a recursive relation governing the dependence of the deformed fields on $\Theta$ to all orders\footnote{Note the difference in conventions used in \cite{Seiberg:1999vs} and \cite{Chamseddine:2000si}. The following change of notation in \cite{Seiberg:1999vs}
\begin{equation}
	\begin{split}
		&A_i \mapsto i\omega_\mu \\
		&F_{ij} \mapsto iR_{\mu\nu}\\
		&\left\{\cdot,\cdot\right\}\mapsto\left\{\cdot,\cdot\right\}^{AB} \text{(explicitly writing $\mathrm{SO}(4,1)$ indices)}
	\end{split}
	\label{8}
\end{equation}
leads to the formulas in Chamseddine's \cite{Chamseddine:2000si}.}, expressed as
\begin{equation}
	\delta \hat{\omega}_\mu^{AB} = -\frac{i}{4}\Theta^{\nu\rho}\left\{\hat{\omega}_\nu \stackrel{\star}{,} \partial_\rho \hat{\omega}_\mu + \hat{R}_{\rho\mu}\right\}^{AB}
	\label{13}
\end{equation}
which is solved recursively by plugging in the following expansion
\begin{equation}
	\hat{\omega}^{AB}_\mu= {\omega}^{AB}_\mu -i \Theta^{\nu\rho} \omega^{AB}_{\mu\nu\rho} + \Theta^{\nu\rho}\Theta^{\lambda\tau}\omega^{AB}_{\mu\nu\rho\lambda\tau} + \mathcal{O}(\Theta^3) \equiv \omega^{AB}_\mu + \omega_\mu^{AB(1)} + \omega_\mu^{AB(2)} + \mathcal{O}(\Theta^3)\;.
	\label{14}
\end{equation}
The solution of \eqref{13} up to $\mathcal{O}(\Theta^2)$ is obviously
\begin{equation}
	\hat{\omega}_\mu^{AB} = \omega^{AB}_\mu -\frac{i}{4}\Theta^{\nu\rho}\left\{{\omega}_\nu , \partial_\rho {\omega}_\mu + {R}_{\rho\mu}\right\}^{AB}\;,
	\label{15}
\end{equation}
from which we read off
\begin{equation}
    \omega^{AB}_{\mu\nu\rho} = \frac{1}{4}\left\{{\omega}_\nu , \partial_\rho {\omega}_\mu + {R}_{\rho\mu}\right\}^{AB}
\label{spincon1}
\end{equation}
whereas the second order contribution equals the following expression:\footnote{Obtained by expanding the recursion (\ref{13}) twice.}
\begin{equation}
\begin{split}
		\omega_\mu^{(2)AB} =& -\frac{i}{4}\Theta^{\nu\rho}\eta_{CD}\left[\omega_\nu^{(1)AC}\left(\partial_\rho\omega_\mu^{DB} + R_{\rho\mu}^{DB}\right) + \left(\partial_\rho\omega_\mu^{AC} + R_{\rho\mu}^{AC} \right) \omega_\nu^{(1)DB}    \right]\\
		&-\frac{i}{4}\Theta^{\nu\rho}\eta_{CD}\left[ \omega_\nu^{AC}\left(\partial_\rho\omega_\mu^{(1)DB} + R_{\rho\mu}^{(1)DB}\right) + \left(\partial_\rho\omega_\mu^{(1)AC} + R_{\rho\mu}^{(1)AC} \right) \omega_\nu^{DB}  \right]\\
		&-\frac{i}{4}\Theta^{\nu\rho}\eta_{CD}\left[\omega_\nu^{AC}\star^{(1)}\left(\partial_\rho\omega_\mu^{DB} + R_{\rho\mu}^{DB}\right) +\left(\partial_\rho\omega_\mu^{AC} + R_{\rho\mu}^{AC} \right)\star^{(1)} \omega_\nu^{DB} \right]\;,
\end{split}
\label{16}
\end{equation}
where $\star^{(1)}$ denotes the $\Theta^1$ component of the $\star$-product. It is easy to see that the first line in \eqref{16} simplifies to
\begin{equation}
	 -\frac{i}{4}\Theta^{\nu\rho}\left\{\omega_\nu^{(1)},\left(\partial_\rho\omega_\mu + R_{\rho\mu}\right)  \right\}^{AB}\;,
\end{equation}
and the second line simplifies to
\begin{equation}
	 -\frac{i}{4}\Theta^{\nu\rho}\left\{\omega_\nu,\left(\partial_\rho\omega_\mu^{(1)} + R_{\rho\mu}^{(1)}\right)  \right\}^{AB}\;,
\end{equation}
while the last line simplifies to the following commutator
\begin{equation}
	\begin{split}
		& -\frac{i}{4}\Theta^{\nu\rho}\eta_{CD}\left(\frac{i}{2}\right)\Theta^{\lambda\tau}\left[\partial_\lambda\omega_\nu^{AC}\partial_\tau\left(\partial_\rho\omega_\mu^{DB} + R_{\rho\mu}^{DB}\right)  + \partial_\lambda\left(\partial_\rho\omega_\mu^{AC} + R_{\rho\mu}^{AC} \right)\partial_\tau\omega_\nu^{DB}\right] \\
		&= \frac{1}{8}\Theta^{\nu\rho}\eta_{CD}\Theta^{\lambda\tau}\left[\partial_\lambda\omega_\nu^{AC}\partial_\tau\left(\partial_\rho\omega_\mu^{DB} + R_{\rho\mu}^{DB}\right)  - \partial_\tau\left(\partial_\rho\omega_\mu^{AC} + R_{\rho\mu}^{AC} \right)\partial_\lambda\omega_\nu^{DB}\right] \\
		&= \frac{1}{8}\Theta^{\nu\rho}\Theta^{\lambda\tau}\left[\partial_\lambda\omega_\nu, \partial_\tau\left(\partial_\rho\omega_\mu + R_{\rho\mu}\right)\right]^{AB}\;.
	\end{split}
\end{equation}
\noindent Altogether, a more compact form of the second order correction to the gauge field is as follows:
\begin{equation}
\begin{split}
	\Theta^{\nu\rho}\Theta^{\lambda\tau}\omega_{\mu\nu\rho\lambda\tau}^{AB} = & -\frac{i}{4}\Theta^{\nu\rho}\left\{\omega_\nu^{(1)},\left(\partial_\rho\omega_\mu + R_{\rho\mu}\right)  \right\}^{AB} -\frac{i}{4}\Theta^{\nu\rho}\left\{\omega_\nu,\left(\partial_\rho\omega_\mu^{(1)} + R_{\rho\mu}^{(1)}\right)  \right\}^{AB} \\
 &+\frac{1}{8}\Theta^{\nu\rho}\Theta^{\lambda\tau}\left[\partial_\lambda\omega_\nu, \partial_\tau\left(\partial_\rho\omega_\mu + R_{\rho\mu}\right)\right]^{AB}\;.
\end{split}
\end{equation}
Further trivial simplification leads to the following form of the correction
\begin{equation}
\begin{split}
		\Theta^{\nu\rho}\Theta^{\lambda\tau}\omega_{\mu\nu\rho\lambda\tau}^{AB} = & \frac{1}{16}\Theta^{\nu\rho}\Theta^{\lambda\tau} \Bigl(-\Bigl\{\Bigl\{\omega_\lambda,\bigl(\partial_\tau\omega_\nu + R_{\tau\nu}\bigr)\Bigr\},\bigl(\partial_\rho\omega_\mu + R_{\rho\mu}\bigr)\Bigr\}^{AB} \\
  &\quad  - \Bigl\{\omega_\nu, \partial_\rho \Bigl\{\omega_\lambda,\bigl(\partial_\tau\omega_\mu + R_{\tau\mu}\bigr)\Bigr\} \Bigr\}^{AB} +2 \Bigl[\partial_\lambda\omega_\nu, \partial_\tau\bigl(\partial_\rho\omega_\mu + R_{\rho\mu}\bigr)\Bigr]^{AB} \Bigr) \\
  & -\frac{i}{4}\Theta^{\nu\rho}\Bigl\{\omega_\nu,R_{\rho\mu}^{(1)} \Bigr\}^{AB}.
\end{split}
\label{21}
\end{equation}
In order to simplify further, we make use of the formula for the first order correction to the curvature tensor $\hat{R}_{\mu\nu}^{(1)AB}$ expressed in terms of the commutative fields (equation (3.6) in \cite{Seiberg:1999vs})  
\begin{equation}
	\hat{R}_{\mu\nu}^{(1)AB} = \frac{i}{4}\Theta^{\alpha\beta} \left(2\left\{R_{\mu\alpha},R_{\nu\beta}\right\}^{AB} - \left\{\omega_\alpha, D_\beta R_{\mu\nu} + \partial_\beta R_{\mu\nu}\right\}^{AB}\right)
	\label{11}
\end{equation}
where, very importantly, the covariant derivative is given as\footnote{This covariant derivative convention leads from applying the change of convention, described above in \eqref{8}, to the covariant derivative formula (3.26) in Seiberg-Witten paper \cite{Seiberg:1999vs}. Looking at (2.26) in \cite{Chaichian:2007dr}, we can see that the covariant derivative is given as an anticommutator instead of a commutator, which can not be correct. It will turn out that the second order correction formula (2.24) \cite{Chaichian:2007dr} is actually completely correct, but they give an incorrect expression for the covariant derivative in (2.26) and they incorrectly derive (2.29) form (2.24). The wrong derivation from \cite{Chaichian:2007dr}'s (2.24) to (2.29) is partially responsible for the incorrectness of the tetrad field components, while the rest of the "incorrectness" is due to the incorrect covariant derivative convention (2.26) in \cite{Chaichian:2007dr}}
\begin{equation}
	D_\beta R_{\mu\nu}^{AB} = \partial_\beta R_{\mu\nu}^{AB} + \bigl[\omega_\beta, R_{\mu\nu}\bigr]^{AB}\;.
	\label{12}
\end{equation}
By plugging in the relation \eqref{11} into \eqref{21}, the second order contribution expression for the $\mathrm{SO}(4,1)$ spin connection is,
\begin{equation}
	\begin{split}
		\omega_{\mu\nu\rho\lambda\tau}^{AB} = \frac{1}{16}\Bigl(&-\Bigl\{\Bigl\{\omega_\lambda,\bigl(\partial_\tau\omega_\nu + R_{\tau\nu}\bigr)\Bigr\},\bigl(\partial_\rho\omega_\mu + R_{\rho\mu}\bigr)\Bigr\}^{AB} \\
  & - \Bigl\{\omega_\nu, \partial_\rho \Bigl\{\omega_\lambda,\bigl(\partial_\tau\omega_\mu + R_{\tau\mu}\bigr)\Bigr\} \Bigr\}^{AB}  +2 \Bigl[\partial_\lambda\omega_\nu, \partial_\tau\bigl(\partial_\rho\omega_\mu + R_{\rho\mu}\bigr)\Bigr]^{AB}  \\
  & + \Bigl\{\omega_\nu,2\bigl\{R_{\rho\lambda},R_{\mu\tau}\bigr\}\Bigr\}^{AB} - \Bigl\{\omega_\nu,\Bigl\{\omega_\lambda, D_\tau R_{\rho\mu} + \partial_\tau R_{\rho \mu}\Bigr\}\Bigr\}^{AB}\Bigr)\;.
	\end{split}
\label{22}
\end{equation}

\subsection{Correction to tetrads}
In order to obtain the correction for the tetrad fields, we need to make the contraction $k\rightarrow 0$ to revert back to the Poincar\'{e} group. This amounts to setting $B=5$ and $A=a\in\left\{0,1,2,3\right\}$ in \eqref{22} (equation in (2.24) in \cite{Chaichian:2007dr}) dividing by $k$ and taking the limit $\lim_{k\rightarrow 0}$. The formula (2.29) in \cite{Chaichian:2007dr} has all correct factors and all the terms that are included are correct, but it is missing a term. To see this, let us study the term 
\begin{equation}
	\delta\omega^{AB(2)}_\mu \supset \frac{1}{16}\Theta^{\nu\rho}\Theta^{\lambda\tau}(-1) \left\{\omega_\nu,\left\{\omega_\lambda, D_\tau R_{\rho\mu} + \partial_\tau R_{\rho \mu}\right\}\right\}^{AB}
\end{equation}
that appears in the spin connection's NC correction. Setting $B=5$ and $A=a$, we obtain after expanding the anticommutators the following
\begin{equation}
	\begin{split}
		&- \frac{1}{16}\Theta^{\nu\rho}\Theta^{\lambda\tau} \left\{\omega_\nu,\left\{\omega_\lambda, D_\tau R_{\rho\mu} + \partial_\tau R_{\rho \mu}\right\}\right\}^{a5} \\ 
    =	&- \frac{1}{16}\Theta^{\nu\rho}\Theta^{\lambda\tau} \eta_{cc}\eta_{dd}\Bigl[\omega_\nu^{ac}\Bigl(\omega_\lambda^{cd}\bigl(D_\tau R^{d5}_{\rho\mu} + \partial_\tau R^{d5}_{\rho_\mu}\bigr) + \bigl(D_\tau R^{cd}_{\rho\mu} + \partial_\tau R^{cd}_{\rho_\mu}\bigr) \omega_\lambda^{d5} \Bigr) \\
    & \quad + \Bigl\{\omega_\lambda, D_\tau R_{\rho\mu} + \partial_\tau R_{\rho \mu}\Bigr\}^{ac}\omega_\nu^{c5} \Bigr] \\
		 = &- \frac{1}{16}\Theta^{\nu\rho}\Theta^{\lambda\tau}\eta_{cc}\eta_{dd}\Bigl[\omega_\nu^{ac}\omega_\lambda^{cd}\bigl(D_\tau R^{d5}_{\rho\mu} + \partial_\tau R^{d5}_{\rho_\mu}\bigr) + \omega^{ac}_\nu\bigl(D_\tau R^{cd}_{\rho\mu} + \partial_\tau R^{cd}_{\rho_\mu}\bigr) \omega_\lambda^{d5} \\
   & \quad + \Bigl\{\omega_\lambda, D_\tau R_{\rho\mu} + \partial_\tau R_{\rho \mu}\Bigr\}^{ac}\omega_\nu^{c5} \Bigr]\;.
	\end{split}
\label{24}
\end{equation}
Comparing this to (2.29) in \cite{Chaichian:2007dr}, we see that the second and third terms can indeed be found\footnote{Up to an obvious typo $\frac{1}{32}$ which should be $\frac{1}{16}$} as the second and third terms in (2.29), while the first term is absent. The authors in \cite{Chaichian:2007dr} did not include the first term 
\begin{equation} \label{additionalterm}
	\frac{1}{16}\Theta^{\nu\rho}\Theta^{\lambda\tau}(-1)\left[\omega_\nu^{ac}\omega_\lambda^{cd}\left(D_\tau R^{d5}_{\rho\mu} + \partial_\tau R^{d5}_{\rho_\mu}\right)\right]
\end{equation}
because they argued that the torsion $R^{a5}$ vanishes, so they probably assumed that the $(d5)$ components of its derivatives should also vanish. This is true for the partial derivative, but looking at the covariant derivative more closely,
\begin{equation}
	D_\tau R^{d5}_{\rho\mu} = \partial_\tau R^{d5}_{\rho\mu} + \left[\omega_\tau, R_{\rho \mu}\right]^{d5} = \partial_\tau R^{d5}_{\rho\mu} + \eta_{cd} \omega_\tau^{ac}R^{d5}_{\rho\mu} - \eta_{cd}R^{ac}_{\rho\mu}\omega^{d5}_\tau = - \eta_{cd}R^{ac}_{\rho\mu}\omega^{d5}_\tau = k\eta_{cd}e^{d}_\tau R^{ca}_{\rho\mu}\;,
\label{26}
\end{equation}
we can see that the covariant derivative's $(d5)$ component does not vanish identically, so when factoring out one power of $k$ in the derivation of tetrad corrections $k \cdot e^a_\mu = \omega^{a5}_\mu $, the term \eqref{26} will actually contribute to the overall tetrad field correction. All in all, the following term should be added to the equation (2.29) in \cite{Chaichian:2007dr},
\begin{equation}
	-\frac{1}{16}\eta_{cd}\eta_{bm}\eta_{fg} \omega^{ac}_\lambda\omega^{db}_\nu e^{f}_\rho R^{gm}_{\tau\mu} .
\label{missed term}
\end{equation}
To summarise, the NC correction to the tetrad fields reads,
\begin{equation} \label{tetradcorr}
    \hat e _\mu ^{a} (x, \Theta)= e _\mu ^{a} (x)-i \Theta ^{\nu \rho} e _{\mu \nu \rho} ^{a} (x) +\Theta ^{\nu \rho} \Theta ^{\lambda \tau} e _{\mu \nu \rho \lambda \tau} ^{a} (x)+\mathcal{O}(\Theta ^3)
\end{equation}
where
\begin{align}
e_{\mu \nu \rho }^{a}
&= \frac{1}{4}\Bigl[
   \omega_{\nu}^{a\,c}\,\partial_{\rho}\,e_{\mu}^{d}
   \;+\;\bigl(\partial_{\rho}\,\omega_{\mu}^{a\,c} + R_{\rho \mu}^{a\,c}\bigr)\,e_{\nu}^{d}
\Bigr]\;\eta_{c\,d}, 
\label{3.11} 
\\
e_{\mu \nu \rho \lambda \tau }^{a}
&= \frac{1}{16}\Bigl[
   2\,\Bigl\{R_{\tau \nu},\,R_{\mu \rho}\Bigr\}^{ab}\,e_{\lambda}^{c}
   \;-\;\omega_{\lambda}^{a\,b}\,\Bigl(D_{\rho}\,R_{\tau \mu}^{c\,d}
     \;+\;\partial_{\rho}\,R_{\tau \mu}^{c\,d}\Bigr)\,e_{\nu}^{m}\,\eta_{d\,m}
   -\,\Bigl\{\omega_{\nu},\,\bigl(D_{\rho}\,R_{\tau \mu}
     + \partial_{\rho}\,R_{\tau \mu}\bigr)\Bigr\}^{ab}\,e_{\lambda}^{c}
     \nonumber \\
&\quad
   \;-\;\partial_{\tau}\,\Bigl\{\omega_{\nu},\,\bigl(\partial_{\rho}\,\omega_{\mu}
     + R_{\rho \mu}\bigr)\Bigr\}^{a\,b}\,e_{\lambda}^{c}
   -\,\omega_{\lambda}^{a\,b}\,\partial_{\tau}\Bigl(
       \omega_{\nu}^{c\,d}\,\partial_{\rho}\,e_{\mu}^{m}
       + \bigl(\partial_{\rho}\,\omega_{\mu}^{c\,d}
         + R_{\rho \mu}^{c\,d}\bigr)\,e_{\nu}^{m}
     \Bigr)\,\eta_{d\,m}
     \nonumber \\
&\quad
   \;+\;2\,\partial_{\nu}\,\omega_{\lambda}^{a\,b}\,
        \partial_{\rho}\partial_{\tau}\,e_{\mu}^{c}
   -\,2\,\partial_{\rho}\Bigl(\partial_{\tau}\,\omega_{\mu}^{a\,b}
     + R_{\tau \mu}^{a\,b}\Bigr)\,\partial_{\nu}\,e_{\lambda}^{c}
   \;-\;\Bigl\{\omega_{\nu},\,\bigl(\partial_{\rho}\,\omega_{\lambda}
     + R_{\rho \lambda}\bigr)\Bigr\}^{a\,b}\,\partial_{\tau}\,e_{\mu}^{c}
\nonumber \\
&\quad
   \;-\,\Bigl(\partial_{\tau}\,\omega_{\mu}^{a\,b}
     + R_{\tau \mu}^{a\,b}\Bigr)\,\Bigl(
       \omega_{\nu}^{c\,d}\,\partial_{\rho}\,e_{\lambda}^{m}
       + \bigl(\partial_{\rho}\,\omega_{\lambda}^{c\,d}
         + R_{\rho \lambda}^{c\,d}\bigr)\,e_{\nu}^{m}\,\eta_{d\,m}
     \Bigr)
\Bigr]\;\eta_{b\,c}
\nonumber \\
&\quad
\; - \frac{1}{16}\,\omega_{\lambda}^{a\,c}\,\omega_{\nu}^{d\,b}\,e_{\rho}^{f}\,
   R_{\tau \mu}^{g\,m}\,\eta_{c\,d}\,\eta_{f\,g}\,\eta_{b\,m}\,.
\label{3.12}
\end{align}
The flawed expression for the second order correction of the tetrad field, which misses the last term in \eqref{3.12}, and the resulting NC metric have been referenced in subsequent literature \cite{Chaichian:2007dr, Chaichian:2007we, Mukherjee:2007fa,Linares:2019gqf, Touati:2023ubi, Touati:2023cxy, Touati:2022kuf, Touati:2022zbm, Touati:2021eem, Zhao:2023uam, Heidari:2023egu, Heidari:2023bww,Chamseddine:2000si, Touati:2024kmv, AraujoFilho:2025viz}.
\subsection{Inverse NC tetrads}
Analogous to the commutative relation $e^{\mu} _{ a} e _{\mu } ^{b}= \delta ^b _a $, the inverse of the NC corrected tetrads under normal product multiplication can be defined as $\hat{e}^{\mu} _{ a} \hat{e} _{\mu } ^{b}= \delta ^b _a $. However, we are interested in the inverse $\hat{e}^{\mu} _{\star a}$ under $\star$-product multiplication , which is defined as \cite{Chamseddine:2000si}
\begin{equation} \label{starinv}
    \hat{e}^{\mu} _{\star a}\star \hat{e} _{\mu } ^{b}= \delta ^b _a . 
\end{equation}
The $\star$-inverse, $\hat{e}^{\mu} _{\star a}$ can be expanded to second order in $\Theta$ as,
\begin{equation} \label{invtetradcorr}
    \hat e ^\mu _{\star a} (x, \Theta)= e ^\mu _{a} (x)-i \Theta ^{\rho \sigma} e ^{\mu}_{a \rho \sigma} (x) +\Theta ^{\rho \sigma} \Theta ^{\lambda \tau} e ^{\mu}_{a \rho \sigma \lambda \tau} (x)+\mathcal{O}(\Theta ^3).
\end{equation}
Substituting \eqref{tetradcorr} and \eqref{invtetradcorr} in \eqref{starinv}, it can be shown that
\begin{equation} \label{invtetradcomp}
    \begin{split}
        e_{a\rho \sigma}^{\mu } =&-e_{b}^{\mu }\bigl( e_{a}^{\kappa }e_{\kappa 
\rho \sigma }^{b}+\frac{1}{2}\partial _{\rho }e_{a}^{\kappa }\partial _{\sigma
}e_{\kappa }^{b}\bigr),  \\
e_{a \rho \sigma \lambda \tau}^{\mu } =&e_{b}^{\mu }\bigl( e_{a\rho \sigma
}^{\nu }e_{\nu \lambda \tau }^{b}-e_{a}^{\nu }e_{\nu \rho
\sigma \lambda \tau}^{b}-\frac{1}{8}\partial _{\rho }\partial _{\lambda
}e_{a}^{\nu }\partial _{\sigma }\partial _{\tau }e_{\nu }^{b}+\frac{1}{2}\partial _{\rho }e_{a}^{\nu }\partial _{\sigma }e_{\nu \lambda \tau
}^{b}+\frac{1}{2}\partial _{\sigma }e_{\nu}^{b}\partial _{\rho }e_{a\lambda
\tau }^{\nu }\bigr).
    \end{split}
\end{equation}
For given commutative tetrads, the $\star$-inverse can be calculated by using \eqref{invtetradcomp} (along with \eqref{3.11} and \eqref{3.12}) in \eqref{invtetradcorr}. Note that in our expression for the $\star$-inverse the coefficients and signs of some terms differ from those in \cite{Chamseddine:2000si}.

\subsection{Correction to spin connection}
For completeness, we provide the NC correction to the spin connection. Starting from \eqref{14}, and setting $A=a$ and $B=b$, and using \eqref{spincon1} and \eqref{22}, we obtain 
\begin{equation} \label{spincorr}
    \hat \omega _\mu ^{a b } (x, \Theta)= \omega _\mu ^{a b} (x)-i \Theta ^{\nu \rho} \omega _{\mu \nu \rho} ^{a b} (x) +\Theta ^{\nu \rho} \Theta ^{\lambda \tau} \omega _{\mu \nu \rho \lambda \tau} ^{a b} (x)+\mathcal{O}(\Theta ^3)
\end{equation}
where
\begin{equation}
	\omega_{\mu \nu \rho}^{ab} (x) = \frac{1}{4}\left\{{\omega}_\nu , \partial_\rho {\omega}_\mu + {R}_{\rho\mu}\right\}^{ab}\;,
\end{equation}
and
\begin{equation}
	\begin{split}
		\omega_{\mu\nu\rho\lambda\tau}^{ab} = &\frac{1}{16}\biggl[-\left\{\left\{\omega_\lambda,\left(\partial_\tau\omega_\nu + R_{\tau\nu}\right)\right\},\left(\partial_\rho\omega_\mu + R_{\rho\mu}\right)\right\}^{ab} \\
  & - \left\{\omega_\nu, \partial_\rho \left\{\omega_\lambda,\left(\partial_\tau\omega_\mu + R_{\tau\mu}\right)\right\} \right\}^{ab}  +2 \left[\partial_\lambda\omega_\nu, \partial_\tau\left(\partial_\rho\omega_\mu + R_{\rho\mu}\right)\right]^{ab}  \\
  & + \left\{\omega_\nu,2\left\{R_{\rho\lambda},R_{\mu\tau}\right\}\right\}^{ab} - \left\{\omega_\nu,\left\{\omega_\lambda, D_\tau R_{\rho\mu} + \partial_\tau R_{\rho \mu}\right\}\right\}^{ab}   \biggr].
	\end{split}
\label{22b}
\end{equation}
Contrast to the tetrad case, there is no additional term arising as in \eqref{additionalterm}. For a given set of commutative tetrads, one can first calculate the commutative spin connections and field tensors and then use \eqref{spincorr} to calculate the NC spin connections. For some combinations of twists and commutative tetrad fields, the NC spin connection is not necessarily antisymmetric.

\subsection{Correction to Field Tensor}
The correction to the field tensor/curvature tensor can be obtained using the following expression, which follows from \eqref{1} by simply setting \(A = a\) and \(B = b\) , 
\begin{equation}
	\hat{R}^{ab}_{\mu\nu} =  \partial_\mu \hat{\omega}_\nu^{ab} - \partial_\nu \hat{\omega}_\mu^{ab} + \eta_{cd}\left(\hat{\omega}_\mu^{ac} \star\hat{\omega}_\nu^{db} - \hat{\omega}_\nu^{ac} \star \hat{\omega}_\mu^{db}\right)\;.
	\label{1b}
\end{equation}
Using the NC corrected spin connection \eqref{spincorr} and the $\star$-product \eqref{starproduct2}, the correction to the field tensor up to second order in the expansion parameter \(\Theta\) is given by
\begin{equation}
    \hat{R}_{\mu \nu }^{ab}=R_{\mu \nu }^{ab}-i\Theta ^{\rho \sigma }R_{\mu
\nu \rho \sigma }^{ab}+\Theta ^{\rho \sigma }\Theta ^{\lambda \tau }R_{\mu \nu
\rho \sigma \lambda \tau }^{ab}+O(\Theta ^{3}),
\end{equation}
where \footnote{The last two terms of $R_{\mu \nu \rho \sigma \lambda \tau }^{ab}$ were missing in \cite{Chamseddine:2000si} (the upper Latin indices ruin the symmetry in the lower indices) and the factor \(\frac{1}{8}\) in the fifth term is corrected.}
\begin{eqnarray*}
R_{\mu \nu \rho \sigma }^{ab} &=&\partial _{\mu }\omega _{\nu \rho \sigma
}^{ab}+ \eta _{cd} \left( \omega _{\mu }^{ac}\omega _{\nu \rho \sigma }^{db}+\omega _{\mu \rho
\sigma }^{ac}\omega _{\nu }^{db}+\frac{1}{2}\partial _{\rho }\omega _{\mu
}^{ac}\partial _{\sigma }\omega _{\nu }^{db} \right) -\mu \leftrightarrow \nu,  \\
R_{\mu \nu \rho \sigma \lambda \tau }^{ab} &=&\partial _{\mu }\omega _{\nu
\rho \sigma \lambda \tau  }^{ab}+\eta _{cd}\left( \omega _{\mu }^{ac}\omega _{\nu \rho \sigma \lambda \tau }^{db}+\omega _{\mu \rho \sigma \lambda \tau}^{ac}\omega _{\nu
}^{db}-\omega _{\mu \rho \sigma }^{ac}\omega _{\nu \lambda \tau }^{db} + \frac{1}{8}\partial _{\rho }\partial _{\lambda }\omega _{\mu
}^{ac}\partial _{\sigma }\partial _{\tau }\omega _{\nu }^{db} \right. \\
&& \left. -\frac{1}{2} \partial _{\rho}\omega _{\mu }^{ac} \partial _{\sigma }\omega _{\nu \lambda \tau}^{db}-\frac{1}{2} \partial _{\rho}\omega _{\mu \lambda \tau}^{ac} \partial _{\sigma }\omega _{\nu }^{db}\right)-\mu
\leftrightarrow \nu.
\end{eqnarray*}
It is not clear whether one can associate a geometric structure with this correction, and therefore, one should not expect that the curvature scalar constructed from the gauge picture matches the Ricci scalar obtained from the NC metric (defined in the following section) \eqref{metricdef}.

\subsection{NC scalar curvature}
In the commutative case for the gauge theory of gravity, one can consider the Ricci scalar expressed in terms of gauge quantities
\begin{equation}
R = e_a^{\mu} R^{ab}_{\mu\nu} e_b^{\nu}.
\end{equation}
This expression can be generalized to the following NC Ricci scalar \cite{Chamseddine:2000si}, 
\begin{equation}
\hat{R} = \hat{e}^{\mu} _{\star a} \star \hat{R}^{ab}_{\mu\nu} \star \hat{e}^{\nu \dagger} _{\star b}.
\label{ncscalardef}
\end{equation}
It is worth commenting that different orderings of fields in \eqref{ncscalardef} amount to different results for a generic twist. In case the twist is Killing (constructed from Killing vectors of the tetrad field), the ordering is irrelevant. Finally, by inspecting \eqref{ncscalardef} we can see that $\hat{R}$ is not manifestly real, because the NC curvature tensor is not manifestly real, so we can see that
\begin{equation}
    \hat{R}^* = \hat{e}^{\mu} _{\star a} \star \hat{R}^{*ab}_{\mu\nu} \star \hat{e}^{\nu \dagger} _{\star b}
\end{equation}
need not equal $\hat{R}$. In all of the cases that are studied in this paper, $\hat{R}$ does indeed turn out to be real, but we nevertheless propose a new definition of the NC Ricci scalar which is manifestly real
\begin{equation}
    \hat{R} = \frac{1}{2}\left[ \hat{e}^{\mu} _{\star a} \star \hat{R}^{ab}_{\mu\nu} \star \hat{e}^{\nu \dagger} _{\star b} + \hat{e}^{\mu} _{\star a} \star \hat{R}^{*ab}_{\mu\nu} \star \hat{e}^{\nu \dagger} _{\star b} \right]\;,
\end{equation}
but it equals \eqref{ncscalardef} in all of the studied diagonal metric deformations.

\section{Noncommutative corrections to Schwarzschild metric}\label{sec3}
Having derived the NC corrections to the tetrad fields, it is straightforward to compute the corresponding corrections to the metric. The deformed metric is defined\footnote{Note that just like with the scalar curvature, there are infinitely many a priori equally motivated metric deformations which all reproduce the commutative metric in the commutative limit $\Theta\rightarrow 0$.} as \cite{Chaichian:2007we}
\begin{equation}
    \hat g _{\mu \nu} (x, \Theta) = \frac{1}{2} \eta _{a b} \left( \hat e_\mu ^a \star \hat e _\nu ^{b\dagger}  + \hat e_\nu ^b \star \hat e _\mu ^{a\dagger}  \right) \;,
    \label{metricdef}
\end{equation}
where the complex conjugate of the tetrad fields is defined as
\begin{equation} \label{tetradconjugate}
    \hat e _\mu ^{a\dagger} (x, \Theta)= e _\mu ^{a} (x)
    + i \Theta ^{\nu \rho} e _{\mu \nu \rho} ^{a} (x)
    + \Theta ^{\nu \rho} \Theta ^{\lambda \tau} e _{\mu \nu \rho \lambda \tau} ^{a} (x)
    + \mathcal{O}(\Theta ^3)\;.
\end{equation}
It is worth noting that the metric is real by definition. Finally, note that the NC metric depends on the choice of the starting commutative tetrad, i.e., deforming commutatively equivalent tetrads can result in different metrics (see, e.g. \cite{Touati:2023cxy}, the result of deforming a non-diagonal tetrad with the $r-\theta$ twist and compare with our \eqref{NCrtheta1}). In the expression \eqref{metricdef}, we utilize the star product up to the second order in $\Theta$
\begin{equation} \label{starproduct2}
    f\star g = fg
    - \frac{i}{2}\,\Theta ^{\mu \nu} \partial _\mu f \,\partial _\nu g
    + \frac{1}{8}\,\Theta ^{\nu \rho} \Theta ^{\lambda \tau}\,
    \partial _{\nu } \partial _{\lambda} f \,\partial _\rho \partial _\tau g
\end{equation}
to incorporate all the second-order contributions alongside the Seiberg-Witten contributions\footnote{We note that only the first-order $\star$-product is considered in \cite{Chaichian:2007dr} while evaluating the correction to the metric components.}.

In this paper, we use Drinfeld twists that are Abelian,
\begin{equation}
\mathcal{F}=\exp \left(i\,\Theta^{\alpha\beta}\,V^{1}_{\alpha}\otimes V^{2}_{\alpha}\right)
\quad\text{with}\quad\left[V^1,V^2\right] = 0,
\end{equation}
and of the Moyal type. The Drinfeld twist element is used to construct the $\star$-product in the following way, starting from a general commutative algebra $\left(\mathcal{A},1,m\right)$\footnote{Here, $\mathcal{A}$ is the algebra's vector space, $1$ is the identity element for the multiplication, and $m$ is the commutative multiplication map $m(a\otimes b) = a\cdot b$. For example, this can be applied to the pointwise algebra of Minkowski spacetime $(C^\infty(\mathbb{R}^{1,3}),1,m)$.}:
\begin{equation}
    f\star g = m \left(\mathcal{F}^{-1}f\otimes g\right)\;.
\end{equation}
Because it is a Drinfeld twist, the deformed symmetries form a Hopf algebra, and the Abelian condition is a technical simplification that uses commuting vector fields to build the twist. Finally, we say that the twist is of the Moyal type if it reproduces the $\star$-product \eqref{moyalstar}. We call $\mathcal{F}$ a Killing twist if both $V^{1}_{\alpha}$ and $V^{2}_{\alpha}$ are Killing vectors of the commutative metric we want to deform, and a semi-Killing twist if only one of them is such. Since we want to deform static and spherically symmetric metrics, it is natural to use the spherical basis and vector fields $V^{1,2}_{\alpha}\in\{\partial_t, \partial_r, \partial_\theta, \partial_\varphi\}$. Therefore, we have 6 possible combinations: $(\partial_r, \partial_\theta)$, $(\partial_t, \partial_r)$, $(\partial_t, \partial_\theta)$, $(\partial_t, \partial_\varphi)$, $(\partial_r, \partial_\varphi)$, and $(\partial_\theta, \partial_\varphi)$.

In the following subsections, we will deform the spherically symmetric tetrad solutions of the commutative theory for each of these twists and calculate the respective NC metrics. In general, all our results are first expressed in terms of $A(r)$, defined in the ansatz \eqref{ansatz1}, and are therefore applicable to any spherically symmetric black hole spacetime in 4 dimensions. We will examine in detail the properties of the metric for different twists, as inconsistencies in the interpretation of horizon structure and related aspects have been noted in the literature. Broadly, the different twists we use can be divided into two categories based on the horizon structure and curvature singularities they yield. The first category includes twists that do not involve the radial coordinate vector field $\partial_r$, where the physical interpretations remain consistent. The second category comprises twists involving the radial vector field $\partial_r$, where conventional interpretations become less straightforward. Apart from this, when there is a twist generated by the axial angle vector field $\partial_\varphi$, the metric becomes off-diagonal, gaining a nonvanishing component \(g_{r\theta}\).

In this section, we present the NC deformation induced by various twists to the Schwarzschild metric,
\begin{equation}
   ds^2 =  g_{\mu \nu}\,dx^{\mu}\,dx^{\nu}
   = -\left(1- \frac{2M}{r}\right)\, dt^2
   + \left(1- \frac{2M}{r}\right)^{-1} dr^2
   + r^2\left(d\theta^2 + \sin^2 \theta\, d\varphi^2\right),
\end{equation}
where $M$ is the mass of the black hole. In Appendices~\ref{app1}, \ref{app2}, and \ref{app3}, we present the NC corrections to the Reissner-Nordstr\"{o}m, Reissner-Nordstr\"{o}m-de Sitter, and BTZ black hole solutions, respectively.

\subsection{($t, \varphi$) twist}
The commutative Schwarzschild solution possesses two Killing vectors, \(\partial_t\) and \(\partial_\varphi\). Consequently, it is natural to investigate the twist \((t, \varphi)\) to see whether it induces any NC deformations. In contrast to other methods (see \cite{Ciric:2017rnf, DimitrijevicCiric:2018blz, DimitrijevicCiric:2019hqq, Gupta:2022oel, Herceg:2023pmc} and references therein), we find that in our approach, this Killing twist does indeed generate NC deformations. Notably, it also introduces a non-diagonal component into the deformed metric. The Killing \((t, \varphi)\) twist is defined by
\begin{equation}
\Theta ^{\mu \nu }=\left(
\begin{array}{cccc}
0 & 0 & 0 & \Theta \\
0 & 0 & 0 & 0 \\
0 & 0 & 0 & 0 \\
-\Theta & 0 & 0 & 0
\end{array}
\right), \quad \mu ,\nu =0,1,2,3,
\label{4dtphi}
\end{equation}
which leads to the single non-vanishing commutation relation between the coordinates
\begin{equation}
    [t\stackrel{\star}{,} \varphi] = i \, \Theta.
\end{equation}
(This twist is often called the angular twist and has been studied in \cite{Ciric:2017rnf, DimitrijevicCiric:2018blz, DimitrijevicCiric:2019hqq, Gupta:2022oel,Juric:2022bnm,Hrelja:2024tgj}.) The non-zero components of the tetrad $\hat e _\mu ^{a}$ are
\begin{eqnarray}
\hat{e}_{0}^{0}&=&A-\frac{1}{16}\,\bigl(A^{5}\,{A}^{\prime}{}^{2}
-2\,r\,A^{4}\,{A}^{\prime \prime \prime}\bigr)\,\sin ^2 \theta  \,\Theta ^{2}
+O(\Theta {^{3}})\,,  \label{4dtphib}\\
\hat{e}_{0}^{3}&=&-\frac{i}{4}\, A^3\,{A}^{\prime}\,\sin \theta \,\Theta
+O(\Theta {^{3}}),  \cr
\hat{e}_{1}^{1}&=&\frac{1}{A}+\frac{1}{8}\,A^{2}\bigl(-A\,{A}^{\prime}{}^{2}
-r\,{A}^{\prime}{}^{3}
+A^{2}\,{A}^{\prime \prime}
-2\,r\,A\,{A}^{\prime} {A}^{\prime \prime}\bigr)\,\sin ^2 \theta\,\Theta^{2}
+O(\Theta {^{3}}),  \cr
\hat{e}_{1}^{2}&=&\frac{1}{8}\,A^{3}\,A^{\prime \prime}\,\cos \theta \,\sin \theta\,\Theta^2
+O(\Theta {^{3}}),  \cr
\hat{e}_{2}^{1}&=&\frac{1}{8}\,A^{3}\,{A}^{\prime}\,\bigl(A-r\, A^{\prime}\bigr)\,\cos \theta \,\sin \theta\,\Theta^2
+O(\Theta {^{3}}),  \cr
\hat{e}_{2}^{2}&=&r-\frac{1}{8}\,A^{3}\,A^{\prime}\,\Bigl[-1+A \,\sin ^2 \theta\, \bigl(A+r\,{A}^{\prime}\bigr)\Bigr]\,\Theta ^{2}+O(\Theta {^{3}}),  \cr
\hat{e}_{3}^{0}&=&-\frac{i}{4}\,r\, A^2\,{A}^{\prime}\,\sin ^2\theta \,\Theta +O(\Theta {^{3}}),  \cr
\hat{e}_{3}^{3}&=&r\sin \theta +\frac{1}{16}\,A^{3}\,{A}^{\prime} \Bigl[ 2\,\cos ^2 \theta
+\sin ^2 \theta \,\bigl( 2A^{2}-r\,A\,{A}^{\prime}\bigr)\Bigr] \sin \theta \,\Theta ^{2}
+O(\Theta {^{3}}),
\nonumber
\end{eqnarray}
where ${A}^{\prime },\,{A}^{\prime \prime },\,{A}^{\prime \prime
\prime }$ are the first, second, and third derivatives of $A(r)$,
respectively. The deformed metric $\hat g _{\mu \nu} (x, \Theta)$ is obtained by using the definition \eqref{metricdef}. Up to second order in $\Theta$, the metric is off-diagonal, and its non-zero components are
\begin{eqnarray}
\hat{g}_{00}\left( {x,\Theta }\right) &=&-A^{2}
+\frac{1}{16}\,\bigl(
3\,A^{6}\,{ A}^{\prime }{}^{2}
-4\,r\,A^{5}\,{A}^{\prime }{}^{3}\bigr)\,\sin ^2 \theta \,{\Theta }^{2}
+O({\Theta ^{3}}),\cr
\hat{g}_{11}\left( {x,\Theta }\right)
&=&\frac{1}{A^{2}}
-\frac{1}{4}\,\Bigl(A^2\,A^{\prime}{}^{2}
+r\,A\,A^{\prime}{}^{3}
-A^{3}A^{\prime \prime}
-2\,r\,A^{2}\,A^{\prime}\,A^{\prime \prime}\Bigr)\,\sin ^2 \theta \, {\Theta }^{2}
+O({\Theta ^{3}}), \label{NCtphi1}\\
\hat{g}_{12}\left( {x,\Theta }\right) &=&  \hat{g}_{21}\left( {x,\Theta }\right)
=\frac{1}{16}\,A^2 \Bigl(r\,A^{\prime}{}^{2}
+A\,A^{\prime}
+r\,A\,A^{\prime \prime}\Bigr)\,\sin (2 \theta)\,\Theta ^{2}
+O({\Theta ^{3}}), \cr
\hat{g}_{22}\left( {x,\Theta }\right)
&=&r^{2}
+\frac{1}{4}\,\Bigl(r\,A^3\,A^{\prime}
-\bigl(r\,A^{5}\,A^{\prime}+r^2\,A^4\,A^{\prime}{}^{2}\bigr)\,\sin ^2 \theta \Bigr)\,\Theta ^{2}
+O(\Theta ^{3}), \cr
\hat{g}_{33}\left( {x,\Theta }\right)
&=&r^{2}\,\sin ^{2}\theta
+\frac{1}{16}\,\Bigl(4\,r\,A^3\,A^{\prime}\,\frac{\cos ^2 \theta}{\sin ^2 \theta}
+4\,r\,A^{5}\,A^{\prime}
-3\,r\,A^{4}\,A^{\prime}{}^{2}\Bigr)\,\sin ^{4}\theta  \,\Theta ^{2}
+O({\Theta ^{3}}).
\nonumber
\end{eqnarray}
Having calculated the metric $\hat{g}$ in the $t-\varphi$ twist, we can use it to calculate the Ricci and Kretschmann scalars using the standard differential geometry identities
\begin{equation}
\begin{split}
R(\hat{g}) &= \hat{g}^{\mu\rho}\,\hat{g}^{\rho\sigma}\,R_{\mu\nu\rho\sigma},\\
K(\hat{g}) &= R_{\mu\nu\rho\sigma}\, R^{\mu\nu\rho\sigma},
\end{split}
\end{equation}
where $R_{\mu\nu\rho\sigma}$ is the Riemann tensor calculated from $\hat{g}_{\mu\nu}$. On the other hand, one can also calculate the Ricci scalar stemming from the NC gauge theory, as defined in \eqref{ncscalardef},
\begin{equation}
\hat{R} = \hat{e}_a^{\mu} \star \hat{R}^{ab}_{\mu\nu} \star \hat{e}_b^{\nu\dagger},
\end{equation}
and for the $t-\varphi$ twist, it is expressed in terms of $A(r)$ as
\begin{equation}
\begin{split}
\hat{R} = &\, - 2 A A'' 
- 2 (A')^{2} 
- \frac{8 A A'}{r} 
- \frac{2 A^{2}}{r^{2}} 
+ \frac{2}{r^{2}}  
+ \Biggl[ 
        \cos^{2} \theta \Bigl(
        - \frac{2 A^{3} (A')^{3}}{r} 
        + \frac{7 A^{4} (A')^{2}}{8 r^{2}} 
        + \frac{A^{5} A'}{2 r^{3}} 
        - \frac{A^{3} A'}{2 r^{3}}
    \Bigr) \\
        &\,+ \sin^{2} \theta \Bigl(
        - \frac{r A^{5} A' (A'')^{2}}{2} 
        - \frac{r A^{4} (A')^{3} A''}{2} 
        + \frac{A^{6} (A'')^{2}}{4} 
        - \frac{15 A^{5} (A')^{2} A''}{8} 
        - \frac{5 A^{4} (A')^{4}}{4} \\
        &\,+ \frac{11 A^{6} A' A''}{8 r} 
        + \frac{A^{5} (A')^{3}}{4 r} 
        - \frac{A^{4} A' A''}{r} 
        - \frac{A^{3} (A')^{3}}{r} 
        + \frac{A^{4} (A')^{2}}{2 r^{2}} 
        + \frac{A^{5} A'}{4 r^{3}} 
        - \frac{A^{3} A'}{4 r^{3}}
    \Bigr) 
\Biggr] \Theta^{2} + \mathcal{O}(\Theta ^3).
\end{split}
\end{equation}

The NC corrections to the Schwarzschild metric are obtained by substituting
\begin{equation} \label{Asch_eqn}
    A^2= \left(1- \frac{2M}{r}\right)
\end{equation}
into \eqref{NCtphi1}. The NC-corrected Schwarzschild metric components under the ($t,\varphi$) twist are
\begin{equation}\label{nccomp1}
    \begin{split}
        \hat{g}_{tt}\left( {x,\Theta }\right)
        &= -\biggl(1 - \frac{2 M}{r}\biggr)
        + \frac{M^2 \bigl(-2 M + r\bigr)\bigl(-30 M + 9 r\bigr) \sin ^2 \theta}{48 r^6}\,\Theta^2
        + \mathcal{O}(\Theta^3),
        \\
        \hat{g}_{rr}\left( {x,\Theta }\right)
        &= \biggl(1 - \frac{2 M}{r}\biggr)^{-1}
         -  \frac{M \bigl(11 M^2 - 10 M r + 2 r^2\bigr) \sin ^2 \theta}{4\,r^4\,\bigl(-2 M + r\bigr)}\,\Theta^2
         + \mathcal{O}(\Theta^3),
        \\
        \hat{g}_{r\theta}\left( {x,\Theta }\right)
        &= \hat{g}_{\theta r}\left( {x,\Theta }\right)
        =  - \frac{M\,\cos \theta\,\sin \theta}{8\,r^2}\,\Theta^2
        + \mathcal{O}(\Theta^3),
        \\
        \hat{g}_{\theta\theta}\left( {x,\Theta }\right)
        &= r^2
        -  \frac{M \bigl(2 M -  r\bigr)\bigl(r + (M -  r)\,\sin ^2 \theta\bigr)}{4\,r^3}\,\Theta^2
        + \mathcal{O}(\Theta^3),
        \\
        \hat{g}_{\varphi\varphi}\left( {x,\Theta }\right)
        &= r^2 \sin ^2 \theta
        -  \frac{M \bigl(2 M -  r\bigr)\bigl(-11 M + 8 r + 11 M\,\cos(2 \theta)\bigr) \sin ^2 \theta}{32\,r^3}\,\Theta^2
        + \mathcal{O}(\Theta^3).
    \end{split}
\end{equation}
Next, let us examine the geometric properties of the spacetime described by this deformed metric.

\vspace{1em}
\noindent
\textbf{Spherical Symmetry:} 
At first glance, it is apparent that the original spherical symmetry is broken due to the NC correction. A spacetime manifold is spherically symmetric if it is invariant under the action of the rotation group $SO(3)$, whose surfaces of transitivity are two-dimensional. The metric on any of these two-dimensional surfaces must have the form
\begin{equation}
    ds^2 = r^2\left( d\theta^2 + \sin^2 \theta \, d\varphi^2 \right),
\end{equation}
where $\theta$ and $\varphi$ are the usual angular coordinates, and $r$ is the radius of curvature of the 2-sphere~\cite{Carter:2009nex}. In other words, a spherically symmetric manifold has three Killing vectors that are the same as those on $S^2$~\cite{Carroll:2004st}. These three Killing fields satisfy
\begin{equation}
\begin{split}
    [R,S] &= T, \qquad
    [S,T] = R, \qquad
    [T,R] = S.
\end{split}
\end{equation}
This is not the case for the NC-deformed metric \eqref{nccomp1}.

\vspace{1em}
\noindent
\textbf{Horizon Structure:} 
The metric has singularities at \(r=0\) and \(r=2M\), mirroring the commutative Schwarzschild case (these points correspond to \(1/g_{rr} = 0\)). This indicates that the locations of the physical singularity at \(r=0\) and the coordinate singularity at \( r=2 M \) --- the event horizon of the black hole --- remain unaffected by the NC corrections. The definition of the event horizon using the radial velocity of light also yields the same result. In reference~\cite{Chaichian:2007we}, this point was incorrectly interpreted, leading to the conclusion that the event horizon is deformed due to the NC correction\footnote{Although the authors in \cite{Chaichian:2007we} consider a \((r, \theta)\) twist, we find that our argument still holds, as we will see in the following subsections. We note that the misinterpretation of the event horizon location even persists in later works. A recently published article also contains a misleading result on horizon structure~\cite{Zhao:2023uam}, using a \((t, \theta)\) twist.}.

Light rays follow curves satisfying \(ds^2 = 0\). In the radial direction (with \(\theta\) and \(\varphi\) fixed), we have
\[
\frac{dr}{dt} = \pm \sqrt{-\frac{g_{tt}}{g_{rr}}},
\]
where the positive sign corresponds to ingoing null geodesics (as \(t\) increases, \(r\) decreases), and the negative sign to outgoing null geodesics. The velocity is zero when \(g_{tt} = 0\) or \(1/g_{rr} = 0\). As in the commutative Schwarzschild case, the surfaces where \(1/g_{rr} = 0\) and \(g_{tt} = 0\) occur at the same radii. This is further supported by our calculation of NC curvature scalars below~\eqref{tphiricci}, \eqref{tphikretch}, \eqref{tphiriccihat}.

\vspace{1em}
\noindent
\textbf{Killing Fields and Surface Gravity:} 
Before examining the Killing fields of the NC metric, it is instructive to revisit the general understanding of Killing fields in static and stationary spacetimes. A Killing horizon is formally defined as a null hypersurface generated by a Killing vector, namely, a null surface whose generators coincide with the orbits of a one-parameter group of isometries. A Killing field \(\xi^a\) is normal to a Killing horizon. Consequently, a Killing horizon can be viewed as a null hypersurface on which the norm of a Killing vector field vanishes. We also recall general results (the so-called ``rigidity theorems'') that will be useful for understanding the NC results. The first theorem, due to Carter~\cite{Carter:2009nex, Wald:1999vt}, states that for a static black hole spacetime, the Killing field \((\partial_t)^a\) must be normal to the horizon, while for a stationary axisymmetric black hole spacetime, there exists a Killing field of the form \(\xi^a = (\partial_t)^a + \Omega (\partial_\varphi)^a\), which is normal to the event horizon. Here, \(\Omega\) is the angular velocity of the event horizon. An important aspect of Carter's theorem for our purposes is that it does not rely on any field equations, which allows us to investigate its applicability in a NC metric (not derived from field equations). The second result, due to Hawking, directly proves that the event horizon of any stationary black hole must be a Killing horizon, though this depends on field equations~\cite{Hawking:1973uf, Sudarsky:1992ty, Wald:1999vt}.

Returning to the NC metric, it possesses two Killing vector fields: \((\partial_t)^a\) and \((\partial_\varphi)^a\), corresponding to time translation and axisymmetry, respectively. The surface where \(g_{tt} = 0\) is the Killing horizon. The metric is static, so \(\Omega = 0\), and the event horizon is a Killing horizon, in line with Carter's theorem. Although this is straightforward, we emphasize it because for twists involving the radial coordinate vector field~\(\partial_r\), these arguments do not hold, as we will discuss in subsequent sections.

Another essential feature of any Killing horizon is the surface gravity, defined by
\begin{equation}\label{surfeqn}
    \kappa^2 = -\frac{1}{2} (\nabla_a \xi_b)(\nabla^a \xi^b).
\end{equation}
For the metric components~\eqref{nccomp1}, we obtain
\begin{equation}
    \kappa^2 = \frac{M^2}{r^4} + \frac{154 M^5 \sin^2 \theta - 117 M^4 r \sin^2 \theta + 20 M^3 r^2 \sin^2 \theta}{16 r^9} \Theta^2 + O\left(\Theta^3\right),
\end{equation}
which has the correct commutative limit (\(\Theta \to 0\)), giving \(\kappa = 1/(4M)\) for the commutative Schwarzschild case at \(r=2M\). Remarkably, in the NC case at \(r=2M\), the \(\Theta^2\) term vanishes, and the surface gravity remains \(\kappa = 1/(4M)\). Hence, \(r=2M\) is a Killing horizon even in the NC scenario, and the surface gravity is well-defined for the \((t, \varphi)\) twisted NC spacetime.

\vspace{1em}
\noindent
\textbf{Curvature Scalars:} 
We observe that the metric exhibits a singularity at \(r=0\), as in the commutative case. Conventionally, to check whether this divergence is due to a genuine singularity or merely a coordinate artifact, one can inspect scalar invariants. In the commutative case, both \(R\) and \(R\indices{^\mu ^\nu} R\indices{_\mu _\nu}\) vanish, prompting one to consider the Kretschmann scalar \(K = R\indices{_\mu _\nu _\rho _\sigma} R\indices{^\mu ^\nu ^\rho ^\sigma}\). For the NC metric~\eqref{nccomp1}, we compute the curvature invariants and obtain
\begin{equation}\label{tphiricci}
R=\frac{\bigl[3 M^2 r \bigl(-75 M + 17 r + \cos(2 \theta) (-57 M + 33 r)\bigr) + 450 M^4 \sin^2\theta \bigr]}{16 r^8} \Theta^2  + \mathcal{O}(\Theta^3),
\end{equation}
which clearly exhibits a leading-order NC correction that diverges at \(r=0\). This confirms that \(r=0\) is indeed a true singularity. Unlike in the commutative case, the curvature scalar here is not zero, so the divergence at \(r=0\) is evident without needing to compute \(R\indices{^\mu ^\nu} R\indices{_\mu _\nu}\) or \(R\indices{_\mu _\nu _\rho _\sigma} R\indices{^\mu ^\nu ^\rho ^\sigma}\).

The scalar \(R\indices{^\mu ^\nu} R\indices{_\mu _\nu}\) vanishes at the leading (quadratic) order \(\mathcal{O}(\Theta^2)\) of NC corrections. The Kretschmann scalar is 
\begin{equation}\label{tphikretch}
K = \frac{48 M^2}{r^6} + \frac{3 M^3 \left[1032 M^2 \sin^2 \theta + r \cos(2\theta) (434 M - 81 r) + r (41 r - 322 M)\right]}{4 r^{11}} \Theta^2 + \mathcal{O}(\Theta^3),
\end{equation}
which also confirms the true singularity at \(r=0\) and reduces to the commutative result as \(\Theta \to 0\). Note that the NC term diverges more rapidly than the commutative term. The NC scalar curvature obtained from~\eqref{ncscalardef} is
\begin{equation}\label{tphiriccihat}
\begin{split}
  \hat{R} = \frac{\Theta ^2 M^2 \left(\cos (2 \theta ) \left(83 M^2-(51 M) r-3 r^2\right)-83 M^2+(23 M) r+r^2\right)}{16 r^8}+O\left(\Theta ^3\right),
\end{split}
\end{equation}
which differs from the Ricci scalar~\eqref{tphiricci}.

\subsection{($t, \theta$) twist}
For the semi-Killing ($t, \theta$) twist, $\Theta^{\mu \nu}$ is given by:
\begin{equation}
\Theta^{\mu \nu} = \begin{pmatrix}
0 & 0 & \Theta & 0 \\
0 & 0 & 0 & 0 \\
-\Theta & 0 & 0 & 0 \\
0 & 0 & 0 & 0
\end{pmatrix},\quad \mu,\nu = 0,1,2,3.
\label{4dttheta}
\end{equation}
This choice produces the following commutation relation between the coordinates:
\begin{equation}
    [t \stackrel{\star}{,} \theta] = i \, \Theta.
\end{equation}

The non-zero components of the tetrad $\hat{e}_\mu^a$ are
\begin{eqnarray}
\hat{e}_{0}^{0} &=& A-\frac{1}{16}\,\left( A^{5} A'^2 - 2\,r\,A^{4}\,A^{\prime\prime\prime}
\right)\Theta^{2} + O(\Theta^{3})\,,  \label{4dtthetab}\\
\hat{e}_{2}^{0} &=& -\frac{i}{4}\,r\,A^{2}\,A' \,\Theta + O(\Theta^{3}),  \cr
\hat{e}_{1}^{1} &=& \frac{1}{A}+\frac{1}{8}\,A^{2}\bigl(-A'^2 A - r\,A'^3 + A^2 A'' - 2\,r\,A\,A'\,A''\bigr)\,\Theta^{2} + O(\Theta^{3}),  \cr
\hat{e}_{0}^{2} &=& -\frac{i}{4}\,A^3\,A' \,\Theta + O(\Theta^{3}),  \cr
\hat{e}_{2}^{2} &=& r+\frac{1}{16}\,\bigl(2\,A^{5}\,A' - r\,A^{4}\,A'^2\bigr)\,\Theta^{2} + O(\Theta^{3}),  \cr
\hat{e}_{3}^{3} &=& r\,\sin\theta - \frac{1}{8}\,A^{3}\,A'\bigl(-2 + A^{2} + r\,A\,A'\bigr)\sin\theta\,\Theta^{2} + O(\Theta^{3}), \nonumber
\end{eqnarray}
where $A', A'', A^{\prime\prime\prime}$ are the first, second, and third derivatives of $A(r)$, respectively.  
The deformed metric $\hat{g}_{\mu \nu}(x, \Theta)$ is obtained using the definition \eqref{metricdef}. Up to second order in $\Theta$, the metric is diagonal, and its non-zero components are
\begin{equation}
    \begin{split}
\hat{g}_{00}(x,\Theta) &= -A^{2} + \frac{1}{16}\bigl(3A^{6}A'^2 - 4\,r\,A^{5}A^{\prime\prime\prime}\bigr)\Theta^{2} + O(\Theta^{4}),\\
\hat{g}_{11}(x,\Theta) &= \frac{1}{A^{2}} - \frac{1}{4}\bigl(A^2 A'^2 + r\,A\,A'^3 - A^3 A'' + 2\,r\,A^2 A' A''\bigr)\Theta^{2} + O(\Theta^{4}), \label{NCttheta1}\\
\hat{g}_{22}(x,\Theta) &= r^{2} + \frac{1}{16}\bigl(4\,r\,A^5 A' - 3\,r^2 A^{4}A'^2 \bigr)\Theta^{2} + O(\Theta^{4}), \\
\hat{g}_{33}(x,\Theta) &= r^{2}\,\sin^{2}\theta + \frac{1}{4}\bigl(2\,r\,A^3 A' - r\,A^{5}A' - r\,A^{4}A'^2 \bigr)\sin^{2}\theta \,\Theta^{2} + O(\Theta^{4}).
    \end{split}
\end{equation}
%
The NC-corrected Schwarzschild metric components under the ($t, \theta$) twist are obtained by substituting \eqref{Asch_eqn} in \eqref{NCttheta1}:
\begin{equation}
    \begin{split}
        \hat{g}_{00}(x,\Theta) &= -\left(1 - \frac{2M}{r}\right)
        + \frac{M^2(-2M + r)(-30M + 9r)}{48\,r^6}\,\Theta^2
        + \mathcal{O}(\Theta^3), \\
        \hat{g}_{11}(x,\Theta) &= \left(1 - \frac{2M}{r}\right)^{-1}
        - \frac{M (11M^2 - 10Mr + 2r^2)}{4\,r^4(-2M + r)}\,\Theta^2
        + \mathcal{O}(\Theta^3), \\
        \hat{g}_{22}(x,\Theta) &= r^2
        + \frac{M(11M - 4r)(2M - r)}{16\,r^3}\,\Theta^2
        + \mathcal{O}(\Theta^3), \\
        \hat{g}_{33}(x,\Theta) &= r^2 \sin^2\theta
        + \frac{M(-2M + r)(M + r)\,\sin^2\theta}{4\,r^3}\,\Theta^2
        + \mathcal{O}(\Theta^3).
    \end{split}
\end{equation}

\vspace{1em}
\noindent
\textbf{Properties of the deformed spacetime:}
As before, we analyze the spacetime properties. Although the deformed metric is diagonal, the qualitative observations remain the same as in the $(t, \varphi)$ twist. Due to the $(t, \theta)$ twist, the spherical symmetry of the system is broken. The deformed metric exhibits two singularities: $r=0$, which is a true singularity, and $r=2M$, the event horizon (also a Killing horizon). The location of the horizon is not affected by the NC deformation.

The surface gravity is calculated using \eqref{surfeqn} and is given by:
\begin{equation}
    \kappa^2 = \frac{M^2}{r^4}
    + \frac{\left(154 M^5 - 117 M^4 r + 20 M^3 r^2\right)}{16 r^9} \Theta^2
    + \mathcal{O}\left(\Theta^3\right).
\end{equation}
At $r=2M$, the $\Theta^2$ term vanishes, and the surface gravity is $\kappa = 1/(4M)$, indicating that there is no NC correction to the surface gravity.

The curvature scalars are also calculated as follows:
\begin{equation}
\begin{split}
R &= \frac{3 M^2 (75 M^2 - 47 M r + 7 r^2)}{8 r^8} \Theta^2 + \mathcal{O}(\Theta^3), \\
R^{\mu \nu} R_{\mu \nu} &= 0, \\
R_{\mu \nu \rho \sigma} R^{\mu \nu \rho \sigma} &= \frac{48 M^2}{r^6} 
+ \frac{6 M^3 (129 M^2 - 90 M r + 14 r^2)}{r^{11}} \Theta^2 
+ \mathcal{O}(\Theta^3).
\end{split}
\end{equation}
By examining the divergence of these curvature scalars, we confirm that $r=0$ is a physical singularity and $r=2M$ is a coordinate singularity. The Ricci scalar $R$ has a leading NC term, and in the Kretschmann scalar, the NC term diverges more rapidly than the commutative term. Finally, the NC scalar curvature, calculated from \eqref{ncscalardef}, is
\begin{equation}
  \hat{R} =  \frac{ M^2 \left(-83 M^2+29 M r+2 r^2\right)}{8 r^8} \Theta^2 \, .
\end{equation}

\subsection{($\theta, \varphi$) twist}

For the \((\theta, \varphi)\) noncommutativity, the parameter \(\Theta^{\mu \nu}\) is given by
\begin{equation}
\Theta^{\mu \nu } \;=\; \begin{pmatrix}
0 & 0 & 0 & 0\\
0 & 0 & 0 & 0\\
0 & 0 & 0 & \Theta\\
0 & 0 & -\Theta & 0
\end{pmatrix}
,\quad \mu ,\nu =0,1,2,3.  \label{4dthetaphi}
\end{equation}
This choice leads to the following commutation relation between the coordinates:
\begin{equation}
    [\theta \stackrel{\star}{,} \varphi] = i \, \Theta. 
\end{equation}
This twist was also studied in \cite{Wang:2008ut} and \cite{Sun:2010nas}. The non-zero components of the tetrad \(\hat e_\mu^a\) are
\begin{eqnarray}
\hat{e}_{0}^{0}&=&A+\frac{1}{32}\,r\, A^2\, A^{\prime }\left( -5 + 11\, \cos (2\theta) + 8\,A^2\,\sin ^2\theta \right)\,\Theta ^{2}+O(\Theta {^{3}})\,,   \cr
\hat{e}_{1}^{1}&=&\frac{1}{A}+\frac{1}{32}\,r\, A^{\prime }\left( -5 + 11\, \cos (2\theta) + 8\,A^2\,\sin ^2\theta \right)\,\Theta ^{2}+O(\Theta {^{3}})\,,  \cr
\hat{e}_{1}^{2}&=&\frac{1}{4}\,r\,A\,A^{\prime}
\,\cos \theta \sin \theta\,\Theta^2 +O(\Theta {^{3}}) ,  \cr
\hat{e}_{2}^{2}&=&r-\frac{1}{16}\,r\,\bigl( -1 + A^2 \bigr)\,\bigl( -5 + 11\, \cos (2\theta) + A^2\,\sin ^2\theta \bigr)\,\Theta ^{2}+O(\Theta {^{3}})\,, \label{4dthetaphib}\\
\hat{e}_{2}^{3}&=&-\frac{i}{4}\, r\, \bigl( -1 + A^2 \bigr) \sin  \theta \,\Theta +O(\Theta {^{3}}) ,  \cr
\hat{e}_{3}^{1}&=&-\frac{i}{2}\,r\, A\, \cos \theta \,\sin \theta \,\Theta +O(\Theta {^{3}}) ,  \cr
\hat{e}_{3}^{2}&=&-\frac{i}{8}\,r \,\bigl( -1 + 3\, \cos (2\theta) + 2 \, {A}^{2}\,\sin ^2\theta \bigr)\,\Theta +O(\Theta {^{3}}) ,  \cr
\hat{e}_{3}^{3}&=&r\sin \theta +\frac{1}{32}\,r\,\sin \theta \,\Bigl( -9 + 33\, \cos (2\theta) + 8\,A^2\bigl(1 + 4\, \cos (2\theta)\bigr) - 2\, A^4\,\sin ^2\theta \Bigr)\,\Theta ^{2}+O(\Theta {^{3}})\,. \nonumber
\end{eqnarray}
Here, \(A',\,A'',\,A'''\) are the first, second, and third derivatives of \(A(r)\), respectively. The deformed metric \(\hat g_{\mu \nu} (x, \Theta)\) is obtained from the definition~\eqref{metricdef}. Up to second order in \(\Theta\), the metric is off-diagonal, and its non-zero components are
\begin{eqnarray}
\hat{g}_{00}\bigl(x,\Theta\bigr) &=& -A^{2}-\frac{1}{16}\,r\, A^3\, A^{\prime }\Bigl( -5 + 11\, \cos (2\theta) + 8\,A^2\,\sin ^2\theta \Bigr)\,\Theta ^{2}+O(\Theta {^{3}})\,, \cr
\hat{g}_{11}\bigl(x,\Theta\bigr)
&=& \frac{1}{A^{2}}+ \frac{A^{\prime }}{16\,A}\,r\, \bigl( -5 + 11\, \cos (2\theta) + 8\,A^2\,\sin ^2\theta \bigr)\,\Theta ^{2}+O(\Theta {^{3}})\,, \label{NCthetaphi1}\\
\hat{g}_{12}\bigl(x,\Theta\bigr)  &=&  \hat{g}_{21}\bigl(x,\Theta\bigr) \;=\; \frac{1}{4}\,r^2\,A\,A^{\prime}\,\cos \theta\,\sin  \theta \,\Theta ^{2}+O(\Theta^{3}), \cr
\hat{g}_{22}\bigl(x,\Theta\bigr) &=& r^{2}-\frac{1}{32}\,r^2\,\bigl( -1 + A^2 \bigr)\,\bigl( 1 - 21\, \cos (2\theta) + 2\,A^2\,\sin ^2\theta \bigr)\,\Theta ^{2}+O(\Theta {^{3}})\,,  \cr
\hat{g}_{33}\bigl(x,\Theta\bigr) &=& r^{2}\,\sin ^{2}\theta 
+ \frac{1}{128}\,r^2\,\Bigl[ 41 - 108\,\cos (2\theta) + 75\,\cos (4\theta) \cr
&&\qquad\qquad + 8\, A^2\Bigl( 9 + 37\, \cos (2\theta)\,\sin ^2\theta - 8\, A^4\, \sin ^4\theta \Bigr)\Bigr] \,\Theta ^{2}+O(\Theta ^{3}). \nonumber
\end{eqnarray}
%
The NC-corrected Schwarzschild metric components under the \((\theta , \varphi )\) twist are obtained by substituting \eqref{Asch_eqn} into \eqref{NCthetaphi1},
\begin{equation}
    \begin{split}
        \hat{g}_{00}\bigl(x,\Theta\bigr) &= -\left( 1 - \frac{2 M}{r} \right) + \frac{M (2 M -  r)  \bigl(-8 M -  r + (8 M + 7 r) \cos 2 \theta \bigr)}{16 r^3} \Theta^2 + \mathcal{O}(\Theta^3), \\
        \hat{g}_{11}\bigl(x,\Theta\bigr) &= \left( 1 - \frac{2 M}{r} \right)^{-1} + \frac{M  \bigl(8 M + r -  (8 M + 7 r) \cos 2 \theta \bigr)}{16 (2 M -  r) r} \Theta^2 + \mathcal{O}(\Theta^3),\\
        \hat{g}_{12}\bigl(x,\Theta\bigr) &= \hat{g}_{21}\bigl(x,\Theta\bigr) = \frac{1}{8} M  \sin 2 \theta  \Theta^2 + \mathcal{O}(\Theta^3), \\
        \hat{g}_{22}\bigl(x,\Theta\bigr) &= r^2 + \frac{1}{8} M\bigl(- M + r + (M - 11 r) \cos 2 \theta \bigr)  \Theta^2  + \mathcal{O}(\Theta^3), \\
        \hat{g}_{33}\bigl(x,\Theta\bigr) &= r^2 \sin^2 \theta 
        + \frac{1}{32}\biggl[\,2 (2 M^2 - 30 M r + r^2) \cos 2 \theta \\
        &\qquad\quad + M \bigl(-3 M + 22 r -  (M - 38 r) \cos 4 \theta \bigr)\biggr]  \Theta^2  + \mathcal{O}(\Theta^3).
    \end{split}
\end{equation}

\vspace{1em}
\noindent
\textbf{Properties of the deformed spacetime:} The \((\theta, \varphi)\)-twisted metric is off-diagonal, and its spacetime properties are qualitatively similar to those of the \((t, \varphi)\) and \((t, \theta)\) twists. The spherical symmetry of the system is broken by the NC deformation. The deformed metric exhibits two singularities: \(r=0\), which is a true singularity, and \(r=2M\), the event horizon (also a Killing horizon). The location of the horizon is not altered by the NC deformation. The surface gravity is calculated using \eqref{surfeqn} and is given by
\begin{equation}
    \kappa^2 = \frac{M^2}{r^4} + \frac{M^2 \left[32 M^2 \cos (2\theta) - 32 M^2 - 2 M r \cos (2\theta) + 14 M r - 7 r^2 \cos (2\theta) + r^2\right]}{16 r^6} \Theta^2 + \mathcal{O}\bigl(\Theta^3\bigr) .
\end{equation}
At \(r=2M\), the \(\Theta^2\) term vanishes, and the surface gravity remains \(\kappa = 1/(4M)\), indicating that there is no NC correction to the surface gravity. The curvature scalars are also calculated as follows:
\begin{equation}
\begin{split}
R &= -\frac{\Bigl[3 M \bigl(48 M^2 \sin^2\theta - 11 M r + 41 r^2\bigr) + 3 M r \cos (2\theta) (87 r - 19 M) + r^3 \csc^4\theta\Bigr]}{8 r^5} \Theta^2 + \mathcal{O}\bigl(\Theta^3\bigr), \\
R\indices{^\mu ^\nu} R\indices{_\mu _\nu} &= 0, \\
R\indices{_\mu _\nu _\rho _\sigma} R\indices{^\mu ^\nu ^\rho ^\sigma} &= \frac{48 M^2}{r^6} + \frac{M \Bigl(6 M \cos (2\theta) (13 M - 54 r) + 24 M (M - 6 r) - r^2 \csc^4\theta\Bigr)}{r^7} \Theta^2 + \mathcal{O}\bigl(\Theta^3\bigr).
\end{split}
\end{equation}
As before, by observing the divergence of these curvature scalars, we confirm that \(r=0\) is a physical singularity and \(r=2M\) is a coordinate singularity. The Ricci scalar \(R\) has a leading NC term, and in the Kretschmann scalar, the NC term diverges more rapidly than the commutative term. Finally, the NC scalar curvature, calculated from \eqref{ncscalardef}, is
\begin{equation}
  \hat{R} =   -\frac{\Theta ^2 \Bigl(M \Bigl(\cos (2 \theta ) \bigl(80 M^2-17 M r+114 r^2\bigr)-80 M^2+55 M r+16 r^2 \csc ^2(\theta )+70 r^2\Bigr)\Bigr)}{16 r^5}
  +O\bigl(\Theta ^3\bigr).
\end{equation}

\subsection{($t, r$) twist}
The ($t,r$)-twist is a semi-Killing twist defined by $\Theta ^{\mu \nu}$ given by
\begin{equation}
\Theta ^{\mu \nu }=\left(
\begin{array}{cccc}
0 & \Theta & 0 & 0 \\
-\Theta & 0 & 0 & 0 \\
0 & 0 & 0 & 0 \\
0 & 0 & 0 & 0
\end{array}
\right),\quad \mu,\nu =0,1,2,3.  
\label{4dtr}
\end{equation}
This choice produces the following commutation relation between the coordinates:
\begin{equation}
    [t\stackrel{\star}{,} r] = i \, \Theta. 
\end{equation}
This twist was also studied in \cite{Herceg:2023pmc}, where NC perturbations of the metric were derived, but using a different formalism. The non-zero components of the tetrad $\hat e_\mu^a$ are
\begin{eqnarray}
\hat{e}_{0}^{0}&=&A-\frac{1}{16}\,\left( 20\,A\,{A}^{\prime}{}^{4} + 69\,A^{2}\,{A}^{\prime}{}^{2}\,{A}^{\prime\prime}
+ 12\,A^{3}\,{A}^{\prime\prime}{}^{2} + 12\,{A}^{3}\,{A}^{\prime}\,{A}^{\prime\prime\prime}
\right) \Theta ^{2} + O(\Theta {^{3}})\,, 
\label{4dtrb}\\
\hat{e}_{1}^{0}&=&-\frac{i}{4} {\,{A}^{\prime \prime }}
\,\Theta +O(\Theta {^{3}}) ,  \cr
\hat{e}_{1}^{1}&=&\frac{1}{A}-\frac{1}{16}\,\left( 17\,{A}^{\prime}{}^{2}{A}^{\prime\prime}
+ 5\,A\,{A}^{\prime\prime}{}^{2} + 5\,A\,{A}^{\prime} {A}^{\prime\prime\prime}
\right)\Theta^{2} + O(\Theta {^{3}}),  \cr
\hat{e}_{0}^{1}&=&-\frac{i}{4}\, A\, \left( 3\,{A}^{\prime}{}^{2}+2\,A\,{A}^{\prime\prime}\right)
\,\Theta +O(\Theta {^{3}}) ,  \cr
\hat{e}_{2}^{2}&=&r - \left( A\,{A}^{\prime}{}^{3} + \frac{17}{16}\,{A}^{2}\,{A}^{\prime}\,{A}^{\prime\prime}\right)
\,\Theta ^{2}+O(\Theta {^{3}}) ,  \cr
\hat{e}_{3}^{3}&=&r\sin \theta -\left( A\,{A}^{\prime}{}^{3} + \frac{17}{16}\,{A}^{2}\,{A}^{\prime}\,{A}^{\prime\prime}\right)\sin \theta \,\Theta ^{2}+O(\Theta {^{3}}), \nonumber
\end{eqnarray}
where ${A}^{\prime}$, ${A}^{\prime\prime}$, and ${A}^{\prime\prime\prime}$ are the first, second, and third derivatives of $A(r)$, respectively. The deformed metric $\hat g_{\mu\nu}(x,\Theta)$ is obtained using the definition~\eqref{metricdef}. Up to second order in $\Theta$, the metric remains diagonal, and the non-zero components are
\begin{eqnarray} 
\hat{g}_{00}\left( x,\Theta \right) &=& -A^{2}
+\frac{1}{16}\,{A}^{2}\bigl(
49\,{A}^{\prime}{}^{4} + 150\,A\,A^{\prime}{}^{2}\,{A}^{\prime \prime } 
+ 28\,A^{2}\,{A}^{\prime\prime}{}^{2}
+ 24\,A^{2}\,{A}^{\prime}\,{A}^{\prime\prime\prime}\bigr)\Theta^{2}
+O\bigl(\Theta^{4}\bigr),\cr
\hat{g}_{11}\left( x,\Theta \right) &=& \frac{1}{A^{2}}
-\frac{1}{16}\,\biggl( \frac{34\,A^{\prime}{}^{2}A^{\prime\prime}}{A}
+11 \,A^{\prime\prime}{}^{2} +10 \,A^{\prime} \,A^{\prime\prime\prime}\biggr)\Theta^{2}
+O\bigl(\Theta^{4}\bigr), 
\label{NCtr1}\\
\hat{g}_{22}\left( x,\Theta \right) &=& r^{2}
-\frac{1}{8}\bigl(
16\,r\,A\,A^{\prime}{}^{3}+17\,r\,A^{2}\,A^{\prime}\,A^{\prime\prime}\bigr)\Theta^{2}
+O\bigl(\Theta^{4}\bigr), \cr
\hat{g}_{33}\left( x,\Theta \right) &=& r^{2}\,\sin^{2}\theta
-\frac{1}{8}\bigl(16\,r\,A\,A^{\prime}{}^{3}+17\,r\,A^{2}\,A^{\prime}\,A^{\prime\prime}\bigr)\sin^{2}\theta\;\Theta^{2}
+O\bigl(\Theta^{4}\bigr). \nonumber
\end{eqnarray}
%
The NC-corrected Schwarzschild metric components under the ($t, r$) twist are obtained by substituting \eqref{Asch_eqn} in \eqref{NCtr1}:
\begin{equation}
    \begin{split}
        \hat{g}_{00}\left( x,\Theta \right) &= -\left( 1 - \frac{2 M}{r} \right) 
        + \frac{M^2 (1111 M^2 - 1068 M r + 256 r^2)}{16 r^7 (-2 M + r)}  \Theta^2 
        + \mathcal{O}(\Theta^3), \\
        \hat{g}_{11}\left( x,\Theta \right) &= \left( 1 - \frac{2 M}{r} \right)^{-1}
        + \frac{M^2 (-351 M^2 + 380 M r - 104 r^2)}{16 r^5 (-2 M + r)^3} \Theta^2 
        + \mathcal{O}(\Theta^3),\\
        \hat{g}_{22}\left( x,\Theta \right) &= r^2 
        -  \frac{M^2 (67 M - 34 r)}{8 r^4 (-2 M + r)} \Theta^2 
        + \mathcal{O}(\Theta^3), \\
        \hat{g}_{33}\left( x,\Theta \right) &= r^2 \sin ^2 \theta 
        -  \frac{M^2 (67 M - 34 r)  \sin^2\theta}{8 r^4 (-2 M + r)} \Theta^2 
        + \mathcal{O}(\Theta^3). 
    \end{split}
\end{equation}

\vspace{1em}
\noindent
\textbf{Properties of the Deformed Spacetime:} 
The \((t,r)\) twisted metric is diagonal, and the spherical symmetry of the system is preserved in the NC deformation\footnote{Due to the twist being defined without a prefered direction, unlike twists containing, e.g. $\partial_\phi$ or $\partial_\theta$}. The spacetime properties differ significantly from those of the previously discussed twists because this twist involves the radial coordinate vector field $\partial_r$. The deformed metric exhibits two singularities: \(r=0\) and \(r=2M\) (these points correspond to \(1/g_{rr} = 0\)). This indicates that the locations of the physical singularity at \(r=0\) and the coordinate singularity, which corresponds to the event horizon of the black hole, are not changed by the NC corrections. However, unlike in the commutative Schwarzschild case and the previously discussed deformed metrics, the surfaces where \(1/g_{rr} = 0\) and \(g_{tt} = 0\) occur at different radii (i.e., the event horizon and the Killing horizon are distinct). The former has no NC corrections, while the latter does. In other words, in the Schwarzschild metric, the event horizon is the surface where the purely temporal component \(g_{tt}\) changes sign from positive to negative, and this occurs at a different surface in this deformed metric (\(g_{tt} = 0\) surfaces, which are not singularities).

The surface gravity is calculated using \eqref{surfeqn} and is given by
\begin{equation} 
\kappa^2 = \frac{M^2}{r^4} 
+ \frac{\left(-3523 M^6 + 5064 M^5 r - 2419 M^4 r^2 + 384 M^3 r^3\right)}{4 r^{10} (2 M - r)^2} \Theta^2 
+ \mathcal{O}\bigl(\Theta^3\bigr). 
\end{equation}
At \(r=2M\), the \(\Theta^2\) term diverges, making the definition of surface gravity unclear. This indicates that the twist in the radial direction renders the usual \(r\) coordinate unsuitable for analyzing the properties of the deformed spacetime. This issue is also evident from the curvature scalars, which are calculated as follows:
\begin{equation} 
\begin{split} 
R &= \frac{M^2 \left(60816 M^4 - 112574 M^3 r + 77605 M^2 r^2 - 23580 M r^3 + 2660 r^4\right)}{8 r^9 (r - 2 M)^3} \Theta^2 + \mathcal{O}\bigl(\Theta^3\bigr), \\ 
R\indices{^\mu ^\nu} R\indices{_\mu _\nu} &= 0, \\ 
R\indices{_\mu _\nu _\rho _\sigma} R\indices{^\mu ^\nu ^\rho ^\sigma} 
&= \frac{48 M^2}{r^6} 
+ \frac{M^3 \left(141852 M^4 - 269720 M^3 r + 191683 M^2 r^2 - 60318 M r^3 + 7088 r^4\right)}{r^{12} (r - 2 M)^3} \Theta^2 
+ \mathcal{O}\bigl(\Theta^3\bigr). 
\end{split} 
\end{equation}
By observing the divergence of the curvature scalars, we confirm that \(r=0\) is a physical singularity. However, the divergence at \(r=2M\) again suggests that the radial coordinate is not suitable for analyzing the properties of the deformed spacetime. The Ricci scalar \(R\) has a leading NC term, and in the Kretschmann scalar, the NC term diverges more rapidly than the commutative term. Finally, the NC scalar curvature, calculated from \eqref{ncscalardef}, is given by
\begin{equation}
  \hat R =  \frac{\Theta ^2 M^3 \left(-1667 M^2 + 1672 M r - 416 r^2\right)}{4 r^9 (r-2 M)^2} + O\bigl(\Theta ^3\bigr).
\end{equation}


\subsection{($r , \theta$) twist}
The $(r,\theta)$ twist is non-Killing and is defined by $\Theta ^{\mu \nu}$ as
\begin{equation}
\Theta ^{\mu \nu }=\left(
\begin{array}{cccc}
0 & 0 & 0 & 0 \\
0 & 0 & \Theta  & 0 \\
0 & -\Theta & 0 & 0 \\
0 & 0 & 0 & 0
\end{array}
\right) ,\quad \mu ,\nu =0,1,2,3.  \label{4drtheta}
\end{equation}
This choice leads to the following commutation relation between the coordinates:
\begin{equation}
    [r\stackrel{\star}{,} \theta] = i \, \Theta. 
\end{equation}
In fact, this twist was previously explored in reference \cite{Chaichian:2007we} to derive NC corrections to the Schwarzschild black hole. However, due to the identification of new correction terms in \eqref{3.12}, we revisit this case to include the previously overlooked contributions. The new non-zero components of the tetrad $\hat{e} _\mu ^{a}$ are
\begin{eqnarray}
\hat{e}_{0}^{0}&=&A+\frac{1}{4}\,\left( 2\,r\,{A}^{\prime
}{}^{3}+5\,r\,A\,{A}^{\prime }\,{A}^{\prime \prime
}+r\,A^{2}\,{A}^{\prime \prime \prime }+2\,A\, {A}^{\prime
}{}^{2}+A^{2}\,{A}^{\prime \prime }\right) \Theta ^{2}+O(\Theta {^{3}})\,,  \label{4drthetab}\\
\hat{e}_{1}^{1}&=&\frac{1}{A}+\frac{{A}^{\prime \prime}}{4}\,\Theta
^{2}+O(\Theta {^{3}}),  \cr
\hat{e}_{2}^{1}&=&-\frac{i}{4}\left( A+2\,r\,{A}^{\prime }\right)
\,\Theta +O(\Theta {^{3}}) ,  \cr
\hat{e}_{2}^{2}&=&r+\frac{1}{4}\,\left( 2A\,{A}^{\prime
}+3\,r\,{A}^{\prime }{}^{2}+3\,r\,A\,{A}^{\prime \prime }\right)
\Theta ^{2}+O(\Theta {^{3}}) ,  \cr
\hat{e}_{3}^{3}&=&r\sin \theta -\frac{i}{4}\left( \cos \theta
\right) \Theta +\frac{1}{4}\,\left( 2r\,{A}^{\prime }{}^{2}+r A
{A}^{\prime \prime }+2A{A}^{\prime }-\frac{{A}^{\prime }}{A}\right)
\sin \theta \,\Theta ^{2}+O(\Theta {^{3}}), \nonumber
\end{eqnarray}
where ${A}^{\prime },\,{A}^{\prime \prime },\,{A}^{\prime \prime
\prime }$ are the first, second, and third derivatives of $A(r)$,
respectively. The deformed metric $\hat g _{\mu \nu} (x, \Theta)$ is obtained by using the definition \eqref{metricdef}. Up to second order in $\Theta$, the metric is diagonal, and its non-zero components are
\begin{eqnarray}
\hat{g}_{tt}\left( x,\Theta \right) &=&-A^{2}-\frac{1}{2}\,\left(
2\,r\,A\,{ A}^{\prime }{}^{3}+r\,A^{3}\,{A}^{\prime \prime \prime
}+A^{3}\,{A}^{\prime \prime }+2\,A^{2}\,{A}^{\prime
}{}^{2}+5\,r\,A^{2}\,{A}^{\prime }\,{A} ^{\prime  \prime }\right)
{\Theta }^{2}+O\left(\Theta ^{4}\right) ,\cr
\hat{g}_{rr}\left( x,\Theta \right)
&=&\frac{1}{A^{2}}+\frac{1}{2}\,\frac{{ A}^{\prime \prime
}}{A}\,{\Theta }^{2}+O\left(\Theta ^{4}\right) , \label{NCrtheta1}\\
\hat{g}_{\theta\theta}\left( x,\Theta \right) &=&r^{2}+\frac{1}{16}\,\left(
A^{2}+20\,r\,A\,{A}^{\prime }+28\,r^{2}\,{A}^{\prime
}{}^{2}+24\,r^{2}A\,{A} ^{\prime \prime }\right) {\Theta }^{2}+O
\left(\Theta ^{4}\right) , \cr
\hat{g}_{\varphi\varphi}\left( x,\Theta \right) &=&r^{2}\,\sin ^{2}\theta
\cr
&+&\frac{1 }{16}\left[4\,\left(4\,r\,A\,{A}^{\prime
}-\,2r\frac{{A}^{\prime }}{A} +2\,r^{2}\,A\,{A}^{\prime \prime
}+4\,r^{2}\,{A}^{\prime }{}^{2} +1\right) \sin ^{2}\theta +5\cos
^{2}\theta \right] {\Theta }^{2}+O\left(\Theta ^{4}\right) . \nonumber
\end{eqnarray}
%
The NC-corrected Schwarzschild metric components under the $(r, \theta)$ twist are obtained by substituting \eqref{Asch_eqn} into \eqref{NCrtheta1},
\begin{equation}
    \begin{split}
        \hat{g}_{tt}\left( x,\Theta \right) &= -\left(1 - \frac{2 M}{r} \right)+ \frac{M (11 M - 4 r)}{2 r^4}  \Theta^2 + \mathcal{O}(\Theta^3), \\
        \hat{g}_{rr}\left( x,\Theta \right) &= \left(1 -  \frac{2 M}{r}\right)^{-1} + \frac{M (3 M - 2 r) }{2 r^2 (r-2 M)^2} \Theta^2 + \mathcal{O}(\Theta^3), \\
        \hat{g}_{\theta\theta}\left( x,\Theta \right) &= r^2 + \left(\frac{1}{16} -  \frac{2 M}{r} + \frac{M}{8 (r-2 M )}\right) \Theta^2 + \mathcal{O}(\Theta^3), \\
        \hat{g}_{\varphi\varphi}\left( x,\Theta \right) &= r^2 \sin ^2\theta + \left(\frac{5 \cos ^2\theta }{16}+\frac{\sin ^2\theta  \left(2 M^2-4 M r+r^2\right)}{4 r (r-2 M)}\right)\Theta ^2 +\mathcal{O}\left(\Theta ^3\right).
    \end{split}
\end{equation}

\vspace{1em}
\noindent
\textbf{Properties of the Deformed Spacetime:} The $(r,\theta)$ twisted metric is diagonal, and the spherical symmetry of the system is broken due to the NC deformation. As in the case of the $(t, r)$ twist, the radial coordinate is unsuitable for analyzing certain spacetime properties. The deformed metric exhibits two singularities: $r=0$ and $r=2M$ (these points correspond to $1/g_{rr} = 0$). Hence, the locations of the physical singularity at $r=0$ and the coordinate singularity at $r=2M$ (the black hole event horizon) are not shifted by the NC corrections. Similar to the $(t, r)$ twisted metric, the surfaces where $1/g_{rr} = 0$ and $g_{tt} = 0$ occur at different radii (i.e., the event horizon and the Killing horizon do not coincide). The former is unaffected by NC corrections, while the latter is not.

The surface gravity, calculated using \eqref{surfeqn}, is
\begin{equation}
    \kappa^2 = \frac{M^2}{r^4} + \frac{\left(-6 M^2 r^2 + 31 M^3 r - 37 M^4\right)}{r^7 (2 M - r)} \Theta^2 + \mathcal{O}\left(\Theta^3\right).
\end{equation}
At $r=2M$, the $\Theta^2$ term diverges, making the definition of surface gravity unclear. The curvature scalars are computed as follows:
\begin{equation}
\begin{split}
R &= \frac{1}{8 r^6} \left[\frac{4 M^3 (12 M - 7 r)}{(r - 2 M)^2} + 10 M r \csc^2\theta + M (215 r - 744 M) - 5 r^2 \csc^4\theta - r^2\right] \Theta^2 + \mathcal{O}\left(\Theta^3\right), \\
R\indices{^\mu ^\nu} R\indices{_\mu _\nu} &= 0, \\
R\indices{_\mu _\nu _\rho _\sigma} R\indices{^\mu ^\nu ^\rho ^\sigma} &= \frac{48 M^2}{r^6} + \frac{1}{2 r^9} M \biggl[-\frac{2 \left(1848 M^4 - 2392 M^3 r + 975 M^2 r^2 - 120 M r^3 + r^4\right)}{(r - 2 M)^2} \\
& \quad + 5 r \csc^2\theta (3 r - 8 M) - 10 r^2 \csc^4\theta\biggr] \Theta^2 + \mathcal{O}\left(\Theta^3\right).
\end{split}
\end{equation}
As before, examining the divergence of these curvature scalars shows that $r=0$ is a physical singularity. The Ricci scalar $R$ gains a leading NC contribution, and in the Kretschmann scalar, the NC term diverges faster than the commutative term. Finally, the NC scalar curvature is calculated from \eqref{ncscalardef} as
\begin{equation}
  \hat{R} =  \frac{ M \left(224 M^3-4 M^2 r \csc ^2(\theta )-384 M^2 r+209 M r^2-36 r^3\right)}{4 r^6 (r-2 M)^2} \, \Theta ^2+O\left(\Theta ^3\right).
\end{equation}


\subsection{($r, \varphi$) twist}
The semi-Killing \((r, \varphi)\) twist is defined by \(\Theta ^{\mu \nu}\) given by
\begin{equation}
\Theta ^{\mu \nu }=\left(
\begin{array}{cccc}
0 & 0 & 0 & 0 \\
0 & 0 & 0  &  \Theta \\
0 & 0 & 0 & 0 \\
0 & -\Theta & 0 & 0
\end{array}
\right) ,\quad \mu ,\nu =0,1,2,3.  \label{4drphi}
\end{equation}
This choice produces the following commutation relation between the coordinates:
\begin{equation}
    [r\stackrel{\star}{,} \varphi] = i \, \Theta. 
\end{equation}
This twist was also used in~\cite{Herceg:2023zlk}. The non-zero components of the tetrad \(\hat e _\mu ^{a}\) are
\begin{eqnarray}
\hat{e}_{0}^{0}&=&A+\frac{1}{4}\,\left( {2\, r\, \,{A}^{\prime
}{}^{3}+2\,\,A\, {A}^{\prime }{}^{2}+\,5\, r\,\,A\, {A}^{\prime }{A}^{\prime \prime}+\, A^2\, {A}^{\prime \prime  }\, +\, r\, A^2\,A^{\prime \prime \prime}} \right)\, \sin ^2 \theta  \,\,\Theta ^{2}+O(
\Theta {^{3}})\,,  \label{4drphib}\\
\hat{e}_{1}^{1}&=&\frac{1}{A}+\frac{1}{4}\,{A}^{\prime \prime }\, \sin ^2 \theta\,\Theta
^{2}+O( \Theta {^{3}}),  \cr
\hat{e}_{1}^{2}&=&\frac{1}{8 \, {A}^{2}}{\,\left(- A^{\prime}{}^{2}+\,A\, A^{\prime \prime}\right)}
\,\cos \theta \sin \theta\,\Theta^2 +O( \Theta {^{3}}) ,  \cr
\hat{e}_{2}^{1}&=&\frac{1}{8}{\,{A}^{3}\, A^{\prime \prime }}
\,\cos \theta \sin \theta\,\Theta^2 +O( \Theta {^{3}}) ,  \cr
\hat{e}_{2}^{2}&=&r+\,\frac{1}{8}\, \left[- \frac{A^{\prime}}{A}+2 \, \sin ^2 \theta\, \left( {2\, {A}\, A^{\prime }+2\, r\,{A}^{\prime  }{}^{2}\,+\, r\, A\, A^{\prime \prime}}\right) \right]
\,\Theta ^{2}+O( \Theta {^{3}}) ,  \cr
\hat{e}_{2}^{3}&=&-\frac{i}{4}\, \cos \theta \,\Theta +O( \Theta {^{3}}) ,  \cr
\hat{e}_{3}^{1}&=&-\frac{i}{4} \left(\, A\,+\,2\,r\,{A}^{\prime} \, \right)\,\sin ^2\theta \,\Theta +O( \Theta {^{3}}) ,  \cr
\hat{e}_{3}^{2}&=&-\frac{i}{4}\, \cos \theta \,\sin \theta \,\Theta +O( \Theta {^{3}}) ,  \cr
\hat{e}_{3}^{3}&=&r\sin \theta +\,\frac{1}{8 \, A}\, \left[ {-\, \cos ^2 \theta \, A^{\prime} +\,2\, A\, \sin ^2 \theta \, \left(\,3\,r\, A^{\prime}{}^{2}\,+\,2\,A\,{A}^{\prime } +3\, r\, A\, A^{\prime \prime}\right) }\right] \sin \theta \,\, \Theta ^{2}+O(\Theta {^{3}}), \nonumber
\end{eqnarray}
where \({A}^{\prime },\,{A}^{\prime \prime },\,{A}^{\prime \prime
\prime }\) are the first, second and third derivatives of \(A(r)\),
respectively. The deformed metric \(\hat g _{\mu \nu} (x, \Theta) \) is obtained by using the definition~\eqref{metricdef}. Up to second order in \(\Theta\), the metric is off-diagonal, and the non-zero components are
\begin{eqnarray}
\hat{g}_{00}\left( {x,\Theta }\right) &=&-A^{2}-\frac{1}{2}\,\left(\,2\,
{{A}^{2}\,{ A}^{\prime }{}^{2}+2\,r\,A\,{A}^{\prime }{}^{3}\,+\, A^3\,{A}^{\prime \prime}\, + \, 5 r\, A^2\, {A}^{\prime }\, {A}^{\prime \prime}+\,r\,A^3\,{A}^{\prime \prime \prime }}\right)\,\sin ^2 \theta
\,{\Theta }^{2}+O( {\Theta ^{3}}) ,\cr
\hat{g}_{11}\left( {x,\Theta }\right)
&=&\frac{1}{A^{2}}+\, \frac{A^{\prime \prime} }{2\,A}\, \sin ^2 \theta \, {\Theta }^{2}+O( {\Theta ^{3}}) , \label{NCrphi1}\\
\hat{g}_{12}\left( {x,\Theta }\right) \,&=&  \hat{g}_{21}\left( {x,\Theta }\right) =-\frac{1}{8}\,
{ \, \frac{1}{A^2} \left( \, r\,{A}^{\prime}{}^{2} + \, A\, A^{\prime}\,+\, r \, A \, {A}^{\prime \prime}  \right) \,\cos \theta \, \sin  \theta } \,{\Theta }^{2}+O(
{\Theta ^{3}}) , \cr
\hat{g}_{22}\left( {x,\Theta }\right) &=&r^{2}+\frac{1}{16}\,\left[ {\, \cos ^2 \theta \,- \,4\,r\,\frac{A^{\prime}}{A}+16 \left( \, r\,{A}\,{A}^{\prime} +\, r^2  \, {A}^{\prime} {}^{2} +\frac{1}{2}\,r^2 \, A\, A^{\prime \prime}\right) \, \sin ^2 \theta }\right] \,{\Theta }^{2}+O(
{\Theta ^{3}}) , \cr
\hat{g}_{33}\left( {x,\Theta }\right) &=&r^{2}\,\sin ^{2}\,{\theta
}\\
&&+\frac{1 }{16}\,\left[ {\left(\, 1-\, \frac{4\, r\, A^{\prime}}{A} \right) \, \frac{\cos ^2 \theta}{\sin ^2 \theta}\,+ A^2 +\,20\,r\,{A}\,{A}^{\prime}+\,28\,r^2\,\,{A}^{\prime}{}^{2}} \,+\, 24 \,r^2 \, A\, A^{\prime \prime} \right] \,\sin ^{4}\,\theta  \,{\Theta }^{2}+O( {\Theta ^{3}}). \nonumber
\end{eqnarray}
%
The NC-corrected Schwarzschild metric components under \((r, \varphi )\) twist are obtained by substituting~\eqref{Asch_eqn} in~\eqref{NCrphi1},
\begin{equation}
    \begin{split}
        \hat{g}_{00}\left( {x,\Theta }\right) &= -\left( 1 - \frac{2 M}{r} \right) + \frac{M (11 M - 4 r)  \sin ^2 \theta}{2 r^4} \Theta^2 + \mathcal{O}(\Theta^3), \\
        \hat{g}_{11}\left( {x,\Theta }\right) &= \left( 1 - \frac{2 M}{r} \right)^{-1} + \frac{M (3 M - 2 r)  \sin ^2 \theta}{2 r^2 (-2 M + r)^2} \Theta^2 + \mathcal{O}(\Theta^3), \\
        \hat{g}_{12}\left( {x,\Theta }\right) &= \hat{g}_{21}\left( {x,\Theta }\right) = \frac{M  \cos \theta  \sin \theta }{-16 M r + 8 r^2} \Theta^2 + \mathcal{O}(\Theta^3), \\
        \hat{g}_{22}\left( {x,\Theta }\right) &= r^2 + \frac{1}{16}  \Bigl[\cos ^2 \theta + \frac{4 M \bigl(r - 2 M \sin ^2 \theta \bigr)}{(2 M -  r) r}\Bigr] \Theta^2 + \mathcal{O}(\Theta^3), \\
        \hat{g}_{33}\left( {x,\Theta }\right) &= r^2 \sin ^2 \theta + \frac{\bigl(32 M^2 - 19 M r + r^2 + M (-32 M + 13 r) \cos 2 \theta \bigr) \sin ^2 \theta}{16 r (-2 M + r)} \Theta^2  + \mathcal{O}(\Theta^3).
    \end{split}
\end{equation}

\vspace{1em}
\noindent
\textbf{Properties of the Deformed Spacetime:} The \((r,\varphi)\) twisted metric is off-diagonal, and the spherical symmetry of the system is broken due to the NC deformation. As in the cases of the \((t, r)\) and \((r,\theta)\) twists, the radial coordinate is not suitable for analyzing some spacetime properties. The deformed metric has two singularities at \(r=0\) and \(r=2M\) (these points correspond to \(1/g_{rr} = 0\)). This indicates that the locations of the physical singularity at \(r=0\) and the coordinate singularity, corresponding to the event horizon of the black hole, are not changed by the NC corrections. Similar to the \((t, r)\) and \((r,\theta)\) twisted metrics, the surfaces where \(1/g_{rr} = 0\) and \(g_{tt} = 0\) occur at different radii (i.e., the event horizon and the Killing horizons are distinct). The former has no NC corrections, while the latter does.

The surface gravity is calculated using~\eqref{surfeqn} and is given by:
\begin{equation}
    \kappa^2 = \frac{M^2}{r^4}+\frac{ \left(40 M^4 \sin^2\theta - 33 M^3 r \sin^2\theta + 6 M^2 r^2 \sin^2\theta \right)}{r^7 (2 M - r)} \Theta^2 + \mathcal{O}\left(\Theta^3\right).
\end{equation}
At \(r=2M\), the \(\Theta^2\) term diverges, making the definition of the surface gravity unclear. The curvature scalars are calculated as follows:
\begin{equation}
\begin{split}
R &= \frac{1}{16 r^6 (r - 2 M)^2}\left[1536 M^4 \sin^2\theta + r \left(-672 M^3 + M \cos(2\theta) \left(960 M^2 - 448 M r + 85 r^2\right) \right. \right. \\
&\left. \left. \quad + M r (56 M + 55 r) - 4 r^3 \cos^2\theta\right)\right] \Theta^2 + \mathcal{O}\left(\Theta^3\right), \\
R\indices{^\mu ^\nu} R\indices{_\mu _\nu} &= 0, \\
R\indices{_\mu _\nu _\rho _\sigma} R\indices{^\mu ^\nu ^\rho ^\sigma} &= \frac{48 M^2}{r^6} + \frac{M}{4 r^9 (r - 2 M)^2} \left[20064 M^4 \sin^2\theta + 48 M^3 r (305 \cos(2\theta) - 269) \right. \\
&\left. \quad + 14 M^2 r^2 (383 - 487 \cos(2\theta)) + 8 M r^3 (124 \cos(2\theta) - 89) - r^4 (\cos(2\theta) - 5)\right] \Theta^2 + \mathcal{O}\left(\Theta^3\right).
\end{split}
\end{equation}
By observing where these scalars diverge, we confirm that \(r=0\) is a physical singularity. The Ricci scalar \(R\) has a leading NC term, and in the Kretschmann scalar, the NC term diverges more rapidly than the commutative term. Finally, the NC scalar curvature calculated from~\eqref{ncscalardef} is
\begin{equation}
  \hat R = \frac{\Theta ^2 M \left(224 M^3-340 M^2 r+\cos (2 \theta ) \left(-224 M^3+196 M^2 r-48 M r^2+r^3\right)+172 M r^2-31 r^3\right)}{8 r^6 (r-2 M)^2}+O\left(\Theta ^3\right).
\end{equation}


\section{Final Remarks}\label{sec4}

In this paper, we have revisited and refined the construction of de Sitter \((\mathrm{SO}(4,1))\) theory (which is not a gauge theory of the DS group) and its contraction to the Poincar\'e \((\mathrm{ISO}(3,1))\) gauge theory of gravity, presenting a consistent framework for both the commutative and noncommutative (NC) cases. In particular, we solved the equations of motion in the de Sitter theory for a spherically symmetric gauge field configuration, taking the contraction parameter \(k\) explicitly into account. We showed that although contracting the de Sitter solution 
\[
\omega_\mu^{a5} = k\,e_\mu^a,
\]
yields a metric of the form 
\begin{equation}
    g_{\mu\nu} \;=\; \eta_{ab}\,e^a_\mu\, e^b_\nu 
    \;=\;
    \operatorname{diag} \left( -A^2(r), \frac{1}{A^2(r)}, r^2, r^2\sin^2\theta \right)
    \qquad\text{with}\quad
    A^2(r)\;=\;1-\frac{\alpha}{r}+k^2r^2,
\label{ee contraction conclusion2}
\end{equation}
the so called cosmological term \(k^2 r^2\) does \emph{not} survive in a way that genuinely reproduces Einstein gravity with a cosmological constant. In effect, the \(k^2\)-dependence remains embedded in the de Sitter curvature \(R_{\mu\nu}^{AB}\) and does not behave as an ordinary cosmological constant term in the Einstein equations. Contrary to \cite{Chaichian:2007dr}, we argued that correctly introducing a nonzero \(\Lambda\) upon contraction requires explicit source terms of the form \(\Lambda\,g_{\mu\nu}\), much like other matter-energy contributions. This observation clarifies that one cannot simply rely on the gauge-theoretic Wigner--\.{I}n\"{o}n\"{u} contraction and retain a nonzero cosmological constant.

\medskip

After laying out the commutative setup, we turned to the NC extension. In doing so, we systematically applied the Seiberg--Witten (SW) map to the de Sitter gauge fields \(\omega_\mu^{AB}\) and then contracted them to obtain the NC Poincar\'e gauge theory. This de Sitter-based approach places all gauge fields \(\omega^{AB}_\mu\) on an equal footing, making the SW manipulations more transparent. Focusing only on the components that survive the contraction to Poincar\'e gauge theory (and thus to Einstein gravity), we obtained the SW expansion of the spin connection and tetrads, thereby achieving our goal of constructing the NC gauge fields that obey NC Poincar\'e transformations. Along the way, we discovered that prior works on this topic-including \cite{Chamseddine:2000si,Chaichian:2007dr, Chaichian:2007we, Mukherjee:2007fa, Linares:2019gqf, Touati:2023ubi, Touati:2023cxy, Touati:2022kuf, Touati:2022zbm, Touati:2021eem, Zhao:2023uam, Heidari:2023egu, Heidari:2023bww, Touati:2024kmv, AraujoFilho:2025viz}-systematically missed or miswrote certain second-order NC corrections. Specifically, we identified:
\begin{itemize}
    \item An overlooked second-order SW contribution to the tetrad field \(\hat{e}_\mu^a\), encapsulated in Eq.~\eqref{missed term}.
    \item Overuse of the Moyal star product only to first order while examining second-order corrections to the gauge fields.
    \item Incorrect overall numerical factors, such as confusing \(\tfrac{1}{16}\) with \(\tfrac{1}{32}\), leading to distorted coefficients in the second-order expansions.
\end{itemize}
These omissions and/or sign/factor errors, found in all prior investigations, have real physical impact, as they alter the NC-corrected tetrad, metric, curvature, thus also the horizon structure and many other properties in black hole solutions. Since one of the main motivations for studying NC Poincar\'e gauge theory is precisely to probe short-distance quantum-spacetime phenomena (at or near the Planck scale), there is no margin for significant theoretical error. 

\medskip
With these issues clarified, we provided explicit examples for the NC Schwarzschild geometry under various Moyal twists, constructed from mutually commuting coordinate vector fields (e.g., \(\partial_t \wedge \partial_\varphi\), \(\partial_r \wedge \partial_\theta\), etc.). We found that when the radial vector field is involved in the twist, the event horizon determined by \(g_{rr}^{-1}=0\) generically no longer coincides with the Killing horizon given by \(g_{tt}=0\). By contrast, purely angular or time-angular twists leave the Schwarzschild horizon's radial location unchanged, preserving the usual identification of the event horizon. In all twists, except for $(t,r)$, the underlying spherical symmetry is broken by the noncommutativity and the curvature invariants acquire finite corrections at \(\mathcal{O}(\Theta^2)\). Additionally, to exemplify the generality of the approach, we computed NC metric corrections for Reissner-Nordstr\"{o}m, Reissner-Nordstr\"{o}m-de~Sitter, and BTZ-type black holes.

\medskip
We would like to point out that the procedure of deforming commutative gauge fields is not restricted to the specific tetrad form adopted in \eqref{ansatz1}. Any tetrad field can be used, for example by applying the Gram-Schmidt procedure to construct a local orthonormal frame for an arbitrary metric solution of the Einstein field equations. Our approach thus applies beyond diagonal metrics, offering a broad, systematic method for introducing noncommutative effects in a wide array of gravitational solutions.

\medskip

Beyond the calculations presented here, these corrected SW expansions and NC metrics open the door to further applications, including a variety of black hole physics questions at the intersection of quantum gravity and high-energy regimes. One can, for instance, analyze near-horizon solutions to the Klein--Gordon equation, entropy (along the lines of \cite{Juric:2022bnm, Gupta:2022oel, Hrelja:2024tgj, Anacleto:2020efy, Anacleto:2020zfh, Anacleto:2015kca}), greybody factors and quasinormal modes (QNMs), or study how short-distance effects of noncommutativity might alter geodesics, shadows, or the Lyapunov exponent and chaotic properties of the black hole. (See, for instance, \cite{Heidari:2023egu} for recent developments in related contexts, and \cite{C:2024cnk, Campos:2021sff, Anacleto:2019tdj} for investigations along similar lines.) Moreover, as most astrophysical objects are rotating, it would be particularly interesting to apply this formalism to spinning black hole solutions and examine how the NC corrections manifest in the rotating case. These avenues may yield distinctive signatures of NC deformations that could be tested in future gravitational-wave or electromagnetic observations of compact objects.

\medskip

Altogether, our results provide a more reliable foundation for using NC gauge theory as a model of quantum-gravity-inspired corrections in black hole spacetimes. By correcting long-standing oversights in the SW expansions and clarifying how the cosmological constant behaves under contraction, we hope to enable more robust and precise analyses of NC effects in classical gravitational solutions. As quantum tests of general relativity continue to draw interest, having a well-founded theoretical framework for introducing noncommutativity in spacetime can help illuminate new short-distance phenomena beyond the reach of standard approaches.


\vspace{5mm}
\noindent\textbf{Acknowledgment}\\ 
 This research was supported by the Croatian Science Foundation Project  IP-2020-02-9614
"\textit{Search for Quantum spacetime in Black Hole QNM spectrum and Gamma Ray Bursts}". Author F.P. thanks Nikola Herceg for his comments on spherical symmetry.

\appendix
\section{Noncommutative corrections to Reissner-Nordstr{\"o}m metric}\label{app1}

In this appendix, we calculate the NC corrections to the Reissner-Nordstr{\"o}m metric,
\begin{equation}
   ds^2 =  g_{\mu \nu} \, dx^{\mu} dx^{\nu} 
   = -\left(1- \frac{2M}{r}+\frac{Q^2}{r^2}\right) dt^2 
     + \left(1- \frac{2M}{r}+\frac{Q^2}{r^2}\right)^{-1} dr^2 
     + r^2\left(d\theta^2 + \sin^2 \theta \, d\varphi^2\right),
\end{equation}
where \(M\) denotes the mass and \(Q\) represents the charge of the black hole.

\subsection{The (\(t, \varphi\)) twist}
The NC-corrected Reissner-Nordstr{\"o}m metric components under the \((t, \varphi)\) twist are obtained by substituting
\begin{equation} \label{ARN_eqn}
    A^2= \left(1 - \frac{2M}{r} + \frac{Q^2}{r^2}\right)
\end{equation}
into \eqref{NCtphi1}. They are given by
\begin{align*}
\allowdisplaybreaks
        \hat{g}_{00}\left( {x,\Theta }\right) 
        = &  -\left( 1 - \frac{2 M}{r} + \frac{Q^2}{r^2} \right) 
        + \frac{(Q^2 -  M r)^2 \bigl(Q^2 + r (-2 M + r)\bigr) \bigl(7 Q^2 + r (-10 M + 3 r)\bigr) \Theta^2 \sin ^2 \theta}{16 r^{10}} 
        + \mathcal{O}(\Theta^3),\\[6pt]
        \hat{g}_{11}\left( {x,\Theta }\right) 
        = &  \left( 1 - \frac{2 M}{r} + \frac{Q^2}{r^2} \right)^{-1} \\
        & \quad + \frac{\sin ^2 \theta}{4 r^6 \bigl(Q^2 + r (-2 M + r)\bigr)}
        \Bigl[6 Q^6 + 5 Q^4 r (-5 M + 2 r) \\
        & \quad + M r^3 (-11 M^2 + 10 M r - 2 r^2)
        + Q^2 r^2 (31 M^2 - 22 M r + 3 r^2)\Bigr]   \Theta^2 
        + \mathcal{O}(\Theta^3),\\[6pt]
        \hat{g}_{12}\left( {x,\Theta }\right) 
        = &  \hat{g}_{21}\left( {x,\Theta }\right) 
        = - \frac{\bigl(-2 Q^2 r + M (Q^2 + r^2)\bigr)  \cos \theta  \sin \theta}{8 r^4} \Theta^2 
        + \mathcal{O}(\Theta^3),\\[6pt]
        \hat{g}_{22}\left( {x,\Theta }\right) 
        = & r^2 
        -  \frac{(Q^2 -  M r) \bigl(Q^2 + r (-2 M + r)\bigr)  \bigl(r + (M -  r) \sin ^2 \theta\bigr)}{4 r^5} 
        \Theta^2 
        + \mathcal{O}(\Theta^3),\\[6pt]
        \hat{g}_{33}\left( {x,\Theta }\right) 
        = & r^2 \sin ^2 \theta 
        + \frac{(- Q^2 + M r) \bigl(Q^2 + r (-2 M + r)\bigr)  \Bigl(\bigl(7 Q^2 + r (-11 M + 4 r)\bigr)  \sin ^4 \theta + r^2  \sin ^2 (2 \theta)\Bigr)}{16 r^6} 
        \Theta^2 \\
        & \quad + \mathcal{O}(\Theta^3).
 \end{align*}

\subsection{(\(t, \theta\)) twist}
The NC-corrected Reissner-Nordstr{\"o}m metric components under the \((t, \theta)\) twist are obtained by substituting \eqref{ARN_eqn} into \eqref{NCttheta1}:
\begin{align*}
\allowdisplaybreaks
        \hat{g}_{00}\left( {x,\Theta }\right) 
        = &  -\left( 1 - \frac{2 M}{r} + \frac{Q^2}{r^2} \right)
        + \frac{(Q^2 -  M r)^2 \bigl(Q^2 + r (-2 M + r)\bigr) \bigl(7 Q^2 + r (-10 M + 3 r)\bigr) }{16 r^{10}} \Theta^2 
        + \mathcal{O}(\Theta^3),\\[6pt]
        \hat{g}_{11}\left( {x,\Theta }\right) 
        = &  \left( 1 - \frac{2 M}{r} + \frac{Q^2}{r^2} \right)^{-1} \\
        & \quad + \frac{1}{4 r^6 \bigl(Q^2 + r (-2 M + r)\bigr)}
        \Bigl[6 Q^6 + 5 Q^4 r (-5 M + 2 r) + M r^3 (-11 M^2 + 10 M r - 2 r^2)\\
        & \quad + Q^2 r^2 (31 M^2 - 22 M r + 3 r^2)\Bigr] \Theta^2  
        + \mathcal{O}(\Theta^3),\\[6pt]
        \hat{g}_{22}\left( {x,\Theta }\right) 
        = & r^2 
        + \frac{(- Q^2 + M r) \bigl(Q^2 + r (-2 M + r)\bigr) \bigl(7 Q^2 + r (-11 M + 4 r)\bigr)}{16 r^6} \Theta^2 
        + \mathcal{O}(\Theta^3),\\[6pt]
        \hat{g}_{33}\left( {x,\Theta }\right) 
        = &  r^2 \sin ^2 \theta 
        + \frac{(M + r) (- Q^2 + M r) \bigl(Q^2 + r (-2 M + r)\bigr)  \sin ^2 \theta}{4 r^5} \Theta^2 
        + \mathcal{O}(\Theta^3).
    \end{align*}

\subsection{(\(\theta, \varphi\)) twist}
The NC-corrected Reissner-Nordstr{\"o}m metric components under the \((\theta, \varphi)\) twist are obtained by substituting \eqref{ARN_eqn} into \eqref{NCthetaphi1}. They are given below:
\begin{align*}
\allowdisplaybreaks
        \hat{g}_{00}\left( {x,\Theta }\right) 
        = &  -\left( 1 - \frac{2 M}{r} + \frac{Q^2}{r^2} \right) 
          -  \frac{1}{16 r^6}
        \Bigl[(Q^2 -  M r) \bigl(Q^2 + r (-2 M + r)\bigr) \\
        & \quad \times \bigl(-4 Q^2 + r (8 M + r) + (4 Q^2 -  r (8 M + 7 r)) \cos(2 \theta)\bigr)\Bigr] \Theta^2 
        + \mathcal{O}(\Theta^3),\\[6pt]
        \hat{g}_{11}\left( {x,\Theta }\right) 
        = &  \left( 1 - \frac{2 M}{r} + \frac{Q^2}{r^2} \right)^{-1} 
         + \frac{1}{16}
        \Bigl[(Q^2 -  M r) \Bigl(\frac{5 - 11 \cos(2 \theta)}{Q^2 + r (-2 M + r)} 
        -  \frac{8 \sin ^2 \theta}{r^2}\Bigr)\Bigr] \Theta^2
        + \mathcal{O}(\Theta^3),\\[6pt]
        \hat{g}_{12}\left( {x,\Theta }\right) 
        = &  \hat{g}_{21}\left( {x,\Theta }\right) 
        = \frac{(- Q^2 + M r) \sin 2 \theta}{8 r} \Theta^2 
        + \mathcal{O}(\Theta^3),\\[6pt]
        \hat{g}_{22}\left( {x,\Theta }\right) 
        = & r^2 
        -  \frac{(Q^2 - 2 M r)  \bigl(-10 r^2 \cos 2 \theta 
        + (Q^2 + 2 r (- M + r)) \sin ^2 \theta\bigr)}{16 r^2} \Theta^2 
        + \mathcal{O}(\Theta^3),\\[6pt]
        \hat{g}_{33}\left( {x,\Theta }\right) 
        = & r^2 \sin ^2 \theta 
        + \frac{\Theta^2 }{128 r^2}
        \Bigl[4 \bigl(Q^4 + 2 Q^2 r (-2 M + 15 r) 
        + 2 r^2 (2 M^2 - 30 M r + r^2)\bigr) \cos 2 \theta \\
        & \quad -  (Q^2 - 2 M r) \bigl(3 Q^2 - 6 M r + 44 r^2 
        + (Q^2 - 2 M r + 76 r^2) \cos 4 \theta \bigr)\Bigr] 
        + \mathcal{O}(\Theta^3).
 \end{align*}

\subsection{(\(t, r\)) twist}
The NC-corrected Reissner-Nordstr{\"o}m metric components under the \((t, r)\) twist are obtained by substituting \eqref{ARN_eqn} into \eqref{NCtr1}:
\begin{align*}
\allowdisplaybreaks
        \hat{g}_{00}\left( {x,\Theta }\right) 
        &= -\left( 1 - \frac{2 M}{r} + \frac{Q^2}{r^2} \right) \\
        &\quad + \frac{1}{16 r^{10} \bigl(Q^2 + r (-2 M + r)\bigr)}
        \Bigl[605 Q^8 + 2 Q^6 r (-1616 M + 573 r) \\
        &\qquad + 4 Q^4 r^2 (1497 M^2 - 968 M r + 135 r^2) 
          + M^2 r^4 (1111 M^2 - 1068 M r + 256 r^2) \\
        &\qquad - 2 M Q^2 r^3 (2222 M^2 - 1869 M r + 384 r^2)\Bigr] \Theta^2  
        + \mathcal{O}(\Theta^3),\\[6pt]
        \hat{g}_{11}\left( {x,\Theta }\right) 
        &= \left( 1 -\frac{2 M}{r} + \frac{Q^2}{r^2} \right)^{-1} \\
        &\quad + \frac{1}{16 r^6 \bigl(Q^2 + r (-2 M + r)\bigr)^3} 
        \Bigl[-172 Q^8 + 4 Q^6 (241 M - 96 r) r \\
        &\qquad + Q^4 r^2 (-1856 M^2 + 1356 M r - 219 r^2) 
          + M^2 r^4 (-351 M^2 + 380 M r - 104 r^2)\\
        &\qquad + 2 M Q^2 r^3 (702 M^2 - 665 M r + 156 r^2)\Bigr] \Theta^2  
        + \mathcal{O}(\Theta^3),\\[6pt]
        \hat{g}_{22}\left( {x,\Theta }\right) 
        &= r^2 
        + \frac{(Q^2 -  M r) \bigl(50 Q^4 + M (67 M - 34 r) r^2 
        + Q^2 r (-134 M + 51 r)\bigr) \Theta^2}{8 r^6 \bigl(Q^2 + r (-2 M + r)\bigr)} 
        + \mathcal{O}(\Theta^3),\\[6pt]
        \hat{g}_{33}\left( {x,\Theta }\right) 
        &= r^2 \sin ^2 \theta \\
        &\quad + \frac{(Q^2 -  M r) \bigl(50 Q^4 + M (67 M - 34 r) r^2 
        + Q^2 r (-134 M + 51 r)\bigr)  \sin ^2 \theta}{8 r^6 \bigl(Q^2 + r (-2 M + r)\bigr)} 
        \Theta^2 
        + \mathcal{O}(\Theta^3).
 \end{align*}

\subsection{(\(r , \theta\)) twist}
The components of the NC-corrected metric are obtained by substituting \eqref{ARN_eqn} into \eqref{NCrtheta1} and are given as
\begin{align*}
\allowdisplaybreaks
        \hat{g}_{00}\left( {x,\Theta }\right) 
        &= -\left(1- \frac{2M}{r}+\frac{Q^2}{r^2}\right)
        + \frac{ \left(3 Q^2 r (3 r-10 M)+M r^2 (11 M-4 r)+14 Q^4\right)}{2 r^6}\Theta ^2
        + O\left(\Theta ^3\right),\\[6pt]
        \hat{g}_{11}\left( {x,\Theta }\right) 
        &= \left(1- \frac{2M}{r}+\frac{Q^2}{r^2}\right)^{-1}
        + \frac{ \left(3 Q^2 r (r-2 M)+M r^2 (3 M-2 r)+2 Q^4\right)}{2 r^2 \left(r (r-2 M)+Q^2\right)^2}\Theta ^2
        + O\left(\Theta ^3\right),\\[6pt]
        \hat{g}_{22}\left( {x,\Theta }\right) 
        &= r^2
        + \frac{1}{16}  \left(1-\frac{30 M}{r}+\frac{57 Q^2}{r^2}
        +\frac{4 (M-Q) (M+Q)}{r (r-2 M)+Q^2}\right)\Theta ^2 
        + O\left(\Theta ^3\right),\\[6pt]
        \hat{g}_{33}\left( {x,\Theta }\right) 
        &= r^2 \sin ^2\theta \\
        &\quad + \left(\frac{5 \cos ^2\theta }{16} 
        +\frac{\sin ^2\theta  \left(r^2 \left(2 M^2-4 M r+r^2\right)
        +Q^2 r (5 r-8 M)+4 Q^4\right)}{4 r^2 \left(r (r-2 M)+Q^2\right)}\right)\Theta ^2
        + O\left(\Theta ^3\right).
 \end{align*}

\subsection{(\(r, \varphi\)) twist}
The NC-corrected Reissner-Nordstr{\"o}m metric components under the \((r, \varphi)\) twist are obtained by substituting \eqref{ARN_eqn} into \eqref{NCrphi1}:
\begin{align*}
\allowdisplaybreaks
        \hat{g}_{00}\left( {x,\Theta }\right) 
        = &  -\left( 1 - \frac{2 M}{r} + \frac{Q^2}{r^2} \right) 
        + \frac{\bigl(14 Q^4 + M (11 M - 4 r) r^2 + 3 Q^2 r (-10 M + 3 r)\bigr)  \sin ^2 \theta}{2 r^6} \Theta^2 
        + \mathcal{O}(\Theta^3),\\[6pt]
        \hat{g}_{11}\left( {x,\Theta }\right) 
        = &  \left( 1 -\frac{2 M}{r} + \frac{Q^2}{r^2} \right)^{-1} 
        + \frac{\bigl(2 Q^4 + M (3 M - 2 r) r^2 
        + 3 Q^2 r (-2 M + r)\bigr)  \sin ^2 \theta}{2 r^2 \bigl(Q^2 + r (-2 M + r)\bigr)^2} \Theta^2 
        + \mathcal{O}(\Theta^3),\\[6pt]
        \hat{g}_{12}\left( {x,\Theta }\right) 
        = &  \hat{g}_{21}\left( {x,\Theta }\right) 
        = \frac{(-2 Q^2 + M r)  \cos \theta \sin \theta}{8 r \bigl(Q^2 + r (-2 M + r)\bigr)} \Theta^2 
        + \mathcal{O}(\Theta^3),\\[6pt]
        \hat{g}_{22}\left( {x,\Theta }\right) 
        = & r^2 
        + \frac{1}{16} 
        \Biggl[\cos ^2 \theta 
        + \frac{4 r^2 (Q^2 -  M r) 
        + 8 \bigl(2 Q^4 + M^2 r^2 + Q^2 r (-4 M + r)\bigr) \sin ^2 \theta}{r^2 \bigl(Q^2 + r (-2 M + r)\bigr)}\Biggr] \Theta^2 
        + \mathcal{O}(\Theta^3),\\[6pt]
        \hat{g}_{33}\left( {x,\Theta }\right) 
        = & r^2 \sin ^2 \theta  
        + \frac{1 }{16 r^2 \bigl(Q^2 + r (-2 M + r)\bigr)} 
        \Bigl[r^2 \bigl(5 Q^2 + r (-6 M + r)\bigr) \cos ^2 \theta \sin ^2 \theta \\
        & \quad + \bigl(57 Q^4 
        + 18 Q^2 r (-8 M + 3 r) 
        + r^2 (64 M^2 - 32 M r + r^2)\bigr) \sin ^4 \theta\Bigr] \Theta^2 
        + \mathcal{O}(\Theta^3).
 \end{align*}


\section{Noncommutative corrections to Reissner-Nordstr{\"o}m de-Sitter metric}\label{app2}

In this appendix, we calculate the NC corrections to the Reissner-Nordstr{\"o}m de-Sitter metric, which is given by
\begin{equation}
   ds^2 =  g_{\mu \nu} \, dx^{\mu} dx^{\nu} = - f(r)\, dt^2 + f(r)^{-1}\, dr^2 + r^2\left(d\theta^2 + \sin^2\theta \, d\varphi^2\right),
\end{equation}
with the metric function
\begin{equation}
   f(r)= \left(1 - \frac{2M}{r} + \frac{Q^2}{r^2} - \frac{\Lambda}{3}r^2\right),
\end{equation}
where $\Lambda$ is the cosmological constant.

\subsection{The ($t, \varphi$) twist}
Under the ($t, \varphi$) twist, the NC correction to the Reissner-Nordstr{\"o}m de-Sitter metric is obtained by substituting
\begin{equation} \label{ARNdS_eqn}
    A^2= \left(1 - \frac{2M}{r} + \frac{Q^2}{r^2} - \frac{\Lambda}{3}r^2\right)
\end{equation}
in \eqref{NCtphi1}. The NC-corrected metric components are given by
\begin{align*}
\allowdisplaybreaks
        \hat{g}_{00}\left( {x,\Theta }\right) = &  -\left( 1 - \frac{2 M}{r} + \frac{Q^2}{r^2} -\frac{\Lambda}{3}r^2 \right)   +  \frac{1 }{1296 r^{10}} \Biggl[ \Bigl(3 \bigl(Q^2 + r (-2 M + r)\bigr) -  r^4 \Lambda \Bigr)\\
       & \times (3 Q^2 - 3 M r + r^4 \Lambda)^2 \bigl(21 Q^2 + r (-30 M + 9 r + r^3 \Lambda)\bigr) \sin ^2 \theta \Biggr] \Theta^2 + \mathcal{O}(\Theta^3),\\
        \hat{g}_{11}\left( {x,\Theta }\right) = &  \left( 1 -\frac{2 M}{r} + \frac{Q^2}{r^2} -\frac{\Lambda}{3}r^2 \right)^{-1} \\
        & + \frac{\sin ^2 \theta  }{36 r^6 \Bigl(-3 \bigl(Q^2 + r (-2 M + r)\bigr) + r^4 \Lambda \Bigr)} \Biggl[-162 Q^6 + 27 M r^3 (11 M^2 - 10 M r + 2 r^2) \\
        & + 9 r^6 (-7 M + r) (-3 M + r) \Lambda + 3 r^9 (-3 M + 2 r) \Lambda^2 - 2 r^{12} \Lambda^3 \\
        & + 9 Q^4 r (75 M - 30 r + 14 r^3 \Lambda) + 3 Q^2 r^2 \bigl(-9 (31 M^2 - 22 M r + 3 r^2) \\
        & + 6 r^3 (-17 M + 4 r) \Lambda + 2 r^6 \Lambda^2\bigr)\Biggr] \Theta^2 + \mathcal{O}(\Theta^3),\\
        \hat{g}_{12}\left( {x,\Theta }\right) = &  \hat{g}_{21}\left( {x,\Theta }\right) = - \frac{ \bigl(2 r^5 \Lambda + Q^2 (-6 r + 8 r^3 \Lambda) + 3 M (Q^2 + r^2 - 3 r^4 \Lambda)\bigr) \cos \theta  \sin \theta }{24 r^4} \Theta^2 + \mathcal{O}(\Theta^3), \\
        \hat{g}_{22}\left( {x,\Theta }\right) = & r^2  + \frac{1 }{108 r^5} \Biggl[ (3 Q^2 - 3 M r + r^4 \Lambda) \Bigl(-3 \bigl(Q^2 + r (-2 M + r)\bigr) + r^4 \Lambda \Bigr) \\
        & \times \Bigl(3 r + (3 M - 3 r + 2 r^3 \Lambda) \sin ^2 \theta\Bigr) \Biggr] \Theta^2 + \mathcal{O}(\Theta^3),\\
        \hat{g}_{33}\left( {x,\Theta }\right) = &  r^2 \sin ^2 \theta  -  \frac{1}{432 r^6} \Biggl[ (3 Q^2 - 3 M r + r^4 \Lambda) \Bigl(-3 \bigl(Q^2 + r (-2 M + r)\bigr) + r^4 \Lambda \Bigr)\\
& \times \Bigl(\bigl(-21 Q^2 + r (33 M - 12 r + r^3 \Lambda)\bigr) \sin ^4 \theta - 3 r^2 \sin ^2 (2 \theta)\Bigr) \Biggr] \Theta^2  + \mathcal{O}(\Theta^3).
\end{align*}

\subsection{($t, \theta$) twist}
The NC-corrected Reissner-Nordstr{\"o}m de-Sitter metric components under the ($t, \theta$) twist are obtained by substituting \eqref{ARNdS_eqn} in \eqref{NCttheta1},
\begin{align*}
\allowdisplaybreaks
        \hat{g}_{00}\left( {x,\Theta }\right) = &  -\left( 1 - \frac{2 M}{r} + \frac{Q^2}{r^2} -\frac{\Lambda}{3}r^2 \right)  \\
        & + \frac{\Bigl(3 \bigl(Q^2 + r (-2 M + r)\bigr) -  r^4 \Lambda \Bigr) (3 Q^2 - 3 M r + r^4 \Lambda)^2 \bigl(21 Q^2 + r (-30 M + 9 r + r^3 \Lambda)\bigr)}{1296 r^{10}} \Theta^2  \\ 
&+ \mathcal{O}(\Theta^3),\\
        \hat{g}_{11}\left( {x,\Theta }\right) = &  \left( 1 -\frac{2 M}{r} + \frac{Q^2}{r^2}-\frac{\Lambda}{3}r^2 \right)^{-1} \\
        & + \frac{1 }{36 r^6 \Bigl(-3 \bigl(Q^2 + r (-2 M + r)\bigr) + r^4 \Lambda \Bigr)}\Biggl[-162 Q^6 + 27 M r^3 (11 M^2 - 10 M r + 2 r^2) \\
        & + 9 r^6 (-7 M + r) (-3 M + r) \Lambda + 3 r^9 (-3 M + 2 r) \Lambda^2 - 2 r^{12} \Lambda^3 \\
        & + 9 Q^4 r (75 M - 30 r + 14 r^3 \Lambda) + 3 Q^2 r^2 \bigl(-9 (31 M^2 - 22 M r + 3 r^2)\\
        & + 6 r^3 (-17 M + 4 r) \Lambda + 2 r^6 \Lambda^2\bigr)\Biggr] \Theta^2 + \mathcal{O}(\Theta^3),\\
        \hat{g}_{22}\left( {x,\Theta }\right) = & r^2  \\ 
& -  \frac{ (3 Q^2 - 3 M r + r^4 \Lambda) \Bigl(-3 \bigl(Q^2 + r (-2 M + r)\bigr) + r^4 \Lambda \Bigr) \bigl(-21 Q^2 + r (33 M - 12 r + r^3 \Lambda)\bigr)}{432 r^6} \Theta^2 \\ 
& + \mathcal{O}(\Theta^3), \\
        \hat{g}_{33}\left( {x,\Theta }\right) = &  r^2 \sin ^2 \theta \\ 
& + \frac{ \bigl(3 (M + r) + 2 r^3 \Lambda \bigr) (3 Q^2 - 3 M r + r^4 \Lambda) \Bigl(-3 \bigl(Q^2 + r (-2 M + r)\bigr) + r^4 \Lambda \Bigr) \sin ^2 \theta}{108 r^5} \Theta^2 \\ 
&+ \mathcal{O}(\Theta^3).
\end{align*}
\subsection{($\theta, \varphi$) twist}

The NC-corrected Reissner-Nordstr{\"o}m de-Sitter metric components under the ($\theta, \varphi$) twist are obtained by substituting \eqref{ARNdS_eqn} in \eqref{NCthetaphi1},
\begin{align*}
\allowdisplaybreaks
        \hat{g}_{00}\left( {x,\Theta }\right) = &  -\left( 1 - \frac{2 M}{r} + \frac{Q^2}{r^2} -\frac{\Lambda}{3}r^2 \right)  + \frac{\Theta^2}{432 r^6} \Biggl[ (3 Q^2 - 3 M r + r^4 \Lambda) \Bigl(-3 \bigl(Q^2 + r (-2 M + r)\bigr) + r^4 \Lambda \Bigr) \\
        &\times  \biggl(3 r^2 \Bigl(5 - 11 \cos ^2 \theta\Bigr)  + \bigl(-24 Q^2 + r (48 M + 9 r + 8 r^3 \Lambda)\bigr) \sin ^2 \theta\biggr) \Biggr] + \mathcal{O}(\Theta^3),\\
        \hat{g}_{11}\left( {x,\Theta }\right) = &  \left( 1 -\frac{2 M}{r} + \frac{Q^2}{r^2}-\frac{\Lambda}{3}r^2 \right)^{-1}\\
        & + \frac{1}{48} \Biggl[ (3 Q^2 - 3 M r + r^4 \Lambda) \Biggl(\frac{3 \bigl(-5 + 11 \cos 2 \theta \bigr)}{-3 \bigl(Q^2 + r (-2 M + r)\bigr) + r^4 \Lambda} -  \frac{8 \sin ^2 \theta}{r^2}\Biggr) \Biggr]\Theta^2  + \mathcal{O}(\Theta^3), \\
        \hat{g}_{12}\left( {x,\Theta }\right) = &   \hat{g}_{21}\left( {x,\Theta }\right) = - \frac{ (3 Q^2 - 3 M r + r^4 \Lambda) \sin 2 \theta }{24 r} \Theta^2 + \mathcal{O}(\Theta^3),\\
        \hat{g}_{22}\left( {x,\Theta }\right) = & r^2 -  \frac{ (-3 Q^2 + 6 M r + r^4 \Lambda) \Bigl(30 r^2 \cos 2 \theta + \bigl(-3 Q^2 + r (6 M - 6 r + r^3 \Lambda)\bigr) \sin ^2 \theta\Bigr)}{144 r^2} \Theta^2 + \mathcal{O}(\Theta^3), \\
        \hat{g}_{33}\left( {x,\Theta }\right) = &r^2 \sin ^2 \theta  + \frac{1 }{576 r^2} \Biggl[2 \bigl(9 Q^4 + 18 r^2 (2 M^2 - 30 M r + r^2) + 6 (2 M - 15 r) r^5 \Lambda + r^8 \Lambda^2 \\
        & - 6 Q^2 r (6 M - 45 r + r^3 \Lambda)\bigr) \cos 2 \theta -  \tfrac{1}{2} (3 Q^2 - 6 M r -  r^4 \Lambda) \\
        & \times \Bigl(9 Q^2 - 3 r (6 M - 44 r + r^3 \Lambda) + \bigl(3 Q^2 -  r (6 M - 228 r + r^3 \Lambda)\bigr) \cos 4 \theta \Bigr)\Biggr] \Theta^2 + \mathcal{O}(\Theta^3).
\end{align*}

\subsection{($t, r$) twist}

The NC-corrected Reissner-Nordstr{\"o}m de-Sitter metric components under the ($t, r$) twist are obtained by substituting \eqref{ARNdS_eqn} in \eqref{NCtr1},
\begin{align*}
\allowdisplaybreaks
        \hat{g}_{00}\left( {x,\Theta }\right) &= -\left( 1 - \frac{2 M}{r} + \frac{Q^2}{r^2}-\frac{\Lambda}{3}r^2 \right) \\
        & + \frac{ 1 }{432 r^{10} \Bigl(-3 \bigl(Q^2 + r (-2 M + r)\bigr) + r^4 \Lambda \Bigr)}\Biggl[-49005 Q^8 + 54 Q^6 r (4848 M - 1719 r + 460 r^3 \Lambda)\\
        & - 54 Q^4 r^2 \bigl(8982 M^2 + 810 r^2 + 48 M r (-121 + 31 r^2 \Lambda) + r^4 \Lambda (-365 + 39 r^2 \Lambda)\bigr) \\
        & + 6 Q^2 r^3 \Bigl(59994 M^3 + 27 M^2 r (-1869 + 454 r^2 \Lambda) + r^5 \Lambda \bigl(-540 + r^2 \Lambda (243 - 8 r^2 \Lambda)\bigr) \\
        & + 72 M r^2 \bigl(144 + r^2 \Lambda (-51 + 4 r^2 \Lambda)\bigr)\Bigr) \\
        & + r^4 \Bigl(-89991 M^4 + 108 M^3 r (801 - 203 r^2 \Lambda) + r^8 \Lambda^2 \bigl(-252 + r^2 \Lambda (234 - 49 r^2 \Lambda)\bigr) \\
        & + 48 M r^5 \Lambda \bigl(18 + r^2 \Lambda (9 - 8 r^2 \Lambda)\bigr) - 54 M^2 r^2 \bigl(384 + r^2 \Lambda (-159 + 28 r^2 \Lambda)\bigr)\Bigr)\Biggr] \Theta^2 + \mathcal{O}(\Theta^3), \\
        \hat{g}_{11}\left( {x,\Theta }\right) &= \left( 1 -\frac{2 M}{r} + \frac{Q^2}{r^2} -\frac{\Lambda}{3}r^2 \right)^{-1} \\
        & + \frac{1}{16 r^6 \Bigl(-3 \bigl(Q^2 + r (-2 M + r)\bigr) + r^4 \Lambda \Bigr)^3} 
        \Biggl[4644 Q^8 - 36 Q^6 r (723 M - 288 r + 98 r^3 \Lambda) \\
        & + 3 Q^4 r^2 \bigl(16704 M^2 + 1971 r^2 + 4 r^4 \Lambda (-315 + 59 r^2 \Lambda) + 12 M r (-1017 + 341 r^2 \Lambda)\bigr) \\
        & + 6 Q^2 r^3 \Bigl(-6318 M^3 + 9 M^2 r (665 - 218 r^2 \Lambda) + r^5 \Lambda \bigl(81 + 4 r^2 \Lambda (-13 + 4 r^2 \Lambda)\bigr) \\
        & - 6 M r^2 \bigl(234 + 5 r^2 \Lambda (-26 + 5 r^2 \Lambda)\bigr)\Bigr) \\
        & + r^4 \Bigl(9477 M^4 + r^8 \Lambda^2 (33 - 4 r^2 \Lambda) + 108 M^3 r (-95 + 33 r^2 \Lambda) - 36 M r^5 \Lambda (4 + r^2 \Lambda + r^4 \Lambda^2)\\
        & + 54 M^2 r^2 \bigl(52 + r^2 \Lambda (-33 + 10 r^2 \Lambda)\bigr)\Bigr)\Biggr] \Theta^2  + \mathcal{O}(\Theta^3),  \\
        \hat{g}_{22}\left( {x,\Theta }\right) &= r^2 \\
        & -  \frac{1}{72 r^6 \Bigl(-3 \bigl(Q^2 + r (-2 M + r)\bigr) + r^4 \Lambda \Bigr)} \Bigg[ (3 Q^2 - 3 M r + r^4 \Lambda) \bigl(450 Q^4 + 9 M (67 M - 34 r) r^2 \\
        & + 3 (70 M - 17 r) r^5 \Lambda + 16 r^8 \Lambda^2 - 3 Q^2 r (402 M - 153 r + 70 r^3 \Lambda)\bigr) \Biggr] \Theta^2 + \mathcal{O}(\Theta^3), \\
        \hat{g}_{33}\left( {x,\Theta }\right) & =  r^2 \sin ^2 \theta \\ 
         & -  \frac{1}{72 r^6 \Bigl(-3 \bigl(Q^2 + r (-2 M + r)\bigr) + r^4 \Lambda \Bigr)} \Bigg[ (3 Q^2 - 3 M r + r^4 \Lambda) \bigl(450 Q^4 + 9 M (67 M - 34 r) r^2 \\
         & + 3 (70 M - 17 r) r^5 \Lambda + 16 r^8 \Lambda^2 - 3 Q^2 r (402 M - 153 r + 70 r^3 \Lambda)\bigr) \sin ^2 \theta \Bigg] \Theta^2 + \mathcal{O}(\Theta^3).
\end{align*}

\subsection{($r , \theta$) twist}

The components of the NC-corrected Reissner-Nordstr{\"o}m de-Sitter metric are obtained by substituting \eqref{ARNdS_eqn} in \eqref{NCrtheta1} and are given by
\begin{align*}
\allowdisplaybreaks
        \hat{g}_{00}\left( {x,\Theta }\right) = &  -\left(1 - \frac{2M}{r} + \frac{Q^2}{r^2} - \frac{\Lambda}{3}r^2\right) \\
        & + \frac{ \bigl(126 Q^4 + 9 M (11 M - 4 r) r^2 + 3 r^5 (2 M + r) \Lambda - 4 r^8 \Lambda^2 - 9 Q^2 r (30 M - 9 r + 2 r^3 \Lambda)\bigr)}{18 r^6} \Theta^2 \\ 
& + \mathcal{O}(\Theta^3),\\
        \hat{g}_{11}\left( {x,\Theta }\right)  = & \left(1 - \frac{2M}{r} + \frac{Q^2}{r^2} - \frac{\Lambda}{3}r^2\right)^{-1} \\
        & + \frac{\Bigl(9 \bigl(2 Q^4 + M (3 M - 2 r) r^2 + 3 Q^2 r (-2 M + r)\bigr) - 3 r^4 \bigl(6 Q^2 + r (-6 M + r)\bigr) \Lambda \Bigr)}{2 \Bigl(-3 r \bigl(Q^2 + r (-2 M + r)\bigr) + r^5 \Lambda \Bigr)^2} \Theta^2  \\ 
& + \mathcal{O}(\Theta^3),\\
        \hat{g}_{22}\left( {x,\Theta }\right) = & r^2  + \frac{1 }{48 r^2 \Bigl(-3 \bigl(Q^2 + r (-2 M + r)\bigr) + r^4 \Lambda \Bigr)} \biggl[-513 Q^4 - 9 r^2 (64 M^2 - 32 M r + r^2)\\
& + 6 r^5 (-56 M + 23 r) \Lambda - 49 r^8 \Lambda^2 + 54 Q^2 r (24 M - 9 r + 5 r^3 \Lambda)\biggr] \Theta^2  + \mathcal{O}(\Theta^3), \\
        \hat{g}_{33}\left( {x,\Theta }\right)  = & r^2 \sin ^2 \theta  + \Theta^2 \Biggl[\frac{5}{16} \cos ^2\theta  -  \frac{\sin^2 \theta}{-36 r^2 \bigl(Q^2 + r (-2 M + r)\bigr) + 12 r^6 \Lambda}\Biggl(36 Q^4 \\
        & + 9 r^2 (2 M^2 - 4 M r + r^2) + 3 (8 M - 5 r) r^5 \Lambda + 8 r^8 \Lambda^2 - 3 Q^2 r (24 M - 15 r + 4 r^3 \Lambda)\Biggr) \Biggr] \\ 
& + \mathcal{O}(\Theta^3). \\
\end{align*}

\subsection{($r, \varphi$) twist}
The NC-corrected Reissner-Nordstr{\"o}m de-Sitter metric components under the ($r, \varphi$) twist are obtained by substituting \eqref{ARNdS_eqn} in \eqref{NCrphi1},
\begin{align*}
\allowdisplaybreaks
        \hat{g}_{00}\left( {x,\Theta }\right) = &  -\left( 1 - \frac{2 M}{r} + \frac{Q^2}{r^2} - \frac{\Lambda}{3}r^2 \right)  + \frac{1 }{18 r^6} \Biggl[126 Q^4 + 9 M (11 M - 4 r) r^2 \\
        & + 3 r^5 (2 M + r) \Lambda - 4 r^8 \Lambda^2 - 9 Q^2 r (30 M - 9 r + 2 r^3 \Lambda)\Biggr] \sin ^2 \theta \Theta^2  + \mathcal{O}(\Theta^3),  \\
        \hat{g}_{11}\left( {x,\Theta }\right) = &  \left( 1 -\frac{2 M}{r} + \frac{Q^2}{r^2} -\frac{\Lambda}{3}r^2 \right)^{-1} \\
        & + \frac{1}{2 \Bigl(-3 r \bigl(Q^2 + r (-2 M + r)\bigr) + r^5 \Lambda \Bigr)^2} \Biggl[ 3  \Bigl(6 Q^4 + 3 M (3 M - 2 r) r^2 \\
        & + 9 Q^2 r (-2 M + r) -  r^4 \bigl(6 Q^2 + r (-6 M + r)\bigr) \Lambda \Bigr) \sin ^2 \theta \Biggr] \Theta^2  + \mathcal{O}(\Theta^3), \\
        \hat{g}_{12}\left( {x,\Theta }\right) = &  \hat{g}_{21}\left( {x,\Theta }\right) = \frac{ (-6 Q^2 + 3 M r + 2 r^4 \Lambda) \cos \theta \sin \theta}{24 r \bigl(Q^2 + r (-2 M + r)\bigr) - 8 r^5 \Lambda} \Theta^2 + \mathcal{O}(\Theta^3), \\
        \hat{g}_{22}\left( {x,\Theta }\right) = & r^2 + \frac{1}{48}  \Biggl[3 \cos ^ 2\theta -  \frac{1}{-3 r^2 \bigl(Q^2 + r (-2 M + r)\bigr) + r^6 \Lambda}\Biggr( 12 r^2 (3 Q^2 - 3 M r + r^4 \Lambda) \\
        & + 8 \Bigl(9 \bigl(2 Q^4 + M^2 r^2 + Q^2 r (-4 M + r)\bigr) - 3 r^4 \bigl(2 Q^2 + r (-4 M + 3 r)\bigr) \Lambda + 4 r^8 \Lambda^2\Bigr) \sin ^2 \theta \Biggl) \Biggr] \Theta^2 \\ 
& + \mathcal{O}(\Theta^3), \\
        \hat{g}_{33}\left( {x,\Theta }\right) = & r^2 \sin ^2 \theta  \\ 
& -  \frac{1 }{48 r^2 \Bigl(-3 \bigl(Q^2 + r (-2 M + r)\bigr) + r^4 \Lambda \Bigr)} \Biggl[9 r^2 \bigl(5 Q^2 + r (-6 M + r + r^3 \Lambda)\bigr)\cos ^2 \theta \sin ^2 \theta \\
& + \bigl(513 Q^4 + 9 r^2 (64 M^2 - 32 M r + r^2) + 6 (56 M - 23 r) r^5 \Lambda + 49 r^8 \Lambda^2 \\
& - 54 Q^2 r (24 M - 9 r + 5 r^3 \Lambda)\bigr) \sin ^4 \theta\Biggr] \Theta^2  + \mathcal{O}(\Theta^3).
\end{align*}

\section{Noncommutative corrections to three dimensional nonrotating metric}\label{app3}
In this appendix, we provide the NC corrections to the BTZ and charged BTZ black holes resulting from various twists, using equations \eqref{metricdef} and \eqref{starproduct2}.
\subsection{($r , \varphi$) twist}
For the ($r, \varphi$) twist the non commutativity the parameter $\Theta ^{\mu \nu} $ is given by,
\begin{equation}
\Theta ^{\mu \nu }=\left(
\begin{array}{ccc}
0 & 0 & 0 \\
0 & 0 & \Theta  \\
0 & -\Theta & 0 \\
\end{array}
\right) ,\quad \mu ,\nu =0, 1,2.  \label{3drphitwist}
\end{equation}
This choice produces the following commutation relation between the coordinates:
\begin{equation}
    [r\stackrel{\star}{,} \varphi] = i \, \Theta. 
\end{equation}
The non-zero components of the tetrad $\hat e _\mu ^{a}$ are
\begin{equation}
    \begin{split}
        \hat{e}_{0}^{0} &= A+\frac{1}{4}  \left(2 r A'^3+A^2 \left(r A^{(3)}(r)+A''\right)+A A' \left(5 r A''+2 A'\right)\right)\Theta ^2+O\left(\Theta ^3\right),\\
        \hat{e}_{1}^{1} &= \frac{1}{A}+\frac{1}{4}  A'' \Theta ^2+O\left(\Theta ^3\right),\\
        \hat{e}_{2}^{2} &= r+\frac{1}{4}  \left(3 r A'^2+A \left(3 r A''+2 A'\right)\right) \Theta ^2+O\left(\Theta ^3\right) ,\\
        \hat{e}_{2}^{1} &= -\frac{1}{4} i   \left(2 r A'+A\right)\Theta+O\left(\Theta ^3\right) .
    \end{split}
\end{equation}

The deformed metric $\hat g _{\mu \nu} (x, \Theta) $ is obtained by using the definition \eqref{metricdef}. Up to second order in $\Theta$, the metric is diagonal, and the non-zero components are
\begin{equation}
    \begin{split}
        \hat{g}_{00}\left( {x,\Theta }\right) &= -A^2+ \left(-\frac{1}{2} r A^{(3)}(r) A^3-\frac{1}{2} A^3 A''-A^2 A'^2-r A A'^3-\frac{5}{2} r A^2 A' A''\right)\Theta ^2+O\left(\Theta ^3\right),\\
        \hat{g}_{11}\left( {x,\Theta }\right) &= \frac{1}{A^2}+\frac{ A''}{2 A}\Theta ^2+O\left(\Theta ^3\right), \label{NCrphibtz}\\ 
        \hat{g}_{22}\left( {x,\Theta }\right) &= r^2+ \left(\frac{3}{2} r^2 A A''+\frac{7}{4} r^2 A'^2+\frac{5}{4} r A A'+\frac{A^2}{16}\right)\Theta ^2+O\left(\Theta ^3\right).
    \end{split}
\end{equation}
\subsubsection{BTZ and QBTZ black holes}
The QBTZ metric is given by \cite{Carlip:1995qv, Martinez:1999qi, Clement:1995zt, Unver:2010uw, Juric:2022bnm}
\begin{equation}
    ds^2=-f(r)dt^2+\tfrac{1}{f(r)}dr^2+r^2d\varphi^2
\end{equation}
with the metric function,
\begin{equation}
    f(r)=-M+\frac{r^2}{l^2}-2Q^2\ln \left( \frac{r}{l} \right).
\end{equation}
Here, $M$ represents the mass of the black hole, $Q$ denotes the charge of the black hole, and $l$ is the AdS radius. In the limit $Q\to 0$, we recover the BTZ black hole. To derive the NC corrections to the BTZ black hole, we chose
\begin{equation} \label{Abtz}
    A^2 = -M+\frac{r^2}{l^2}
\end{equation}
and for the QBTZ black hole
\begin{equation} \label{Aqbtz}
    A^2=-M+\frac{r^2}{l^2}-2Q^2\ln \left( \frac{r}{l} \right)
\end{equation}
in our formalism. The NC corrections to the BTZ black hole due to the $(r, \varphi)$ twist are obtained by substituting \eqref{Abtz} into \eqref{NCrphibtz}. The components of the NC corrected metric are given by,
\begin{equation}
    \begin{split}
        \hat{g}_{00}\left( {x,\Theta }\right) &= \left(  M - \frac{r^2}{l^2} \right) + \frac{(l^2 M - 4 r^2) }{2 l^4} \Theta^2 + \mathcal{O}(\Theta^3) , \\
        \hat{g}_{11}\left( {x,\Theta }\right) &= \left( - M + \frac{r^2}{l^2} \right)^{-1} -  \frac{l^2 M }{2 (- l^2 M + r^2)^2} \Theta^2 + \mathcal{O}(\Theta^3) , \\
        \hat{g}_{22}\left( {x,\Theta }\right) &= r^2 + \frac{1}{16} \Biggl[- M + \frac{r^2}{l^2} \left( 45 + \frac{4 r^2}{- l^2 M + r^2}\right) \Biggr] \Theta^2 + \mathcal{O}(\Theta^3).
    \end{split}
\end{equation}

The NC corrections to the QBTZ black hole due to the $(r, \varphi)$ twist are obtained by substituting equation \eqref{Aqbtz} into equation \eqref{NCrphibtz}. The components of the NC corrected metric are as follows,
\begin{equation}
    \begin{split}
        \hat{g}_{00}\left( {x,\Theta }\right) = & \left(  M - \frac{r^2}{l^2} + 2Q^2 \ln \left(\frac{r}{l}\right) \right) , \\
        & -  \frac{(l Q -  r) (l Q + r)  \bigl(l^2 (M -  Q^2) - 4 r^2 + 2 l^2 Q^2 \log \left(\frac{r}{l}\right)\bigr)}{2 l^4 r^2} \Theta^2 + \mathcal{O}(\Theta^3) \\
        \hat{g}_{11}\left( {x,\Theta }\right) &= \left( - M + \frac{r^2}{l^2} -  2Q^2 \ln \left(\frac{r}{l}\right)\right)^{-1} \\
        & -  \frac{l^2 \bigl(l^2 Q^2 (M + Q^2) + (M - 3 Q^2) r^2 + 2 Q^2 (l^2 Q^2 + r^2) \log \left(\frac{r}{l}\right)\bigr)}{2 r^2 \bigl(l^2 M -  r^2 + 2 l^2 Q^2 \log \left(\frac{r}{l}\right)\bigr)^2} \Theta^2  + \mathcal{O}(\Theta^3) , \\
        \hat{g}_{22}\left( {x,\Theta }\right) &= r^2 + \frac{1}{16} \Biggl[- M + 4 Q^2 + \frac{45 r^2}{l^2} - 2 Q^2 \log\left(\frac{r}{l}\right) -  \frac{4 (- l^2 Q^2 + r^2)^2}{l^4 M -  l^2 r^2 + 2 l^4 Q^2 \log \left(\frac{r}{l}\right)}\Biggr] \Theta^2 \nonumber \\ 
& + \mathcal{O}(\Theta^3) .
    \end{split}
\end{equation}
\subsection{($t, r $) twist}
For the ($r, \varphi$) twist the non commutativity the parameter $\Theta ^{\mu \nu} $ is given by,
\begin{equation}
\Theta ^{\mu \nu }=\left(
\begin{array}{ccc}
0 & \Theta & 0 \\
-\Theta & 0 & 0  \\
0 & 0 & 0 \\
\end{array}
\right) ,\quad \mu ,\nu =0, 1,2.  \label{3drthetatwist}
\end{equation}
This choice produces the following commutation relation between the coordinates:
\begin{equation}
    [t\stackrel{\star}{,} r] = i \, \Theta. 
\end{equation}
The non-zero components of the tetrad $\hat e _\mu ^{a}$ are
\begin{equation}
    \begin{split}
        \hat{e}_{0}^{0} &= A-\frac{1}{16}  A \left(12 A^2 A''^2+20 A'^4+ 12 A^{(3)}  A^2 A'+69 A A'^2A''\right) \Theta ^2+O\left(\Theta ^3\right) , \\
        \hat{e}_{1}^{1} &= \frac{1}{A}+\frac{1}{16}  \left(-5 A A''^2-5 A^{(3)}  A A'-17 A'^2 A''\right)\Theta ^2+O\left(\Theta ^3\right) , \\
        \hat{e}_{2}^{2} &= r+ \left(-A A'^3-\frac{17}{16} A^2 A' A''\right)\Theta ^2+O\left(\Theta ^3\right) , \\
        \hat{e}_{1}^{0} &=  -\frac{1}{4} i \Theta  A''+O\left(\Theta ^3\right) , \\
         \hat{e}_{0}^{1} &= -\frac{1}{4}   i A \left(2 A A''+3 A'^2\right) \Theta+O\left(\Theta ^3\right).
    \end{split}
\end{equation}
The deformed metric $\hat g _{\mu \nu} (x, \Theta) $ is obtained by using the definition \eqref{metricdef}. Up to second order in $\Theta$, the metric is diagonal, and the non-zero components are
\begin{equation}
    \begin{split}
        \hat{g}_{00}\left( {x,\Theta }\right) &= -A^2+ \left(\frac{7}{4}  A^4 A''^2+\frac{49}{16}  A^2 A'^4+\frac{3}{2} A^{(3)}  A^4A'+\frac{75}{8}  A^3A'^2 A''\right) \Theta ^2+O\left(\Theta ^3\right) ,\\
        \hat{g}_{11}\left( {x,\Theta }\right) &= \frac{1}{A^2}+\left(-\frac{11}{16} A''^2-\frac{5}{8} A^{(3)}  A'-\frac{17 A'^2 A''}{8 A}\right)\Theta ^2 +O\left(\Theta ^3\right) \label{NCrtbtz} , \\ 
        \hat{g}_{22}\left( {x,\Theta }\right) &= r^2+ \left(-2 r A A'^3-\frac{17}{8}  r A^2 A' A''\right)\Theta ^2+O\left(\Theta ^3\right) .
    \end{split}
\end{equation}
\subsubsection{BTZ black hole}
The NC corrections to the BTZ black hole due to the $(t, r)$ twist are obtained by substituting \eqref{Abtz} in \eqref{NCrtbtz}. The NC corrected metric components are given by,
\begin{equation}
    \begin{split}
        \hat{g}_{00}\left( {x,\Theta }\right) &= \left(  M - \frac{r^2}{l^2} \right) -  \frac{(28 l^4 M^2 - 78 l^2 M r^2 + 49 r^4) }{16 l^6 \left( l^2 M -  r^2 \right)} \Theta^2 + \mathcal{O}(\Theta^3) , \\
        \hat{g}_{11}\left( {x,\Theta }\right) &= \left( - M + \frac{r^2}{l^2} \right)^{-1} + \frac{M (11 l^2 M - 4 r^2) }{16 (l^2 M -  r^2)^3} \Theta^2 + \mathcal{O}(\Theta^3) , \\
        \hat{g}_{22}\left( {x,\Theta }\right) &= r^2 -  \frac{1}{8 l^4}\left(17 r^2 + \frac{r^4}{l^2 M -  r^2}\right) \Theta^2 + \mathcal{O}(\Theta^3) .
    \end{split}
\end{equation}
\subsubsection{QBTZ black hole}
The NC corrections to the QBTZ black hole due to the $(t, r)$ twist are obtained by substituting \eqref{Aqbtz} in \eqref{NCrtbtz}. The NC corrected metric components are given by,
\begin{align*}
\allowdisplaybreaks
        \hat{g}_{00}\left( {x,\Theta }\right) = & \left(  M - \frac{r^2}{l^2} + 2Q^2 \ln \left(\frac{r}{l}\right) \right) + \frac{1}{16 l^6 r^4} \Biggl[2 \bigl(l^6 (-38 M Q^4 + 11 Q^6) + l^4 Q^2 (-4 M + 27 Q^2) r^2 \\
        & - 7 l^2 (2 M + Q^2) r^4 + 25 r^6\bigr) - 8 (19 l^6 Q^6 + 2 l^4 Q^4 r^2 + 7 l^2 Q^2 r^4) \log\left(\frac{r}{l}\right) \\
        & + \frac{(- l^2 Q^2 + r^2)^4}{l^2 M -  r^2 + 2 l^2 Q^2 \log\left(\frac{r}{l}\right)}\Biggr]  \Theta^2 + \mathcal{O}(\Theta^3) , \\
        \hat{g}_{11}\left( {x,\Theta }\right) &= \left( - M + \frac{r^2}{l^2} -  2Q^2 \ln \left(\frac{r}{l}\right)\right)^{-1} \\
        & + \frac{1 }{16 l^2 r^4 \bigl(l^2 M -  r^2 + 2 l^2 Q^2 \log \left(\frac{r}{l}\right)\bigr)^3}  \Biggl[7 (- l^2 Q^2 + r^2)^4 \\
        & + 18 (- l^2 Q^2 + r^2)^2 (l^2 Q^2 + r^2) \Bigl(l^2 M -  r^2 + 2 l^2 Q^2 \log \left(\frac{r}{l}\right)\Bigr) \\
        &+ (31 l^4 Q^4 + 2 l^2 Q^2 r^2 + 11 r^4) \Bigl(l^2 M -  r^2 + 2 l^2 Q^2 \log \left(\frac{r}{l}\right)\Bigr)^2\Biggr] \Theta^2  + \mathcal{O}(\Theta^3) , \\
        \hat{g}_{22}\left( {x,\Theta }\right) & = r^2 + \frac{}{8 l^4 r^2} \Biggl[ (l Q -  r) (l Q + r) \Biggl(17 (l^2 Q^2 + r^2) + \frac{(- l^2 Q^2 + r^2)^2}{l^2 M -  r^2 + 2 l^2 Q^2 \log \left(\frac{r}{l}\right)}\Biggr) \Biggr]  \Theta^2+ \mathcal{O}(\Theta^3) .
    \end{align*}
\subsection{($t, \varphi $) twist}
For the ($t, \varphi$) twist the non commutativity the parameter $\Theta ^{\mu \nu} $ is given by,
\begin{equation}
\Theta ^{\mu \nu }=\left(
\begin{array}{ccc}
0 & 0 & \Theta \\
0 & 0 & 0  \\
-\Theta & 0 & 0 \\
\end{array}
\right) ,\quad \mu ,\nu =0, 1,2.  \label{3dtphitwist}
\end{equation}
This choice produces the following commutation relation between the coordinates:
\begin{equation}
    [t\stackrel{\star}{,} \varphi] = i \, \Theta. 
\end{equation}
The non-zero components of the tetrad $\hat e _\mu ^{a}$ are
\begin{equation}
    \begin{split}
        \hat{e}_{0}^{0} &= A-\frac{1}{16}  \left(A^4 A'^2 \left(A-2 r A'\right)\right) \Theta ^2 +O\left(\Theta ^3\right) , \\
        \hat{e}_{1}^{1} &= \frac{1}{A}+\frac{1}{8} A^2 \left(A \left(A-2 r A'\right) A''-A'^2 \left(r A'+A\right)\right)  \Theta ^2 +O\left(\Theta ^3\right) , \\
        \hat{e}_{2}^{2} &= r+\frac{1}{16}  A^4 A' \left(2 A-r A'\right) \Theta ^2+O\left(\Theta ^3\right) , \\
        \hat{e}_{2}^{0} &= -\frac{1}{4} i   r A^2 A' \Theta +O\left(\Theta ^3\right), \\
         \hat{e}_{0}^{2} &= -\frac{1}{4} i   A^3 A' \Theta +O\left(\Theta ^3\right) .
    \end{split}
\end{equation}

The deformed metric $\hat g _{\mu \nu} (x, \Theta) $ is obtained by using the definition \eqref{metricdef}. Up to second order in $\Theta$, the metric is diagonal, and the non-zero components are
\begin{equation}
    \begin{split}
        \hat{g}_{00}\left( {x,\Theta }\right) &= -A^2+\left(\frac{3}{16} A^6 A'^2-\frac{1}{4} r A^5 A'^3\right) \Theta ^2  +O\left(\Theta ^3\right) , \\
        \hat{g}_{11}\left( {x,\Theta }\right) &= \frac{1}{A^2}+ \left(\frac{1}{4} A^3 A''-\frac{1}{4} A^2 A'^2-\frac{1}{4} r A A'^3-\frac{1}{2} r A^2 A' A''\right) \Theta ^2 +O\left(\Theta ^3\right) , \label{NCtphibtz}\\ 
        \hat{g}_{22}\left( {x,\Theta }\right) &= r^2+ \left(\frac{1}{4} r A^5 A'-\frac{3}{16} r^2 A^4 A'^2\right) \Theta ^2 +O\left(\Theta ^3\right) .
    \end{split}
\end{equation}
\subsubsection{BTZ black hole}
The NC corrections to the BTZ black hole due to the $(t, \varphi)$ twist are obtained by substituting \eqref{Abtz} in \eqref{NCtphibtz}. The NC corrected metric components are given by,
\begin{equation}
    \begin{split}
        \hat{g}_{00}\left( {x,\Theta }\right) &= \left(  M - \frac{r^2}{l^2} \right)  + \frac{r^2 (l^2 M -  r^2) (3 l^2 M + r^2) }{16 l^8} \Theta^2 + \mathcal{O}(\Theta^3) , \\
        \hat{g}_{11}\left( {x,\Theta }\right) &= \left( - M + \frac{r^2}{l^2} \right)^{-1}  + \frac{1}{4} \left(- \frac{2 r^2}{l^4} + \frac{M^2}{- l^2 M + r^2}\right) \Theta^2 + \mathcal{O}(\Theta^3) , \\
        \hat{g}_{22}\left( {x,\Theta }\right) &= r^2 + \frac{(4 l^4 M^2 r^2 - 5 l^2 M r^4 + r^6) }{16 l^6} \Theta^2 + \mathcal{O}(\Theta^3) .
    \end{split}
\end{equation}
\subsubsection{QBTZ black hole}
The NC corrections to the QBTZ black hole due to the $(t, \varphi)$ twist are obtained by substituting \eqref{Aqbtz} in \eqref{NCtphibtz}. The NC corrected metric components are given by,
\begin{align*}
\allowdisplaybreaks
        \hat{g}_{00}\left( {x,\Theta }\right) = & \left(  M - \frac{r^2}{l^2} + 2Q^2 \ln \left(\frac{r}{l}\right) \right)  + \frac{1}{16 l^8 r^2}\Biggl[(- l^2 Q^2 + r^2)^2  \left(l^2 M -  r^2 + 2 l^2 Q^2 \log\left(\frac{r}{l}\right)\right) \\
        & \times \left(l^2 (3 M - 4 Q^2) + r^2 + 6 l^2 Q^2 \log\left(\frac{r}{l}\right)\right) \Biggr]\Theta^2  + \mathcal{O}(\Theta^3) , \\
        \hat{g}_{11}\left( {x,\Theta }\right) = & \left( - M + \frac{r^2}{l^2} -  2Q^2 \ln \left(\frac{r}{l}\right)\right)^{-1}  -  \frac{ 1}{4 l^4 r^2}\Biggl[l^4 M Q^2 + l^2 (M - 5 Q^2) r^2 + 3 r^4 \\
        & + 2 l^2 Q^2 (l^2 Q^2 + r^2) \log\left(\frac{r}{l}\right)  -  \frac{(l^2 Q^2 -  r^2)^3}{l^2 M -  r^2 + 2 l^2 Q^2 \log\left(\frac{r}{l}\right)}\Biggr] \Theta^2  + \mathcal{O}(\Theta^3) , \\
        \hat{g}_{22}\left( {x,\Theta }\right) = & r^2+\frac{1}{16 l^6}\Biggl[ \left(r^2-l^2 Q^2\right) \left(l^2 \left(3 Q^2-4 M\right)-8 l^2 Q^2 \log \left(\frac{r}{l}\right)+r^2\right) \\
        & \times \left(l^2 (-M)-2 l^2 Q^2 \log \left(\frac{r}{l}\right)+r^2\right)\Biggr] \Theta ^2  +O\left(\Theta ^3\right) .
    \end{align*}

\bibliography{BibTex}

\providecommand{\href}[2]{#2}\begingroup\raggedright\begin{thebibliography}{10}

\bibitem{Abbott:2016blz}
{\scshape LIGO Scientific, Virgo} collaboration, \emph{{Observation of Gravitational Waves from a Binary Black Hole Merger}}, \href{https://doi.org/10.1103/PhysRevLett.116.061102}{\emph{Phys. Rev. Lett.} {\bfseries 116} (2016) 061102} [\href{https://arxiv.org/abs/1602.03837}{{\ttfamily 1602.03837}}].

\bibitem{TheLIGOScientific:2016src}
{\scshape LIGO Scientific, Virgo} collaboration, \emph{{Tests of general relativity with GW150914}}, \href{https://doi.org/10.1103/PhysRevLett.116.221101}{\emph{Phys. Rev. Lett.} {\bfseries 116} (2016) 221101} [\href{https://arxiv.org/abs/1602.03841}{{\ttfamily 1602.03841}}].

\bibitem{Abbott:2016nmj}
{\scshape LIGO Scientific, Virgo} collaboration, \emph{{GW151226: Observation of Gravitational Waves from a 22-Solar-Mass Binary Black Hole Coalescence}}, \href{https://doi.org/10.1103/PhysRevLett.116.241103}{\emph{Phys. Rev. Lett.} {\bfseries 116} (2016) 241103} [\href{https://arxiv.org/abs/1606.04855}{{\ttfamily 1606.04855}}].

\bibitem{Akiyama:2019bqs}
{\scshape Event Horizon Telescope} collaboration, \emph{{First M87 Event Horizon Telescope Results. IV. Imaging the Central Supermassive Black Hole}}, \href{https://doi.org/10.3847/2041-8213/ab0e85}{\emph{Astrophys. J. Lett.} {\bfseries 875} (2019) L4} [\href{https://arxiv.org/abs/1906.11241}{{\ttfamily 1906.11241}}].

\bibitem{Akiyama:2019cqa}
{\scshape Event Horizon Telescope} collaboration, \emph{{First M87 Event Horizon Telescope Results. I. The Shadow of the Supermassive Black Hole}}, \href{https://doi.org/10.3847/2041-8213/ab0ec7}{\emph{Astrophys. J.} {\bfseries 875} (2019) L1} [\href{https://arxiv.org/abs/1906.11238}{{\ttfamily 1906.11238}}].

\bibitem{Akiyama:2019fyp}
{\scshape Event Horizon Telescope} collaboration, \emph{{First M87 Event Horizon Telescope Results. V. Physical Origin of the Asymmetric Ring}}, \href{https://doi.org/10.3847/2041-8213/ab0f43}{\emph{Astrophys. J. Lett.} {\bfseries 875} (2019) L5} [\href{https://arxiv.org/abs/1906.11242}{{\ttfamily 1906.11242}}].

\bibitem{EventHorizonTelescope:2022wkp}
{\scshape Event Horizon Telescope} collaboration, \emph{{First Sagittarius A* Event Horizon Telescope Results. I. The Shadow of the Supermassive Black Hole in the Center of the Milky Way}}, \href{https://doi.org/10.3847/2041-8213/ac6674}{\emph{Astrophys. J. Lett.} {\bfseries 930} (2022) L12} [\href{https://arxiv.org/abs/2311.08680}{{\ttfamily 2311.08680}}].

\bibitem{EventHorizonTelescope:2022xqj}
{\scshape Event Horizon Telescope} collaboration, \emph{{First Sagittarius A* Event Horizon Telescope Results. VI. Testing the Black Hole Metric}}, \href{https://doi.org/10.3847/2041-8213/ac6756}{\emph{Astrophys. J. Lett.} {\bfseries 930} (2022) L17} [\href{https://arxiv.org/abs/2311.09484}{{\ttfamily 2311.09484}}].

\bibitem{Vagnozzi:2022moj}
S.~Vagnozzi et~al., \emph{{Horizon-scale tests of gravity theories and fundamental physics from the Event Horizon Telescope image of Sagittarius A}}, \href{https://doi.org/10.1088/1361-6382/acd97b}{\emph{Class. Quant. Grav.} {\bfseries 40} (2023) 165007} [\href{https://arxiv.org/abs/2205.07787}{{\ttfamily 2205.07787}}].

\bibitem{Utiyama:1956sy}
R.~Utiyama, \emph{{Invariant theoretical interpretation of interaction}}, \href{https://doi.org/10.1103/PhysRev.101.1597}{\emph{Phys. Rev.} {\bfseries 101} (1956) 1597}.

\bibitem{Kibble:1961ba}
T.~W.~B. Kibble, \emph{{Lorentz invariance and the gravitational field}}, \href{https://doi.org/10.1063/1.1703702}{\emph{J. Math. Phys.} {\bfseries 2} (1961) 212}.

\bibitem{Stelle:1979aj}
K.~S. Stelle and P.~C. West, \emph{{Spontaneously Broken De Sitter Symmetry and the Gravitational Holonomy Group}}, \href{https://doi.org/10.1103/PhysRevD.21.1466}{\emph{Phys. Rev. D} {\bfseries 21} (1980) 1466}.

\bibitem{MacDowell:1977jt}
S.~W. MacDowell and F.~Mansouri, \emph{{Unified Geometric Theory of Gravity and Supergravity}}, \href{https://doi.org/10.1103/PhysRevLett.38.739}{\emph{Phys. Rev. Lett.} {\bfseries 38} (1977) 739}.

\bibitem{Ivanov:1981wn}
E.~A. Ivanov and J.~Niederle, \emph{{Gauge Formulation of Gravitation Theories. 1. The Poincare, De Sitter and Conformal Cases}}, \href{https://doi.org/10.1103/PhysRevD.25.976}{\emph{Phys. Rev. D} {\bfseries 25} (1982) 976}.

\bibitem{Ivanov:1981wm}
E.~A. Ivanov and J.~Niederle, \emph{{Gauge Formulation of Gravitation Theories. 2. The Special Conformal Case}}, \href{https://doi.org/10.1103/PhysRevD.25.988}{\emph{Phys. Rev. D} {\bfseries 25} (1982) 988}.

\bibitem{Kaku:1977pa}
M.~Kaku, P.~K. Townsend and P.~van Nieuwenhuizen, \emph{{Gauge Theory of the Conformal and Superconformal Group}}, \href{https://doi.org/10.1016/0370-2693(77)90552-4}{\emph{Phys. Lett. B} {\bfseries 69} (1977) 304}.

\bibitem{Fradkin:1985am}
E.~S. Fradkin and A.~A. Tseytlin, \emph{{Conformal Supergravity}}, \href{https://doi.org/10.1016/0370-1573(85)90138-3}{\emph{Phys. Rept.} {\bfseries 119} (1985) 233}.

\bibitem{Chamseddine:1976bf}
A.~H. Chamseddine and P.~C. West, \emph{{Supergravity as a Gauge Theory of Supersymmetry}}, \href{https://doi.org/10.1016/0550-3213(77)90018-9}{\emph{Nucl. Phys. B} {\bfseries 129} (1977) 39}.

\bibitem{Chamseddine:2005td}
A.~H. Chamseddine, \emph{{Applications of the gauge principle to gravitational interactions}}, \href{https://doi.org/10.1142/S0219887806001041}{\emph{Int. J. Geom. Meth. Mod. Phys.} {\bfseries 3} (2006) 149} [\href{https://arxiv.org/abs/hep-th/0511074}{{\ttfamily hep-th/0511074}}].

\bibitem{Manolakos:2019fle}
G.~Manolakos, P.~Manousselis and G.~Zoupanos, \emph{{Four-dimensional Gravity on a Covariant Noncommutative Space}}, \href{https://doi.org/10.1007/JHEP08(2020)001}{\emph{JHEP} {\bfseries 08} (2020) 001} [\href{https://arxiv.org/abs/1902.10922}{{\ttfamily 1902.10922}}].

\bibitem{weyl1952space}
H.~Weyl, \emph{Space, Time, Matter}, Dover Books on Advanced Mathematics. Dover Publications, 1952.

\bibitem{cartan1981theory}
E.~Cartan and A.~Mercier, \emph{The Theory of Spinors}, Dover Books on Mathematics. Dover Publications, 1981.

\bibitem{Ashtekar:1987gu}
A.~Ashtekar, \emph{{New Hamiltonian Formulation of General Relativity}}, \href{https://doi.org/10.1103/PhysRevD.36.1587}{\emph{Phys. Rev. D} {\bfseries 36} (1987) 1587}.

\bibitem{Smolin:2004sx}
L.~Smolin, \emph{{An Invitation to loop quantum gravity}},  in \emph{{3rd International Symposium on Quantum Theory and Symmetries}}, pp.~655--682, 8, 2004, \href{https://doi.org/10.1142/9789812702340_0078}{DOI} [\href{https://arxiv.org/abs/hep-th/0408048}{{\ttfamily hep-th/0408048}}].

\bibitem{Freedman:2012zz}
D.~Z. Freedman and A.~Van~Proeyen, \emph{{Supergravity}}. Cambridge Univ. Press, Cambridge, UK, 5, 2012, \href{https://doi.org/10.1017/CBO9781139026833}{10.1017/CBO9781139026833}.

\bibitem{Snyder:1946qz}
H.~S. Snyder, \emph{{Quantized space-time}}, \href{https://doi.org/10.1103/PhysRev.71.38}{\emph{Phys. Rev.} {\bfseries 71} (1947) 38}.

\bibitem{Connes:1997cr}
A.~Connes, M.~R. Douglas and A.~S. Schwarz, \emph{{Noncommutative geometry and matrix theory: Compactification on tori}}, \href{https://doi.org/10.1088/1126-6708/1998/02/003}{\emph{JHEP} {\bfseries 02} (1998) 003} [\href{https://arxiv.org/abs/hep-th/9711162}{{\ttfamily hep-th/9711162}}].

\bibitem{Douglas:1997fm}
M.~R. Douglas and C.~M. Hull, \emph{{D-branes and the noncommutative torus}}, \href{https://doi.org/10.1088/1126-6708/1998/02/008}{\emph{JHEP} {\bfseries 02} (1998) 008} [\href{https://arxiv.org/abs/hep-th/9711165}{{\ttfamily hep-th/9711165}}].

\bibitem{Cheung:1998nr}
Y.-K.~E. Cheung and M.~Krogh, \emph{{Noncommutative geometry from 0-branes in a background B field}}, \href{https://doi.org/10.1016/S0550-3213(98)00380-0}{\emph{Nucl. Phys. B} {\bfseries 528} (1998) 185} [\href{https://arxiv.org/abs/hep-th/9803031}{{\ttfamily hep-th/9803031}}].

\bibitem{Chu:1998qz}
C.-S. Chu and P.-M. Ho, \emph{{Noncommutative open string and D-brane}}, \href{https://doi.org/10.1016/S0550-3213(99)00199-6}{\emph{Nucl. Phys. B} {\bfseries 550} (1999) 151} [\href{https://arxiv.org/abs/hep-th/9812219}{{\ttfamily hep-th/9812219}}].

\bibitem{Schomerus:1999ug}
V.~Schomerus, \emph{{D-branes and deformation quantization}}, \href{https://doi.org/10.1088/1126-6708/1999/06/030}{\emph{JHEP} {\bfseries 06} (1999) 030} [\href{https://arxiv.org/abs/hep-th/9903205}{{\ttfamily hep-th/9903205}}].

\bibitem{Ardalan:1998ce}
F.~Ardalan, H.~Arfaei and M.~M. Sheikh-Jabbari, \emph{{Noncommutative geometry from strings and branes}}, \href{https://doi.org/10.1088/1126-6708/1999/02/016}{\emph{JHEP} {\bfseries 02} (1999) 016} [\href{https://arxiv.org/abs/hep-th/9810072}{{\ttfamily hep-th/9810072}}].

\bibitem{Seiberg:1999vs}
N.~Seiberg and E.~Witten, \emph{{String theory and noncommutative geometry}}, \href{https://doi.org/10.1088/1126-6708/1999/09/032}{\emph{JHEP} {\bfseries 09} (1999) 032} [\href{https://arxiv.org/abs/hep-th/9908142}{{\ttfamily hep-th/9908142}}].

\bibitem{Banerjee:2007th}
R.~Banerjee, P.~Mukherjee and S.~Samanta, \emph{{Lie algebraic noncommutative gravity}}, \href{https://doi.org/10.1103/PhysRevD.75.125020}{\emph{Phys. Rev. D} {\bfseries 75} (2007) 125020} [\href{https://arxiv.org/abs/hep-th/0703128}{{\ttfamily hep-th/0703128}}].

\bibitem{Chamseddine:2000si}
A.~H. Chamseddine, \emph{{Deforming Einstein's gravity}}, \href{https://doi.org/10.1016/S0370-2693(01)00272-6}{\emph{Phys. Lett. B} {\bfseries 504} (2001) 33} [\href{https://arxiv.org/abs/hep-th/0009153}{{\ttfamily hep-th/0009153}}].

\bibitem{Chaichian:2007dr}
M.~Chaichian, M.~R. Setare, A.~Tureanu and G.~Zet, \emph{{On Black Holes and Cosmological Constant in Noncommutative Gauge Theory of Gravity}}, \href{https://doi.org/10.1088/1126-6708/2008/04/064}{\emph{JHEP} {\bfseries 04} (2008) 064} [\href{https://arxiv.org/abs/0711.4546}{{\ttfamily 0711.4546}}].

\bibitem{Chaichian:2007we}
M.~Chaichian, A.~Tureanu and G.~Zet, \emph{{Corrections to Schwarzschild solution in noncommutative gauge theory of gravity}}, \href{https://doi.org/10.1016/j.physletb.2008.01.029}{\emph{Phys. Lett. B} {\bfseries 660} (2008) 573} [\href{https://arxiv.org/abs/0710.2075}{{\ttfamily 0710.2075}}].

\bibitem{Mukherjee:2007fa}
P.~Mukherjee and A.~Saha, \emph{{Deformed Reissner-Nordstrom solutions in noncommutative gravity}}, \href{https://doi.org/10.1103/PhysRevD.77.064014}{\emph{Phys. Rev. D} {\bfseries 77} (2008) 064014} [\href{https://arxiv.org/abs/0710.5847}{{\ttfamily 0710.5847}}].

\bibitem{Linares:2019gqf}
R.~Linares, M.~Maceda and O.~S\'anchez-Santos, \emph{{Thermodynamical properties of a noncommutative anti\textendash{}de Sitter\textendash{}Einstein-Born-Infeld spacetime from gauge theory of gravity}}, \href{https://doi.org/10.1103/PhysRevD.101.044008}{\emph{Phys. Rev. D} {\bfseries 101} (2020) 044008} [\href{https://arxiv.org/abs/1909.08001}{{\ttfamily 1909.08001}}].

\bibitem{Touati:2023ubi}
A.~Touati and S.~Zaim, \emph{{Schwarzschild black hole surrounded by a cavity and phase transition in the non-commutative gauge theory of gravity}}, \href{https://doi.org/10.1016/j.astropartphys.2024.102988}{\emph{Astropart. Phys.} {\bfseries 161} (2024) 102988} [\href{https://arxiv.org/abs/2311.12186}{{\ttfamily 2311.12186}}].

\bibitem{Touati:2023cxy}
A.~Touati and S.~Zaim, \emph{{Quantum tunneling from Schwarzschild black hole in non-commutative gauge theory of gravity}}, \href{https://doi.org/10.1016/j.physletb.2023.138335}{\emph{Phys. Lett. B} {\bfseries 848} (2024) 138335} [\href{https://arxiv.org/abs/2310.02445}{{\ttfamily 2310.02445}}].

\bibitem{Touati:2022kuf}
A.~Touati and S.~Zaim, \emph{{On Modified First Law of Black hole Thermodynamics in The Non-Commutative Gauge Theory}},  \href{https://arxiv.org/abs/2205.13052}{{\ttfamily 2205.13052}}.

\bibitem{Touati:2022zbm}
A.~Touati and S.~Zaim, \emph{{Thermodynamic properties of Schwarzschild black hole in non-commutative gauge theory of gravity}}, \href{https://doi.org/10.1016/j.aop.2023.169394}{\emph{Annals Phys.} {\bfseries 455} (2023) 169394} [\href{https://arxiv.org/abs/2204.01901}{{\ttfamily 2204.01901}}].

\bibitem{Touati:2021eem}
A.~Touati and S.~Zaim, \emph{{Geodesic equation in non-commutative gauge theory of gravity*}}, \href{https://doi.org/10.1088/1674-1137/ac75ca}{\emph{Chin. Phys. C} {\bfseries 46} (2022) 105101} [\href{https://arxiv.org/abs/2112.01558}{{\ttfamily 2112.01558}}].

\bibitem{Zhao:2023uam}
Y.~Zhao, Y.~Cai, S.~Das, G.~Lambiase, E.~N. Saridakis and E.~C. Vagenas, \emph{{Quasinormal Modes in Noncommutative Schwarzschild black holes}},  \href{https://arxiv.org/abs/2301.09147}{{\ttfamily 2301.09147}}.

\bibitem{Heidari:2023egu}
N.~Heidari, H.~Hassanabadi, A.~A.~A. Filho and J.~Kur\'\i{}uz, \emph{{Exploring Non--commutativity as a Perturbation in the Schwarzschild Black Hole: Quasinormal Modes, Scattering, and Shadows}},  \href{https://arxiv.org/abs/2308.03284}{{\ttfamily 2308.03284}}.

\bibitem{Heidari:2023bww}
N.~Heidari, H.~Hassanabadi, A.~A.~A. Filho, J.~Kriz, S.~Zare and P.~J. Porfirio, \emph{{Gravitational signatures of a non-commutative stable black hole}}, \href{https://doi.org/10.1016/j.dark.2023.101382}{\emph{Phys. Dark Univ.} {\bfseries 43} (2024) 101382} [\href{https://arxiv.org/abs/2305.06838}{{\ttfamily 2305.06838}}].

\bibitem{Touati:2024kmv}
A.~Touati and Z.~Slimane, \emph{{Lyapunov exponents and geodesic stability of Schwarzschild black hole in the non-commutative gauge theory of gravity}},  \href{https://arxiv.org/abs/2405.01743}{{\ttfamily 2405.01743}}.

\bibitem{AraujoFilho:2025viz}
A.~A. Ara\'ujo~Filho, N.~Heidari and A.~\"Ovg\"un, \emph{{Axisymmetric black hole in a non-commutative gauge theory: classical and quantum gravity effects}},  \href{https://arxiv.org/abs/2502.12039}{{\ttfamily 2502.12039}}.

\bibitem{Zet:2003bv}
G.~Zet, V.~Manta and S.~Babeti, \emph{{De Sitter gauge theory of gravitation}}, \href{https://doi.org/10.1142/S0129183103004188}{\emph{Int. J. Mod. Phys. C} {\bfseries 14} (2003) 41}.

\bibitem{zet2006computer}
G.~Zet, V.~Manta, S.~Oancea, I.~Radinschi and B.~Ciobanu, \emph{A computer aided study of de-sitter gauge theory of gravitation}, {\emph{Mathematical and computer modelling} {\bfseries 43} (2006) 458}.

\bibitem{Wiesendanger:1995hm}
C.~Wiesendanger, \emph{{A Poincare gauge theory of gravitation in Minkowski space-time}},  in \emph{{International School of Cosmology and Gravitation: 14th Course: Quantum Gravity}}, pp.~366--381, 5, 1995, \href{https://arxiv.org/abs/gr-qc/9604043}{{\ttfamily gr-qc/9604043}}.

\bibitem{Hersent:2022gry}
K.~Hersent, P.~Mathieu and J.-C. Wallet, \emph{{Gauge theories on quantum spaces}}, \href{https://doi.org/10.1016/j.physrep.2023.03.002}{\emph{Phys. Rept.} {\bfseries 1014} (2023) 1} [\href{https://arxiv.org/abs/2210.11890}{{\ttfamily 2210.11890}}].

\bibitem{Ciric:2017rnf}
M.~D. \'Ciri\'c, N.~Konjik and A.~Samsarov, \emph{{Noncommutative scalar quasinormal modes of the Reissner\textendash{}Nordstr\"om black hole}}, \href{https://doi.org/10.1088/1361-6382/aad201}{\emph{Class. Quant. Grav.} {\bfseries 35} (2018) 175005} [\href{https://arxiv.org/abs/1708.04066}{{\ttfamily 1708.04066}}].

\bibitem{DimitrijevicCiric:2018blz}
M.~Dimitrijevic~Ciric, N.~Konjik, M.~A. Kurkov, F.~Lizzi and P.~Vitale, \emph{{Noncommutative field theory from angular twist}}, \href{https://doi.org/10.1103/PhysRevD.98.085011}{\emph{Phys. Rev. D} {\bfseries 98} (2018) 085011} [\href{https://arxiv.org/abs/1806.06678}{{\ttfamily 1806.06678}}].

\bibitem{DimitrijevicCiric:2019hqq}
M.~Dimitrijevi\'c~\'Ciri\'c, N.~Konjik and A.~Samsarov, \emph{{Noncommutative scalar field in the nonextremal Reissner-Nordstr\"om background: Quasinormal mode spectrum}}, \href{https://doi.org/10.1103/PhysRevD.101.116009}{\emph{Phys. Rev. D} {\bfseries 101} (2020) 116009} [\href{https://arxiv.org/abs/1904.04053}{{\ttfamily 1904.04053}}].

\bibitem{Gupta:2022oel}
K.~S. Gupta, T.~Juri\'c, A.~Samsarov and I.~Smoli\'c, \emph{{Noncommutativity and logarithmic correction to the black hole entropy}}, \href{https://doi.org/10.1007/JHEP02(2023)060}{\emph{JHEP} {\bfseries 02} (2023) 060} [\href{https://arxiv.org/abs/2209.07168}{{\ttfamily 2209.07168}}].

\bibitem{Herceg:2023pmc}
N.~Herceg, T.~Juri\'c, A.~Samsarov and I.~Smoli\'c, \emph{{Metric perturbations in noncommutative gravity}}, \href{https://doi.org/10.1007/JHEP06(2024)130}{\emph{JHEP} {\bfseries 06} (2024) 130} [\href{https://arxiv.org/abs/2310.06038}{{\ttfamily 2310.06038}}].

\bibitem{Juric:2022bnm}
T.~Juri\'c and F.~Po\v{z}ar, \emph{{Noncommutative Correction to the Entropy of Charged BTZ Black Hole}}, \href{https://doi.org/10.3390/sym15020417}{\emph{Symmetry} {\bfseries 15} (2023) 417} [\href{https://arxiv.org/abs/2212.06496}{{\ttfamily 2212.06496}}].

\bibitem{Hrelja:2024tgj}
A.~Hrelja, T.~Juri\'c and F.~Po\v{z}ar, \emph{{Entropy of black holes, charged probes and noncommutative generalization}},  \href{https://arxiv.org/abs/2407.13233}{{\ttfamily 2407.13233}}.

\bibitem{Carter:2009nex}
B.~Carter, \emph{{Republication of: Black hole equilibrium states}}, \href{https://doi.org/10.1007/s10714-009-0888-5}{\emph{Gen. Rel. Grav.} {\bfseries 41} (2009) 2873}.

\bibitem{Carroll:2004st}
S.~M. Carroll, \emph{{Spacetime and Geometry}: {An Introduction to General Relativity}}. Cambridge University Press, 7, 2019, \href{https://doi.org/10.1017/9781108770385}{10.1017/9781108770385}.

\bibitem{Wald:1999vt}
R.~M. Wald, \emph{{The thermodynamics of black holes}}, \href{https://doi.org/10.12942/lrr-2001-6}{\emph{Living Rev. Rel.} {\bfseries 4} (2001) 6} [\href{https://arxiv.org/abs/gr-qc/9912119}{{\ttfamily gr-qc/9912119}}].

\bibitem{Hawking:1973uf}
S.~W. Hawking and G.~F.~R. Ellis, \emph{{The Large Scale Structure of Space-Time}}, Cambridge Monographs on Mathematical Physics. Cambridge University Press, 2, 2023, \href{https://doi.org/10.1017/9781009253161}{10.1017/9781009253161}.

\bibitem{Sudarsky:1992ty}
D.~Sudarsky and R.~M. Wald, \emph{{Extrema of mass, stationarity, and staticity, and solutions to the Einstein Yang-Mills equations}}, \href{https://doi.org/10.1103/PhysRevD.46.1453}{\emph{Phys. Rev. D} {\bfseries 46} (1992) 1453}.

\bibitem{Wang:2008ut}
D.~Wang, R.~B. Zhang and X.~Zhang, \emph{{Quantum deformations of Schwarzschild and Schwarzschild-de Sitter spacetimes}}, \href{https://doi.org/10.1088/0264-9381/26/8/085014}{\emph{Class. Quant. Grav.} {\bfseries 26} (2009) 085014} [\href{https://arxiv.org/abs/0809.0614}{{\ttfamily 0809.0614}}].

\bibitem{Sun:2010nas}
W.~Sun, D.~Wang, N.~Xie, R.~B. Zhang and X.~Zhang, \emph{{Gravitational collapse of spherically symmetric stars in noncommutative general relativity}}, \href{https://doi.org/10.1140/epjc/s10052-010-1342-2}{\emph{Eur. Phys. J. C} {\bfseries 69} (2010) 271} [\href{https://arxiv.org/abs/0910.2777}{{\ttfamily 0910.2777}}].

\bibitem{Herceg:2023zlk}
N.~Herceg, T.~Juri\'c, A.~Samsarov, I.~Smoli\'c and K.~S. Gupta, \emph{{Gravitational probe of quantum spacetime}}, \href{https://doi.org/10.1016/j.physletb.2024.138716}{\emph{Phys. Lett. B} {\bfseries 854} (2024) 138716} [\href{https://arxiv.org/abs/2310.06018}{{\ttfamily 2310.06018}}].

\bibitem{Anacleto:2020efy}
M.~A. Anacleto, F.~A. Brito, B.~R. Carvalho and E.~Passos, \emph{{Noncommutative correction to the entropy of BTZ black hole with GUP}}, \href{https://doi.org/10.1155/2021/6633684}{\emph{Adv. High Energy Phys.} {\bfseries 2021} (2021) 6633684} [\href{https://arxiv.org/abs/2010.09703}{{\ttfamily 2010.09703}}].

\bibitem{Anacleto:2020zfh}
M.~A. Anacleto, F.~A. Brito, S.~S. Cruz and E.~Passos, \emph{{Noncommutative correction to the entropy of Schwarzschild black hole with GUP}}, \href{https://doi.org/10.1142/S0217751X21500287}{\emph{Int. J. Mod. Phys. A} {\bfseries 36} (2021) 2150028} [\href{https://arxiv.org/abs/2010.10366}{{\ttfamily 2010.10366}}].

\bibitem{Anacleto:2015kca}
M.~A. Anacleto, F.~A. Brito, A.~G. Cavalcanti, E.~Passos and J.~Spinelly, \emph{{Quantum correction to the entropy of noncommutative BTZ black hole}}, \href{https://doi.org/10.1007/s10714-018-2344-x}{\emph{Gen. Rel. Grav.} {\bfseries 50} (2018) 23} [\href{https://arxiv.org/abs/1510.08444}{{\ttfamily 1510.08444}}].

\bibitem{C:2024cnk}
S.~J. C., K.~R., K.~Hegde, K.~M. Ajith, S.~Punacha and A.~N. Kumara, \emph{{Perturbations of black holes surrounded by anisotropic matter field}}, \href{https://doi.org/10.1103/PhysRevD.111.064034}{\emph{Phys. Rev. D} {\bfseries 111} (2025) 064034} [\href{https://arxiv.org/abs/2411.11629}{{\ttfamily 2411.11629}}].

\bibitem{Campos:2021sff}
J.~A.~V. Campos, M.~A. Anacleto, F.~A. Brito and E.~Passos, \emph{{Quasinormal modes and shadow of noncommutative black hole}}, \href{https://doi.org/10.1038/s41598-022-12343-w}{\emph{Sci. Rep.} {\bfseries 12} (2022) 8516} [\href{https://arxiv.org/abs/2103.10659}{{\ttfamily 2103.10659}}].

\bibitem{Anacleto:2019tdj}
M.~A. Anacleto, F.~A. Brito, J.~A.~V. Campos and E.~Passos, \emph{{Absorption and scattering of a noncommutative black hole}}, \href{https://doi.org/10.1016/j.physletb.2020.135334}{\emph{Phys. Lett. B} {\bfseries 803} (2020) 135334} [\href{https://arxiv.org/abs/1907.13107}{{\ttfamily 1907.13107}}].

\bibitem{Carlip:1995qv}
S.~Carlip, \emph{{The (2+1)-Dimensional black hole}}, \href{https://doi.org/10.1088/0264-9381/12/12/005}{\emph{Class. Quant. Grav.} {\bfseries 12} (1995) 2853} [\href{https://arxiv.org/abs/gr-qc/9506079}{{\ttfamily gr-qc/9506079}}].

\bibitem{Martinez:1999qi}
C.~Martinez, C.~Teitelboim and J.~Zanelli, \emph{{Charged rotating black hole in three space-time dimensions}}, \href{https://doi.org/10.1103/PhysRevD.61.104013}{\emph{Phys. Rev. D} {\bfseries 61} (2000) 104013} [\href{https://arxiv.org/abs/hep-th/9912259}{{\ttfamily hep-th/9912259}}].

\bibitem{Clement:1995zt}
G.~Clement, \emph{{Spinning charged BTZ black holes and selfdual particle - like solutions}}, \href{https://doi.org/10.1016/0370-2693(95)01464-0}{\emph{Phys. Lett. B} {\bfseries 367} (1996) 70} [\href{https://arxiv.org/abs/gr-qc/9510025}{{\ttfamily gr-qc/9510025}}].

\bibitem{Unver:2010uw}
O.~Unver and O.~Gurtug, \emph{{Quantum singularities in (2+1) dimensional matter coupled black hole spacetimes}}, \href{https://doi.org/10.1103/PhysRevD.82.084016}{\emph{Phys. Rev. D} {\bfseries 82} (2010) 084016} [\href{https://arxiv.org/abs/1004.2572}{{\ttfamily 1004.2572}}].

\end{thebibliography}\endgroup

\end{document}